\documentclass[11pt]{article}
\usepackage{amsmath,amsfonts,amssymb,graphicx,epsfig,pdflscape,multirow,theorem} 
\usepackage{arydshln,cite}
\usepackage{simplewick}
\numberwithin{equation}{section}
\graphicspath{{./eps/}}
\long\def\ignore#1{}

\theorembodyfont{\itshape} 
\theoremheaderfont{\scshape}
\theoremstyle{plain}

\numberwithin{equation}{section}

\newcommand{\nc}{\newcommand}
\nc{\bib}{\bibitem}
\nc{\be}{\begin{equation}}
\nc{\ee}{\end{equation}}
\nc{\nn}{\nonumber\\ }
\nc{\chit}{\raisebox{0.25ex}{$\chi$}}
\nc{\chih}{\raisebox{0.25ex}{$\hat\chi$}}
\nc{\g}{\mathfrak{g}}
\nc{\gh}{\widehat{\mathfrak{g}}}
\nc{\Ac}{\mathcal{A}}
\nc{\Bc}{\mathcal{B}}
\nc{\Acb}{\bar{\mathcal{A}}}
\nc{\Ic}{\mathcal{I}}
\nc{\Mc}{\mathcal{M}}
\nc{\Nc}{\mathcal{N}}
\nc{\Oc}{\mathcal{O}}
\nc{\Qc}{\mathcal{Q}}
\nc{\Vc}{\mathcal{V}}
\nc{\Wc}{\mathcal{W}}
\nc{\Vir}{\mathfrak{Vir}}
\nc{\Virb}{\overline{\mathfrak{Vir}}}
\nc{\pa}{\partial}
\nc{\eps}{\epsilon}
\nc{\Tb}{\bar{T}}
\nc{\Wb}{\bar{W}}
\nc{\cb}{\bar{c}}
\nc{\Ab}{\bar{A}}
\nc{\Bb}{\bar{B}}
\nc{\Cb}{\bar{C}}
\nc{\Jb}{\bar{J}}
\nc{\zb}{\bar{z}}
\nc{\kb}{\bar{k}}
\nc{\C}{\mathbb{C}}
\nc{\WA}{W\!A}

\nc{\Ah}{\widehat{A}}
\nc{\Bh}{\widehat{B}}
\nc{\Ch}{\widehat{C}}
\nc{\Jh}{\widehat{J}}
\nc{\Th}{\widehat{T}}
\nc{\Uh}{\widehat{U}}
\nc{\Vh}{\widehat{V}}
\nc{\Wh}{\widehat{W}}

\nc{\gast}{\!\ast}
\nc{\s}{\;\!\!}

\nc{\La}{\Lambda}
\nc{\Lh}{\hat{\Lambda}}

\nc{\I}{\mathbb{I}}

\newcommand\bea{\begin{eqnarray}}
\newcommand\eea{\end{eqnarray}}

\begin{document}

\topmargin -15mm
\oddsidemargin 05mm

%
%

\title{\mbox{}\vspace{0in}
\bf 
\Huge
Galilean contractions of $W$-algebras
\\[-.2cm]
}
\date{}
\author{}

\maketitle


\begin{center}
{\vspace{-5mm}\LARGE J{\o}rgen Rasmussen,\, Christopher Raymond}
\\[.5cm]
{\em School of Mathematics and Physics, University of Queensland}\\
{\em St Lucia, Brisbane, Queensland 4072, Australia}
\\[.4cm] 
{\tt j.rasmussen\,@\,uq.edu.au}
\qquad
{\tt christopher.raymond\,@\,uqconnect.edu.au}
\end{center}

%
%

\vspace{0.5cm}
\begin{abstract}
Infinite-dimensional Galilean conformal algebras can be constructed by contracting pairs of 
symmetry algebras in conformal field theory, such as $W$-algebras. Known examples include contractions of 
pairs of the Virasoro algebra, its $N=1$ superconformal extension, or the $W_3$ algebra. 
Here, we introduce a contraction prescription of the corresponding operator-product algebras, or equivalently, 
a prescription for contracting tensor products of vertex algebras. With this, we work out the 
Galilean conformal algebras arising from contractions of $N=2$ and $N=4$ superconformal algebras 
as well as of the $W$-algebras $W(2,4)$, $W(2,6)$, $W_4$, and $W_5$. 
The latter results provide evidence for the existence of a whole new class of $W$-algebras 
which we call Galilean $W$-algebras. We also apply the contraction prescription to affine Lie algebras and find 
that the ensuing Galilean affine algebras admit a Sugawara construction. The corresponding central charge
is level-independent and given by twice the dimension of the underlying finite-dimensional Lie algebra.
Finally, applications of our results to the characterisation of structure constants in $W$-algebras are proposed.
\end{abstract}

\newpage
\tableofcontents

\newpage
\section{Introduction}
\label{Sec:Introduction}

The earliest example of a (nonlinear) $W$-algebra is Zamolodchikov's $W_3$ algebra~\cite{Zam85}. 
It arose in a study of additional symmetries in two-dimensional conformal field theories~\cite{BPZ84},
and is generated by the Virasoro generator and a primary field of conformal weight $3$.
The construction introduced a new type of extended conformal symmetry,
and gave hope to classify rational conformal field theories with arbitrary central charge\cite{FS87,MS89}.
The $W_3$ algebra was subsequently studied extensively and a myriad of generalisations were proposed, 
see~\cite{Pop91,BS92,BS95,Wat97} and references therein.

This flurry of activity took place primarily in the physics literature but lasted only a little more than a decade.
The field has witnessed a resurgence of interest in recent years, however, due to significant progress in the 
mathematical understanding and description of $W$-algebras,
see~\cite{DeSK06,Wan11,Ara16} and references therein, and to new applications in physics.

Most of the recent developments in physics have taken place in the context of Vasiliev's higher-spin 
theory~\cite{Vas03,BCIV05}, see also~\cite{BBS12,Gio16}. In particular, it has been 
conjectured~\cite{KP02} that the three-dimensional $O(N)$ vector model at its critical points is 
dual to a Vasiliev theory on AdS$_4$. It has also been argued~\cite{BH86,HenRey10,CFPT10,CFP11} 
that the asymptotic algebra of higher-spin gravity
on AdS$_3$ exhibits a $W$-symmetry. By essentially morphing these ideas and observations, 
it was subsequently conjectured~\cite{GG11} that a particular Vasiliev theory on AdS$_3$ is dual to the
$W_N$ minimal models~\cite{FL88} at large $N$. Studies of this conjecture have led to significant advances in the
understanding of $W_N$ algebras, their representation theory and their large-$N$ limit.

There has also been a renewed 
interest~\cite{ABS97,BC07,BG09,BGMM10,BT10,Bag10,BF12,ST13,KRR14,BB14,DGH14,DGH14a,BO14,BO15,BGPR17} 
in the infinite-dimensional symmetry algebras of asymptotically flat spacetimes at null infinity~\cite{BBM62,Sac62}, 
known as BMS algebras. In $2+1$ dimensions, this algebra is occasionally denoted by BMS$_3$ and is an
extension of the Virasoro algebra by an additional spin-$2$ generator. Here, we shall refer to this algebra
as the {\em Galilean Virasoro algebra}.
Further extensions of the BMS and Galilean Virasoro algebras have also been considered. In particular,
supersymmetric extensions already appeared in~\cite{GP77,AGS86} and have been discussed more recently 
in~\cite{AL09,Sak10,BM10,Man10,BDMT14,BDMT15,AS16,CGOR16,BJMN16,MR16,CT16,BCP16,Man16,BJLMN16,LM16}.
Also related to higher-spin theories, the Galilean Virasoro algebra has been extended by adding 
two spin-$3$ generators, resulting in a higher-spin BMS$_3$ algebra or 
a {\em Galilean $W_3$ algebra}~\cite{ABFGR13,GMPT13,CGOR16a}. 
Other $W$-algebraic extensions have also been considered~\cite{GRR15}.

The Galilean Virasoro (or conformal) algebra can be obtained~\cite{BC07,BG09,HR10,BGMM10,BF12} as 
an In\"on\"u-Wigner contraction~\cite{Seg51,IW53,Sal61,Gil06} of a commuting pair of Virasoro algebras. 
Likewise, the Galilean $W_3$ algebra follows~\cite{ABFGR13,GMPT13} by contracting a pair of $W_3$ algebras.
In these constructions, the contractions are performed in terms of the modes generating the infinite-dimensional
Virasoro and $W_3$ algebras. However, in conformal field theory, it is often more convenient to 
work with the corresponding operator-product algebras (OPAs) given in terms of 
operator-product expansions~\cite{BPZ84,DiFMSbook,Schbook}. 

Here, we thus develop a general contraction prescription, 
which we call {\em Galilean contraction}, of the tensor product
of two copies of the same OPA; they are only allowed to differ in the values of their central charges.
This offers a systematic approach to the construction of infinite-dimensional Galilean algebras.
It also allows us to derive conditions necessary for the existence of such a Galilean algebra
and to determine properties of the ensuing algebras.
As the notions of OPAs and vertex algebras~\cite{Bor86} are intimately related, the approach and results
apply to tensor products of vertex algebras as well. 
The concrete computations and analyses of the operator-product expansions underlying the construction
are conveniently performed using the Mathematica~\cite{Wol15} package developed in~\cite{Thi91,Thi95}.

The layout of the present paper is as follows.
In Section~\ref{Sec:OPA}, we review the notion of an OPA, with emphasis on the role and structure of
quasi-primary fields. 
Some details not readily available in the literature are included.
We introduce star products as an economical way to encode the information stored in
an operator-product expansion, and discuss the OPA pendant to the tensor product of vertex algebras.

In Section~\ref{Sec:Galilean}, we introduce the Galilean contraction prescription of tensor products of pairs of OPAs
and outline some general properties of the ensuing Galilean OPAs. With this, we reproduce the known Galilean
conformal and $N=1$ superconformal algebras 
(see Section~\ref{Sec:GCA} for comments on alternative superconformal extensions).
We also identify a class of OPAs, called simply-extended OPAs, whose Galilean counterparts are readily obtained.

In Section~\ref{Sec:Comp}, we consider various superconformal extensions of the Galilean Virasoro algebra
and demonstrate compatibility between Galilean contractions and certain other operations on OPAs, such
as topological twisting and a variety of known In\"on\"u-Wigner contractions. Affine Lie algebras are examples of
simply-extended OPAs and give rise to Galilean affine Lie algebras. Their contractions are shown to
be compatible with the Sugawara construction, yielding a realisation of the Galilean Virasoro algebra
with central charge given by twice the dimension of the underlying finite-dimensional Lie algebra.

In Section~\ref{Sec:AAG}, we explore the properties of an OPA necessary for the Galilean contraction
prescription to be well-defined. Several general conditions on the structure constants are identified, and
the corresponding structure constants of the ensuing Galilean OPA are determined.

In Section~\ref{Sec:W}, we apply the Galilean contraction prescription to the $W$-algebras
$W_3$, $W(2,4)$, $W(2,6)$, $W_4$, $W_5$, and the infinitely generated $W$-algebra $W_\infty$
and some of its extensions. Details of the $W$-algebras $W(2,6)$, $W_4$ and $W_5$ are reviewed in
Appendix~\ref{Sec:Walg}, and their Galilean counterparts are given in Appendix~\ref{Sec:GW}.
In preparation for the derivation of these new Galilean $W$-algebras, we discuss the vacuum module
of the Galilean Virasoro algebra and work out a number of quasi-primary fields involving the Galilean Virasoro
generators and a primary field.

Section~\ref{Sec:Discussion} contains some concluding remarks.

\section{Operator-product algebras} 
\label{Sec:OPA}

We are interested in symmetry algebras in conformal field theory, such as infinite-dimensional Lie algebras and 
$W$-algebras. For convenience, we combine the algebra generators $A_n^\mu$ into generating fields,
\be
 A^\mu(z)=\sum_{n\,\in-\Delta_{A^\mu}+\mathbb{Z}}\!\!A_n^\mu\,z^{-n-\Delta_{A^\mu}},
\label{Aa}
\ee
where $\mu$ labels the set of fields, and where $\Delta_A$ is the conformal weight of $A$. 
Accordingly, instead of commutator algebras, 
we thus work with the corresponding operator-product algebras (OPAs) formed by the fields.
As already indicated in (\ref{Aa}), we shall assume that an OPA is generated by a set of scaling fields,
see (\ref{scaling}), and that their conformal weights are (half-)integer:
\be
 \Delta\in\tfrac{1}{2}\mathbb{N}_0.
\label{DA}
\ee
As we recall below, the Virasoro field algebra is such an example; it is generated by the single field $T$ of 
conformal weight $\Delta_T=2$. We shall admit both even (or bosonic) and odd (or fermionic) fields.

\subsection{Operator-product expansions} 
\label{Sec:OPE}

The operator-product expansion (OPE) 
of the two fields $A$ and $B$ of conformal weights $\Delta_A$ and $\Delta_B$, respectively, can be written as
\be
 A(z)B(w)=\sum_{n=-\infty}^{\Delta_{A,B}}\frac{[AB]_n(w)}{(z-w)^n}.
\label{AzBw}
\ee
Here, $\Delta_{A,B}$ is some integer satisfying $0\leq\Delta_{A,B}\leq\Delta_A+\Delta_B$, while
$[AB]_n$ are fields of conformal weights
\be
 \Delta_{[AB]_n}=\Delta_A+\Delta_B-n.
\ee
As we shall assume that the identity field $\I$ is the only symmetry generator of conformal weight $0$, 
the OPE (\ref{AzBw}) can be re-expressed as
\be
 A(z)B(w)=\frac{f_{A,B}\I}{(z-w)^{\Delta_A+\Delta_B}}
  +\sum_{n=-\infty}^{\Delta_A+\Delta_B-1}\frac{[AB]_n(w)}{(z-w)^{n}},
\label{AB}
\ee
for some $f_{A,B}\in\mathbb{C}$. Since the identity field $\I$ is of even parity, 
the conservation of parity in an OPE implies that $f_{A,B}=0$ if $A$ and $B$ have different parities.
Adopting a widespread tradition in conformal field theory, we will usually not indicate the identity field explicitly in
an OPE like (\ref{AB}).

The singular part of the OPE (\ref{AzBw}) is referred to as the corresponding contraction and is usually denoted by
\be
 \bcontraction{}{A}{(z)}{B} A(z)B(w)=\sum_{n\,=\,1}^{\Delta_{A,B}}\frac{[AB]_n(w)}{(z-w)^n},
\label{ABcont}
\ee
whereas the nonsingular part is given by the normal-ordered product
\be
 :\!A(z)B(w)\!:=\sum_{n=0}^{\infty}\,[AB]_{-n}(w)(z-w)^n.
\label{:AB:}
\ee
Because the contraction encapsulates all the nontrivial information about the OPE, one often writes
\be
 A(z)B(w)\sim\bcontraction{}{A}{(z)}{B} A(z)B(w),
\ee
simply ignoring the nonsingular part of the OPE. In particular, the commutator relations for the modes of the fields 
$A$ and $B$ are determined by the contraction (\ref{ABcont}) alone.
As we will use the term ``contraction" in more than one context, we shall refer to (\ref{ABcont}) as
an {\em OPE contraction}.

If the OPEs of a set of fields closes on itself in a `sufficiently nice' way, the result is an OPA.
Largely following~\cite{Thi95}, this will be qualified in Section~\ref{Sec:OPAdef}.

\subsection{Algebraic structure} 
\label{Sec:OPAdef}

An OPA $\Ac$ is a $\mathbb{Z}_2$-graded vector space with a distinguished vector (or element) $\I$ 
and an even linear map $\pa:\Ac\to\Ac$.
Moreover, $\Ac$ is equipped with a bilinear operation (or product) for every integer $n\in\mathbb{Z}$,
\be
 [\ \,]_n:\Ac\times\Ac\to\Ac,\qquad (A,B)\mapsto[AB]_n,
\ee
where, for any pair $A,B\in\Ac$, $[AB]_n=0$ for $n$ sufficiently large. These data satisfy the following conditions.
First, $\I$ is the {\em identity} in the sense that
\be
 [\I A]_n=A\,\delta_{n,0}.
\label{identity}
\ee
Second, the map $\pa$ is a derivation and thus follows the Leibniz rule\footnote{It is understood that $\pa$ only 
acts on the symbol to its immediate right, that is, $\pa AB\equiv(\pa A)B$, for example.}
\be
 \pa[AB]_n=[\pa AB]_n+[A\pa B]_n.
\ee
Third, with $|A|$ denoting the parity of $A$, {\em commutativity} of $\Ac$ reads
\be
 [BA]_n=(-1)^{|A|\,|B|}\sum_{\ell\,\geq\,n}\frac{(-1)^\ell}{(\ell-n)!}\,\pa^{\ell-n}[AB]_\ell,\qquad \forall\,n\in\mathbb{Z}.
\label{commutativity}
\ee
Fourth, {\em associativity} of $\Ac$ amounts to the relations
\be
 [A[BC]_m]_n=(-1)^{|A|\,|B|}\,[B[AC]_n]_m+\sum_{\ell\,>\,0}\begin{pmatrix} n-1\\ \ell-1 \end{pmatrix}
  [[AB]_\ell C]_{m+n-\ell},\qquad \forall\,m,n\in\mathbb{Z}.
\label{associativity}
\ee
Finally, $\Ac$ may depend on a set of complex parameters, here referred to as {\em central parameters}.
We will denote such parameters by $c_i$, where $i$ is in some index set $I_\Ac$.

One may regard $\Ac$ as being defined only once values for the central parameters have been fixed.
In that case, two distinct sets of values for the central parameters define inequivalent OPAs. Here, however,
we find it convenient to leave the central parameters undetermined and thus let $\Ac$ depend on them.

The normal ordering, indicated by $(AB)$, of $A,B\in\Ac$ is defined by
\be
 (AB)=[AB]_0,
\ee
in accordance with (\ref{:AB:}). Normal ordering of three or more fields is performed iteratively, right-nesting the
normal-ordered parts such that
\be
 (ABC)=(A(BC)).
\ee
In general, a field $[AB]_n$ is a linear combination of normal-ordered expressions of
fields and derivatives thereof.

In this work, the distinction between composite and non-composite fields will play a crucial role.
We thus say that a field $C$ is a {\em composite field} if it is equal to $[AB]_0$ for some $A,B\neq\I$, except if
$A=B$ (both) odd. Accordingly, {\em non-composite fields} are fields which cannot be written in a way involving
normal ordering without the use of the identity field or a pair of identical odd fields.

A set of fields is a set of {\em generators} of $\Ac$ if every element of $\Ac$ can be written as a linear combination 
of normal-ordered products of the generators (including $\I$) and their derivatives. A generating set of fields 
consists of {\em elementary fields} only, if no field in the set can be constructed in terms of the other generators by 
taking linear combinations, computing derivatives or forming normal-ordered products.

An OPA is {\em conformal} if it contains a field $T$ generating the Virasoro (field) algebra, see (\ref{Vir}).
A field in a conformal OPA is a {\em scaling field} with respect to $T$ if
\be
 [TA]_2=\Delta_AA,\qquad [TA]_1=\pa A.
\label{scaling}
\ee 
Such a scaling field is called {\em quasi-primary} if
\be
 [TA]_3=0,
\ee 
and {\em primary} if 
\be
 [TA]_n=0,\qquad n\geq3.
\ee
We denote by $\Qc_\Ac$ the vector space of quasi-primary fields in the conformal OPA $\Ac$.
We will in general assume that a conformal OPA is generated by a set of quasi-primary fields,
except for the topologically twisted $N=2$ superconformal algebra in Section~\ref{Sec:Top}. 
In fact, all the other conformal OPAs we consider explicitly are generated by a set of {\em primary} fields in 
addition to $T$.

A scaling field $A$ of conformal weight $\Delta_A$ as in (\ref{DA}) admits the mode expansion
\be
 A(z)=\sum_{n\in-\Delta_A+\mathbb{Z}}A_nz^{-\Delta_A-n},
\ee
as in (\ref{Aa}).
It is standard to denote the modes of the Virasoro generator by $L_n$ such that
\be
 T(z)=\sum_{n\in\mathbb{Z}}L_nz^{-2-n}.
\ee

\subsection{Quasi-primary field basis}
\label{Sec:Quasi}

For a general OPA $\Ac$, we denote the set of quasi-primary fields by
\be
 \Qc_\Ac=\mathrm{span}(\Bc_\Ac),
\ee
where $\Bc_\Ac$ is a (formal) basis for $\Qc_\Ac$. In the case of a conformal OPA, one of these quasi-primary fields
is the Virasoro field $T$, and in the case $\Ac=\Vir$, $\Qc_\Ac$ is merely the linear span of the quasi-primary 
descendants of the identity (including the identity itself). As discussed in the following, with focus on $\Vir$, a basis 
for the set of quasi-primary fields can be constructed using the state-field correspondence. Although the results 
should be known to experts, some of the details do not seem to be readily available in the literature.

The Virasoro Verma module $\Vc$, with highest-weight vector $|0\rangle$ of conformal weight $\Delta=0$,
has a submodule generated from the singular vector $L_{-1}|0\rangle$. 
The corresponding quotient module, $\Mc$, is $\mathbb{Z}_+$-graded by $L_0$,
\be
 \Mc=\bigoplus_{n=0}^\infty \Mc_n,
\label{Msum}
\ee
and its conformal character is given by
\be
 \chi(q)=\mathrm{Tr}_{\!_\Mc} q^{L_0}
 =\frac{1-q}{\prod_{n=1}^\infty(1-q^n)}
 =\prod_{n=2}^\infty\frac{1}{1-q^n}
 =\sum_{n=0}^\infty a_nq^n,\qquad
 a_n=\dim\Mc_n. 
\ee
Concretely, the character begins as
\be
 \chi(q)=1+q^2+q^3+2q^4+2q^5+4q^6+4q^7+7q^8+8q^9+12q^{10}
  +\Oc(q^{11}).
\ee

The set
\be
 \Mc_n'=\{v\in\Mc_n\,|\,L_1v=0\}
\label{Mnp}
\ee
consists of the {\em quasi-primary vectors} in $\Mc_n$, 
i.e.~the vectors associated with quasi-primary fields under the state-field correspondence.
With respect to (\ref{Mnp}), the vector space $\Mc_n$ decomposes as
\be
 \Mc_n=\Mc_n'\oplus L_{-1}\Mc_{n-1}.
\label{MMLM}
\ee
To see this, we first show that $\Mc_n\subseteq\Mc_n'+L_{-1}\Mc_{n-1}$. To this end, we let $v\in\Mc_n$ and 
verify that
\be
 L_1\Big(v+\sum_{k=1}^{n-2} b_kL_{-1}^kL_1^kv\Big)=0
\ee
is solved by
\be
 b_k=\frac{(-1)^k}{k!\prod_{j=1}^k(2n+3j-1)},\qquad k=1,\ldots,n-2.
\ee
We can thus write
\be
 v=\Big(\!\underbrace{v+\sum_{k=1}^{n-2} b_kL_{-1}^kL_1^kv}_{\in\Mc_n'}\!\Big)
  +L_{-1}\Big(\!\underbrace{-\sum_{k=1}^{n-2} b_kL_{-1}^{k-1}L_1^kv}_{\in\Mc_{n-1}}\!\Big).
\ee
The converse inclusion, $\Mc_n\supseteq\Mc_n'+L_{-1}\Mc_{n-1}$, is obvious, while the remaining condition,
$\Mc_n'\cap L_{-1}\Mc_{n-1}=0$, 
follows by induction on $n$. Indeed, the induction assumption implies that 
\be
 L_{-1}\Mc_{n-1}=\bigoplus_{k=1}^{n-2}L_{-1}^k\Mc_{n-k}',
\ee
so any nonzero $v\in L_{-1}\Mc_{n-1}$ can be written as $v=\sum_{k=1}^{n-2}a_kL_{-1}^kw_{n-k}$
for some nonzero $w_{n-k}\in\Mc_{n-k}'$ and $a_k$ not all zero. We then have
\be
 L_1v=\sum_{k=1}^{n-2}a_kk(2n-k-1)L_{-1}^{k-1}w_{n-k}\in\bigoplus_{k=1}^{n-2}L_{-1}^{k-1}\Mc_{n-k}'.
\ee
Since $a_k$ not all zero, it follows that $L_1v\neq0$,
so $v\notin\Mc_n'$. Hence, $\Mc_n'\cap L_{-1}\Mc_{n-1}=0$, thereby establishing (\ref{MMLM}).

Now, since the linear map
\be
 L_{-1}:\Mc_{n-1}\to\Mc_n
\ee
is injective (see~\cite{Li96}, for example), it follows that
$\dim(L_{-1}\Mc_{n-1})=\dim\Mc_{n-1}$ for $n>1$, so
\be
 \dim\Mc_n'=a_n-a_{n-1},\qquad n>1.
\ee
In particular,
\be
\begin{array}{c}
 \dim\Mc_0'=1,\quad
 \dim\Mc_1'=0,\quad
 \dim\Mc_2'=1,\quad
 \dim\Mc_3'=0,\quad
 \dim\Mc_4'=1,
\\[.3cm]
 \dim\Mc_5'=0,\quad
 \dim\Mc_6'=2,\quad
 \dim\Mc_7'=0,\quad
 \dim\Mc_8'=3,\quad
 \dim\Mc_9'=1,\quad
 \dim\Mc_{10}'=4.
\end{array}
\ee

In terms of fields, the decomposition (\ref{Msum}) reads
\be
 \Vir=\bigoplus_{n=0}^\infty\Vir_n,
\ee
where $\Vir_n$ consists of the descendant fields of the identity of conformal weight $n$.
The field version of the decomposition (\ref{MMLM}) then reads
\be
 \Vir_n=\Qc_n\oplus\pa\,\Vir_{n-1},
\ee
where $\Qc_n\subset\Qc_\Vir$ denotes the set of quasi-primary fields of conformal weight $n$.
This means, in particular, that a quasi-primary field cannot be written
as a derivative of another field, as pointed out in~\cite{BFKNRV91} citing earlier observations by Nahm. 
Moreover, every normal-ordered product (corresponding to a vector $v\in\Mc_n$)
not equal to the derivative of a quasi-primary field (thus assuming that $v\notin\Mc_n'$)
gives rise to a unique quasi-primary field (corresponding to a vector in $\Mc_n'$)
by addition of some derivative of a quasi-primary field (corresponding to a vector in $L_{-1}\Mc_{n-1}$).
For example, the normal-ordered product $(TT)$ gives rise to the well-known quasi-primary field
\be
 \La^{\s2,2}=(TT)-\frac{3}{10}\pa^2T.
\label{La22}
\ee
Conversely, the absence of a quasi-primary field of weight $5$, for example, follows from
\be
  (L_{-1}L_{-2}^2-L_{-2}L_{-1}L_{-2})|0\rangle=L_{-3}L_{-2}|0\rangle=(-L_{-5}+L_{-2}L_{-1}L_{-2})|0\rangle,
\ee
as these relations correspond to
\be
 \pa(TT)-(T\pa T)=(\pa TT)=-\tfrac{1}{6}\pa^3T+(T\pa T),
\ee
implying that
\be
 (T\pa T)=\pa[\tfrac{1}{2}\La^{\s2,2}+\tfrac{7}{30}\pa^2T],\qquad
 (\pa TT)=\pa[\tfrac{1}{2}\La^{\s2,2}+\tfrac{1}{15}\pa^2T].
\ee

Let $\Bc_n$ denote a basis for $\Qc_n$, where, by construction, $\dim\Qc_n=\dim\Mc_n'$. 
A Hamel basis, $\Bc_\Vir$, for the infinite-dimensional vector space $\Vir$ can thus be obtained as
\be
 \Bc_\Vir=\bigcup_{n=0}^\infty\Bc_n.
\label{BVir}
\ee
In our discussion of $W$-algebras in Section~\ref{Sec:W} and Appendix~\ref{Sec:Walg}, 
we will need explicit bases for $\Bc_n$, for $n\leq10$; for ease of reference, and to emphasise some key features, 
we discuss these bases here. 

First, of weight $6$, we have the five normal-ordered products
\be
 (TTT)\equiv(T(TT)),\quad((TT)T),\quad(\pa^2TT),\quad(\pa T\pa T),\quad(T\pa^2T).
\label{TTT6}
\ee
Since $\dim\Mc_6=4$, these products are not linearly independent. Indeed,
\be
 0=(TTT)-((TT)T)+(2-\tfrac{c}{4})(\pa^2TT)+2(\pa T\pa T)+\tfrac{c}{4}(T\pa^2T), 
\ee
so we could eliminate one of them from our considerations. For completeness, we will not do that.
Instead, the quasi-primary fields corresponding to the five normal-ordered products in (\ref{TTT6}) are 
\be
 \La^{\s2,2,2}=(TTT)-\tfrac{1}{4}\pa^2\La^{\s2,2}-\tfrac{2}{35}\pa^4T,\qquad
 \La^{\s(2,2),2}=((TT)T)-\tfrac{5}{4}\pa^2\La^{\s2,2}-\tfrac{32+7c}{168}\,\pa^4T,
\label{La222}
\ee
and
\begin{align}
 \La^{\s2'',2}&=(\pa^2TT)-\tfrac{5}{18}\pa^2\La^{\s2,2}-\tfrac{1}{42}\pa^4T,
\label{La2''2}
\\[.15cm]
 \La^{\s2',2'}&=(\pa T\pa T)-\tfrac{2}{9}\pa^2\La^{\s2,2}-\tfrac{3}{70}\pa^4T,
\\[.15cm]
 \La^{\s2,2''}&=(T\pa^2T)-\tfrac{5}{18}\pa^2\La^{\s2,2}-\tfrac{4}{21}\pa^4T.
\end{align}
However, these fields are related as
\be
 \La^{\s2,2,2}=\La^{\s(2,2),2},\qquad
 \La^{\s2'',2}=-\La^{\s2',2'}=\La^{\s2,2''},
\ee
leaving only two linearly independent quasi-primary fields of weight $6$, in accordance with $\dim\Mc_6'=2$.
In the following, we shall work with $\La^{\s2,2,2}$ and $\La^{\s2'',2}$.
Likewise, a basis $\Bc_8$ for $\Mc_8'$ is given by
\begin{align}
 \La^{\s2,2,2,2}&=(TTTT)-\tfrac{3}{13}\pa^2\La^{\s2,2,2}-\tfrac{3}{52}\pa^2\La^{\s2'',2}
  -\tfrac{5}{132}\pa^4\La^{\s2,2}-\tfrac{1}{126}\pa^6T, 
\label{TTTT}
\\[.15cm]
 \La^{\s2'',2,2}&=(\pa^2TTT)-\tfrac{5}{39}\pa^2\La^{\s2,2,2}+\tfrac{2}{13}\pa^2\La^{\s2'',2}
  -\tfrac{2}{99}\pa^4\La^{\s2,2}-\tfrac{1}{420}\pa^6T,
\label{T2TT}
\\[.15cm]
 \La^{\s2'''',2}&=(\pa^4TT)-\tfrac{21}{13}\pa^2\La^{\s2'',2}-\tfrac{7}{66}\pa^4\La^{\s2,2}-\tfrac{1}{180}\pa^6T,
\label{T4T}
\end{align}
while a basis $\Bc_9$ for $\Mc_9'$ is given by
\be
 \La^{\s2''',2,2}=(\pa^3TTT)-\tfrac{9}{8}\pa\La^{\s2'',2,2}+\tfrac{3}{16}\pa\La^{\s2'''',2}
  -\tfrac{5}{91}\pa^3\La^{\s2,2,2}+\tfrac{27}{364}\pa^3\La^{\s2'',2}-\tfrac{1}{110}\pa^5\La^{\s2,2}
  -\tfrac{1}{1200}\pa^7T.
\ee
It may seem surprising at first that $(\pa TTTT)\in\Mc_9$, for example, does not give rise to an independent 
quasi-primary field $\La^{\s2',2,2,2}$ in $\Mc_9'$; 
after all, it is built from a normal-ordered product with four copies of $T$. However, so is $\La^{\s2,2,2,2}$, 
and one verifies that
\begin{align}
 \La^{\s2',2,2,2}&=(\pa TTTT)-\tfrac{1}{4}\pa\La^{\s2,2,2,2}
  +\tfrac{9}{32}\pa\La^{\s2'',2,2}+\tfrac{1}{64}\pa\La^{\s2'''',2}
\nonumber\\[.15cm]
  &-\tfrac{4}{91}\pa^3\La^{\s2,2,2}+\tfrac{1}{182}\pa^3\La^{\s2'',2}-\tfrac{43}{7920}\pa^5\La^{\s2,2}
  -\tfrac{1}{1050}\pa^7T
\nonumber\\[.15cm]
 &=-\tfrac{1}{4}\La^{\s2''',2,2}.
\end{align}
Finally, a basis $\Bc_{10}$ for $\Mc_{10}'$ is given by
\begin{align}
 \La^{\s2,2,2,2,2}&=(TTTTT)-\tfrac{5}{36}\pa\La^{\s2''',2,2}-\pa^2\big[\tfrac{15}{68}\La^{\s2,2,2,2}
  +\tfrac{45}{544}\La^{\s2'',2,2}+\tfrac{15}{1088}\La^{\s2'''',2}\big]
\nonumber\\[.15cm]
 &-\pa^4\big[\tfrac{17}{546}\La^{\s2,2,2}+\tfrac{5}{364}\La^{\s2'',2}\big]
  -\tfrac{1}{234}\pa^6\La^{\s2,2}-\tfrac{1}{1155}\pa^8T,
\label{TTTTT}
\\[.15cm]
 \La^{\s2'',2,2,2}&=(\pa^2TTTT)+\tfrac{7}{18}\pa\La^{\s2''',2,2}-\pa^2\big[\tfrac{5}{68}\La^{\s2,2,2,2}
  -\tfrac{27}{272}\La^{\s2'',2,2}+\tfrac{13}{544}\La^{\s2'''',2}\big]
\nonumber\\[.15cm]
 &+\pa^4\big[-\tfrac{1}{78}\La^{\s2,2,2}+\tfrac{1}{130}\La^{\s2'',2}\big]
  -\tfrac{73}{51480}\pa^6\La^{\s2,2}-\tfrac{1}{4950}\pa^8T,
\\[.15cm]
 \La^{\s2'''',2,2}&=(\pa^4TTT)-\tfrac{14}{9}\pa\La^{\s2''',2,2}+\pa^2\big[-\tfrac{63}{68}\La^{\s2'',2,2}
  +\tfrac{39}{136}\La^{\s2'''',2}\big]
\nonumber\\[.15cm]
 &+\pa^4\big[-\tfrac{1}{39}\La^{\s2,2,2}+\tfrac{2}{65}\La^{\s2'',2}\big]
  -\tfrac{119}{25740}\pa^6\La^{\s2,2}-\tfrac{2}{5775}\pa^8T,
\\[.15cm]
 \La^{\s2'''''',2}&=(\pa^6TT)-\tfrac{135}{34}\pa^2\La^{\s2'''',2}-\tfrac{18}{13}\pa^4\La^{\s2'',2}
  -\tfrac{7}{143}\pa^6\La^{\s2,2}-\tfrac{3}{1540}\pa^8T.
\label{T6T}
\end{align}

As indicated, our notation for the quasi-primary fields, such as $\La^{\s2''',2'',2,2}$, reflects the
form of the non-derivative normal-ordered product in its construction, in this example $(\pa^3T\pa^2TTT)$.
When discussing $W$-algebras in subsequent sections, we shall extend this notation to indicate quasi-primary
fields built from normal-ordered products of $W$-algebra generators and their derivatives. In $W_4$, for example,
$\La^{\s2',3'',3,4}$ is the quasi-primary field built from the normal-ordered product
$(\pa T\pa^2WWU)$, where $W$ and $U$ are primary fields of conformal weight $3$ and $4$, respectively.

A particular generalisation of the quasi-primary field $\La^{\s2,2}$ appears in most of the OPAs of our interest,
namely the one built from the normal-ordered product $(TA)$, where $A$ is a primary field of conformal weight 
$\Delta$. The ensuing quasi-primary field is
\be
 \La^{\s2,\Delta}=(TA)-\tfrac{3}{2(1+2\Delta)}\pa^2\!A.
\label{LTA}
\ee
Using this, we obtain the likewise frequently appearing quasi-primary field
\be
 \La^{\s2',\Delta}=(\pa TA)-\tfrac{2}{2+\Delta}\pa\La^{\s2,\Delta}-\tfrac{1}{(1+\Delta)(1+2\Delta)}\pa^3\!A,
\label{LT1A}
\ee
and subsequently
\begin{align}
 \La^{\s2,2,\Delta}&=(TTA)+\tfrac{3}{2(3+\Delta)}\pa\La^{\s2',\Delta}
 -\tfrac{3(1+\Delta)}{(2+\Delta)(5+2\Delta)}\pa^2\La^{\s2,\Delta}
 -\tfrac{3(2+3\Delta)}{4(1+\Delta)(1+2\Delta)(3+2\Delta)}\pa^4\!A,
\label{LTTA}
\\[.15cm]
 \La^{\s2'',\Delta}&=(\pa^2TA)-\tfrac{5}{3+\Delta}\pa\La^{\s2',\Delta}
 -\tfrac{10}{(2+\Delta)(5+2\Delta)}\pa^2\La^{\s2,\Delta}
 -\tfrac{5}{2(1+\Delta)(1+2\Delta)(3+2\Delta)}\pa^4\!A,
\label{LT2A}
\end{align}
and 
\begin{align}
 \La^{\s2',2,\Delta}&=(\pa TTA)
 -\tfrac{2}{4+\Delta}\pa\La^{\s2,2,\Delta}
 +\tfrac{3}{2(4+\Delta)}\pa\La^{\s2'',\Delta}
 -\tfrac{3(-4+\Delta)}{2(3+\Delta)(7+2\Delta)}\pa^2\La^{\s2',\Delta}
\nonumber\\[.15cm]
 & 
 -\tfrac{2(1+2\Delta)}{(2+\Delta)(3+\Delta)(5+2\Delta)}\pa^3\La^{\s2,\Delta}
 -\tfrac{3}{4(1+\Delta)(2+\Delta)(3+2\Delta)}\pa^5\!A,
\label{LT1TA}
\\[.15cm]
 \La^{\s2,2',\Delta}&=(T\pa TA)
 -\tfrac{2}{4+\Delta}\pa\La^{\s2,2,\Delta}
 -\tfrac{3(1+\Delta)}{2(3+\Delta)(7+2\Delta)}\pa^2\La^{\s2',\Delta}
\nonumber\\[.15cm]
 & 
 -\tfrac{7+4\Delta}{(2+\Delta)(3+\Delta)(5+2\Delta)}\pa^3\La^{\s2,\Delta}
 -\tfrac{3}{2(2+\Delta)(1+2\Delta)(3+2\Delta)}\pa^5\!A,
\label{LTT1A}
\\[.15cm]
 \La^{\s2''',\Delta}&=(\pa^3TA)
 -\tfrac{9}{4+\Delta}\pa\La^{\s2'',\Delta}
 -\tfrac{45}{(3+\Delta)(7+2\Delta)}\pa^2\La^{\s2',\Delta}
\nonumber\\[.15cm]
 & 
 -\tfrac{30}{(2+\Delta)(3+\Delta)(5+2\Delta)}\pa^3\La^{\s2,\Delta}
 -\tfrac{9}{2(1+\Delta)(2+\Delta)(1+2\Delta)(3+2\Delta)}\pa^5\!A.
\label{LT3A}
\end{align}

\subsection{Star relations}
\label{Sec:Notation}

Decompositions similar to (\ref{MMLM}) apply to vertex-operator algebras more generally, see~\cite{DLM96}.
In a purely bosonic conformal OPA $\Ac$, the OPE contraction of $A,B\in\Bc_\Ac$ of conformal weights 
$\Delta_A$ and $\Delta_B$, respectively, thus only involves quasi-primary fields and their derivatives. 
This allows us to write
\be
 \bcontraction{}{A}{(z)}{B} A(z)B(w)\sim\sum_{Q\in\Bc_\Ac}C_{A,B}^Q
 \Big(\sum_{n=0}^{\Delta_A+\Delta_B-\Delta_Q}
 \frac{\beta_{\Delta_A,\Delta_B}^{\Delta_Q;n}\pa^{n}Q(w)}{(z-w)^{\Delta_A+\Delta_B-\Delta_Q-n}}\Big),
\label{ABQ}
\ee
where $C_{A,B}^Q$ and $\beta_{\Delta_A,\Delta_B}^{\Delta_Q;n}$ are structure constants. 
Associativity with the Virasoro generator $T$ implies
that
\be
 \beta_{\Delta_A,\Delta_B}^{\Delta_Q;n}=\frac{(\Delta_A-\Delta_B+\Delta_Q)_n}{n!\,(2\Delta_Q)_n},
 \qquad
 (x)_n=\prod_{j\,=\,0}^{n-1}(x+j),
\label{Poch}
\ee
where the Pochhammer symbols $(x)_n$ are meant to reduce to $1$ for $n=0$.
This means that we can encode the information stored in the OPE contraction (\ref{ABQ}) in the
{\em star relation}
\be
 A\ast B\simeq\sum_{Q\in\Bc_\Ac}C_{A,B}^Q\{Q\},
\label{star}
\ee
where the {\em conformal tail}
\be
 \{Q\}=\sum_{n=0}^{\Delta_A+\Delta_B-\Delta_Q}
 \frac{\beta_{\Delta_A,\Delta_B}^{\Delta_Q;n}\pa^{n}Q(w)}{(z-w)^{\Delta_A+\Delta_B-\Delta_Q-n}}
\label{Qtail}
\ee
of $Q$ clearly depends on the OPE contraction it appears in (here (\ref{ABQ})). Subleading terms in conformal tails 
are thus suppressed in a star relation; this is indicated by the $\simeq$ notation in (\ref{star}). Note that a conformal 
tail of the identity $\I$ only contains one term, simply because all derivatives of $\I$ vanish. It is also noted that the 
star relation encapsulating the OPE contraction of $T$ with a primary field $A$ of conformal weight $\Delta$ is 
given by
\be
 T\ast A\simeq \Delta\{A\}.
\ee

Although the structure constants $C_{A,B}^Q$ satisfy
\be
 C_{A,B}^Q=(-1)^{|A||B|}(-1)^{\Delta_A+\Delta_B-\Delta_Q}C_{B,A}^Q,
\label{CABQ}
\ee
the star relation (\ref{star}) is in general only (anti-)symmetric `in appearance'. 
The possible difference (aside from a sign) between the corresponding
OPE contractions, $\bcontraction{}{A}{(z)}{B} A(z)B(w)$ and $\bcontraction{}{B}{(z)}{A} B(z)A(w)$, is 
`hidden' in the conformal tails $\{Q\}$;
this is indeed possible because the structure constants $\beta_{\Delta_A,\Delta_B}^{\Delta_Q;n}$ in
(\ref{Qtail}) are not, in general, symmetric in their lower indices. Of course, 
if $\Delta_A=\Delta_B$, then $\beta_{\Delta_A,\Delta_B}^{\Delta_Q;n}$ {\em is} symmetric and the two OPE 
contractions {\em do} agree (up to signs, as in (\ref{CABQ})). To illustrate, we consider the $W$-algebra $W_5$ 
(see Appendix~\ref{Sec:W5}), where the star products of the two even weight-$4$
quasi-primary fields $\La^{\s2,2}$ and $U$ are given by
\be
 \La^{\s2,2} \ast U=\tfrac{84}{5}\{U\}+8\{\La^{\s2,4}\}+6\{\La^{\s2',4}\},\qquad
 U\ast\La^{\s2,2}=\tfrac{84}{5}\{U\}+8\{\La^{\s2,4}\}-6\{\La^{\s2',4}\}.
\ee

We can also use the structure constants to describe the quasi-primary fields constructed from
the normal-ordered products of pairs of quasi-primary fields. For $Q_1,Q_2\in\Qc_\Ac$, we thus have that
\be
 \Qc_\Ac\ni(Q_1Q_2)
  -\sum_{Q_3\in\Bc_\Ac}C_{Q_1,Q_2}^{Q_3}\beta_{\Delta_1,\Delta_2}^{\Delta_3;\Delta_1+\Delta_2-\Delta_3}
  \pa^{\Delta_1+\Delta_2-\Delta_3}Q_3,
\label{QQQ}
\ee
where $\Delta_i=\Delta_{Q_i}$, $i=1,2,3$.
For $Q_1=Q_2=\La^{\s2,2}$ in $\Vir$, for example, the star relation is given by
\be
 \La^{\s2,2}\ast\La^{\s2,2}\simeq\tfrac{c(22+5c)}{10}\{\I\}+\tfrac{4(22+5c)}{5}\{T\}
  +\tfrac{10(64+5c)}{25}\{\La^{\s2,2}\}+8\{\La^{\s2,2,2}\}+(2+c)\{\La^{\s2'',2}\},
\ee
from which we obtain the quasi-primary field
\be
 \La^{\s(2,2),(2,2)}=(\La^{\s2,2}\La^{\s2,2})-\Big(\tfrac{22+5c}{10800}\pa^6T
  +\tfrac{7(64+5c)}{3960}\pa^4\La^{\s2,2}+\tfrac{14}{13}\pa^2\La^{\s2,2,2}
  +\tfrac{7(2+c)}{52}\pa^2\La^{\s2'',2}\Big).
\label{TT0TT0}
\ee
Decomposing this in terms of the elements of the basis $\Bc_8$ given in (\ref{TTTT})-(\ref{T4T}) yields
\be
 \La^{\s(2,2),(2,2)}=\La^{\s2,2,2,2}+\tfrac{7}{5}\La^{\s2'',2,2}+\tfrac{25c-73}{300}\La^{\s2'''',2}.
\ee
The expression in (\ref{QQQ}) can be generalised to involve more than two quasi-primary fields and to include 
derivatives of such fields, but a general discussion of such constructions of quasi-primary fields is beyond the 
scope of the present work.

\subsection{Tensor products of OPAs} 
\label{Sec:OPAsums}

We recall that an OPA has a unique identity and that it is closed under the computation of derivatives and under the
formation of linear combinations and normal-ordered products. Here, we wish to form the {\em tensor product}
of two OPAs, requiring that it shares these properties. In fact, the following characterisation of such a tensor product
matches the similar properties of the corresponding tensor product of vertex-operator algebras~\cite{LLbook}.

From the two OPAs $\Ac$ and $\Acb$ with identities $\I$ and $\bar\I$, respectively, 
the OPA $\Ac\otimes\Acb$ is thus a $\mathbb{Z}_2$-graded vector space, where we set
\be
 A\otimes\bar{\I}\equiv A,\qquad \I\otimes\bar{B}\equiv\bar{B},\qquad \I\otimes\bar{\I}\equiv\I.
\ee
As indicated, the identity of $\Ac\otimes\Acb$ is simply written $\I$.
For $A,B\in\Ac$ and $\bar{A},\bar{B}\in\Acb$, the $n$-products $[AB]_n$ and $[\bar{A}\bar{B}]_n$ are as in 
$\Ac$ and $\Acb$, respectively.
However, the $n$-product of $A\in\Ac$ and $\bar{B}\in\Acb$ is defined by
\be
 [A\bar{B}]_n=\begin{cases} 0,\ &n>0, \\[.15cm] \frac{1}{(-n)!}[\pa^{-n}A\bar{B}]_0,\ &n\leq0. \end{cases}
\ee
Additional relations follow from requiring that $\Ac\otimes\Acb$ is a commutative and associative OPA.
In particular, it follows that
\be
 [\bar{B}A]_n=\begin{cases} 0,\ &n>0, \\[.15cm] 
   \frac{(-1)^{|A|\,|\bar{B}|}}{(-n)!}[A\pa^{-n}\bar{B}]_0,\ &n\leq0. \end{cases}
\ee
By construction, for all $A,B\in\Ac$, the OPE contraction of $A(z)$ and $\bar{B}(w)$ in $\Ac\otimes\Acb$ is zero:
\be
 A(z)\bar{B}(w)\sim0.
\label{ABb0}
\ee
That is, the corresponding full OPE is nonsingular and is simply given by
\be
 A(z)\bar{B}(w)=\,:\!A(z)\bar{B}(w)\!:.
\label{ABb02}
\ee
Moreover, for $A,B,C\in\Ac$, we have
\be
 A(z)(B\bar{C})(w)\sim \sum_{n=1}^{\Delta_{A,B}}\frac{([AB]_n\bar{C})(w)}{(z-w)^n},
\ee
showing that $\Ac\otimes\Acb$ is closed under OPE contractions.
We refer to $\Ac\otimes\Acb$ as the {\em OPA tensor product} of $\Ac$ and $\Acb$.
Because the mode algebra associated with $\Ac\otimes\Acb$ only depends on the OPE {\em contractions} in 
$\Ac\otimes\Acb$, the mode algebra is a direct sum of the two mode algebras corresponding to $\Ac$ and $\Acb$, 
respectively.

\section{Galilean contractions}
\label{Sec:Galilean}

As discussed in the following, new OPAs can be constructed as contractions of given OPAs.
This resembles the way a Lie algebra can
arise as the result of an In\"on\"u-Wigner contraction~\cite{Seg51,IW53,Sal61,Gil06} of another Lie algebra.

\subsection{Contraction prescription and properties} 
\label{Sec:Prescription}

Our focus is on contractions of OPAs of the form $\Ac\otimes\Acb$, including pairs of the Virasoro field algebra and 
its various extensions.
In this work, the two algebras $\Ac$ and $\Acb$ are taken to be the same, 
up to their central parameters. 
In the case of two copies of the Virasoro algebra, for example, the central charges $c$ and $\cb$ may thus differ.
In the following, a bar over a field $A\in\Ac$ is used to indicate that the ensuing field $\Ab$ is the `same field' as 
$A$, but $\Ab\in\Acb$. The companion to $[AB]_n\in\Ac$ is given by
\be
 \overline{[AB]_n}=[\bar{A}\bar{B}]_n\in\Acb.
\label{ABnbar}
\ee

Now, let $\eps\in\mathbb{C}$. For each field $A\in\Ac$ and its companion $\Ab\in\Acb$, 
we then introduce the linear combinations 
\be
 A_\eps^+=A+\Ab,\qquad
 A_\eps^-=\eps(A-\Ab).
\label{ApAm}
\ee
For $\eps\neq0$, this yields an invertible map on the space of fields in $\Ac\otimes\Acb$, 
with inverse transformation given by
\be
 A=\tfrac{1}{2}\big(A_\eps^++\tfrac{1}{\eps}A_\eps^-\big),\qquad 
 \Ab=\tfrac{1}{2}\big(A_\eps^+-\tfrac{1}{\eps}A_\eps^-\big).
\label{Ainv}
\ee
For $\eps=0$, on the other hand, the map is singular and gives rise to a new algebraic structure. 
The ensuing algebra is generated by the fields (\ref{ApAm}) of $\Ac\otimes\Acb$, as $\eps\to0$. 
If the result is a well-defined OPA, we shall denote it by $\Ac_G$. 
Notationally, this is indicated by 
\be
 \Ac\otimes\Acb\to\Ac_G,\qquad
 A_\eps^\pm\mapsto A^\pm.
\ee
We refer to such a construction as a {\em Galilean contraction} and $\Ac_G$ as a {\em Galilean algebra}.
It is stressed that $\Ac\otimes\Acb$ and $\Ac_G$ are in general {\em non-isomorphic}.
Unsurprisingly, as indicated in Section~\ref{Sec:GCA}, other contractions are also possible, but we shall focus on 
the ``Galilean contraction" above.

For $A,B\in\Ac$, in $\Ac\otimes\Acb$, we have
\begin{align}
 A_\eps^+(z)B_\eps^+(w)
  &=\sum_{n=1}^{\Delta_{A,B}}\frac{[AB]_{n,\eps}^+(w)}{(z-w)^n}\,+:\!A_\eps^+(z)B_\eps^+(w)\!:,
\label{ApeBpe}
 \\[.15cm]
 A_\eps^+(z)B_\eps^-(w)
  &=\sum_{n=1}^{\Delta_{A,B}}\frac{[AB]_{n,\eps}^-(w)}{(z-w)^n}\,+:\!A_\eps^+(z)B_\eps^-(w)\!:,
 \\[.15cm]
 A_\eps^-(z)B_\eps^+(w)
  &=\sum_{n=1}^{\Delta_{A,B}}\frac{[AB]_{n,\eps}^-(w)}{(z-w)^n}\,+:\!A_\eps^-(z)B_\eps^+(w)\!:,
 \\[.15cm]
 A_\eps^-(z)B_\eps^-(w)
  &=\eps^2\sum_{n=1}^{\Delta_{A,B}}\frac{[AB]_{n,\eps}^+(w)}{(z-w)^n}\,+:\!A_\eps^-(z)B_\eps^-(w)\!:,
\label{AmeBme}
\end{align}
where, following (\ref{ApAm}), we have introduced
\be
 [AB]_{n,\eps}^+=[AB]_n+[\Ab\bar B]_n,\qquad [AB]_{n,\eps}^-=\eps\big([AB]_n-[\Ab\bar B]_n\big).
\label{ABnp}
\ee
It follows that OPE contractions in $\Ac_G$ are symmetric under exchange of the sign-index of the generators. 
That is,
\be
 A^+(z)B^-(w)
 \sim A^-(z)B^+(w),
\label{A+B-}
\ee
allowing us to omit detailing OPE contractions of the form $\bcontraction{}{A^-}{(z)}{B^-} A^-(z)B^+(w)$. Moreover,
\be
 A_\eps^-(z)B_\eps^-(w)\sim\eps^2A_\eps^+(z)B_\eps^+(w),
\ee
so if $A_\eps^+(z)B_\eps^+(w)$ is well-defined in the limit $\eps\to0$ (which is required for $\Ac_G$ to be
well-defined), then
\be
 A^-(z)B^-(w)\sim0.
\label{A-B-}
\ee

Although this analysis, based on (\ref{ApeBpe})-(\ref{AmeBme}), allowed us to deduce the properties 
(\ref{A+B-}) and (\ref{A-B-}) of $\Ac_G$, it is, in a sense, deceptively simple. Indeed, the field $[AB]_n\in\Ac$ may 
contain composite fields, in which case the fields $[AB]_{n,\eps}^\pm\in\Ac\otimes\Acb$ contain
composite fields, but as elements of $\Ac_G$, we should be able to express the corresponding
fields $[AB]_n^\pm$ in terms of normal-ordered products of non-composite fields (and their derivatives) in $\Ac_G$. 
This is a highly nontrivial task which will we will return to in Section~\ref{Sec:AAG}.

\subsection{Galilean conformal and superconformal algebras}
\label{Sec:GCA}

The Virasoro (field) algebra $\Ac=\Vir$ is generated by the field $T$ whose OPE contraction with itself is
\be
 T(z)T(w) \sim \frac{c/2}{(z-w)^{4}} + \frac{2T(w)}{(z-w)^{2}} + \frac{\pa T(w)}{z-w},
\label{Vir}
\ee
where $c$ is the central charge (and where we have omitted the identity field $\I$). 
The corresponding star relation is
\be
 T\ast T\simeq \frac{c}{2}\{\I\}+2\{T\}.
\label{Virstar}
\ee
The companion algebra $\Virb$ has central charge $\cb$ and is generated similarly by $\Tb$.

We now apply the Galilean contraction prescription to $\Vir\otimes\Virb$. We therefore introduce
\be
\begin{array}{rll}
 &T^{+}_\eps=T+ \Tb,\quad  &  c^{+}_\eps=c+\cb, 
 \\[.3cm]
 &T^{-}_\eps=\eps(T - \Tb),\quad  &  c^{-}_\eps=\eps(c - \cb),
\end{array}
\ee
whose star relations are worked out to be
\be
 T^+_\eps\gast T^\pm_\eps\simeq\frac{c^\pm_\eps}{2}\{\I\}+2\{T^\pm_\eps\},
 \qquad T^-_\eps\gast T^-_\eps\simeq \eps^2\,T^+_\eps\gast T^+_\eps.
\ee
It then follows that the star relations defining the {\em Galilean conformal algebra} $\Vir_{_G}$ are
\be
 T^+\gast T^\pm\simeq\frac{c^\pm}{2}\{\I\}+2\{T^\pm\},\qquad
  T^-\gast T^-\simeq0.
\label{GVir}
\ee
In particular, $T^+$ is seen to generate a Virasoro subalgebra $\Vir$ with central charge $c^+$. 
In terms of modes, where
\be
 T^\pm(z)=\sum_{n\in\mathbb{Z}}L^\pm_nz^{-2-n},
\ee
the Galilean conformal algebra is defined by
\be
 [L^+_n,L^\pm_m]=(n-m)L^\pm_{n+m}+\frac{c^\pm}{12}n(n^2-1)\delta_{n+m,0},\qquad
 [L^-_n,L^-_m]=0.
\label{Lpm}
\ee
Interest in this algebra has also been shown in the vertex-operator algebra literature, 
where it is referred to as $W(2,2)$~\cite{ZD07,AR16}. 
The role of $c^-$ is discussed at the end of Section~\ref{Sec:Simply}.

Superconformal extensions of the Virasoro algebra go back to~\cite{NS71,Ram71}. In particular, 
the $N=1$ superconformal (field) algebra $\mathcal{S}\Vir$ of central charge $c$ is generated by the Virasoro 
field $T$ and its super-partner $G$, with star relations
\be
 T\ast T\simeq\frac{c}{2}\{\I\}+2\{T\},\qquad T\ast G\simeq\frac{3}{2}\{G\},\qquad
  G\ast G\simeq\frac{2c}{3}\{\I\}+2\{T\}.
\label{N1}
\ee
The corresponding {\em Galilean $N=1$ SCA} $\mathcal{S}\Vir_G$ is generated by $T^\pm$ and $G^\pm$, 
whose nontrivial star relations are given by
\be
 T^+\gast T^\pm\simeq\frac{c^\pm}{2}\{\I\}+2\{T^\pm\},\qquad T^+\gast G^\pm\simeq\frac{3}{2}\{G^\pm\},\qquad
  G^+\gast G^\pm\simeq\frac{2c^\pm}{3}\{\I\}+2\{T^\pm\}.
\label{GN1}
\ee
As in the case of the Galilean conformal algebra, the fields $T^+$ and $G^+$ are recognised as generating 
an $N=1$ superconformal subalgebra of central charge $c^+$.

The Galilean $N=1$ SCA (\ref{GN1}) also appeared in~\cite{AL09,Sak10,Man10}, for example. As it contains
a pair of odd spin-$3/2$ fields, $G^\pm$, it can be interpreted as encoding an $N=(1,1)$ supersymmetry.
However, it is not the only possible superconformal extension of the Galilean Virasoro algebra $\Vir_G$. 
Indeed, the `asymmetric' algebra of~\cite{BDMT14,BDMT15,CGOR16,BJLMN16} 
contains only {\em one} odd spin-$3/2$ field and thus corresponds to an
$N=(1,0)$ supersymmetry. As the `symmetric' algebra above, it can be obtained~\cite{BJLMN16} 
by an In\"on\"u-Wigner contraction, but of the tensor product $\mathcal{S}\Vir\otimes\Vir$.
Another alternative is constructed~\cite{CT16,BCP16,Man16} as a contraction of the symmetric 
product $\mathcal{S}\Vir\otimes\mathcal{S}\Vir$, based on ``inhomogeneous scaling".

\subsection{Simply-extended algebras}
\label{Sec:Simply}

The objective here is to characterise a family of algebras $\Ac$ for which the Galilean contraction $\Ac_G$ of 
$\Ac\otimes\Acb$ is readily determined. 
We say that an OPA $\Ac$ is {\em simply-extended} if  
(i) all fields of the form $[AB]_n$, where $A$ and $B$ are elementary generators, are non-composite, 
(ii) no other OPE structure constants than the ones 
accompanying the identity field can depend on central parameters, and that 
(iii) this dependence is linear.
For example, the Virasoro field algebra and its $N=1$ superconformal extension are simply-extended. 
In fact, an OPA is simply-extended precisely if its mode algebra is an infinite-dimensional Lie (super)algebra. 
This suggests that we, alternatively, could refer to such OPAs as {\em Lie algebraic} or as being {\em of Lie type}.

Now, assuming that $\Ac$ is simply-extended implies that the fields $[AB]_n$ in the OPE (\ref{AB}) 
are non-composite and independent of the central parameters, while the dependence of $f_{A,B}$ on the central 
parameters is linear. Furthermore, with $\Ab$ and $\bar B$ denoting the $\Acb$-companions
to the $\Ac$-fields on the lefthand side of (\ref{AB}), we have that
$[\Ab\bar B]_n$ is the companion $\overline{[AB]_n}$ of $[AB]_n$, see (\ref{ABnbar}),
while the structure constants $\bar f_{\Ab,\bar B}$ and $f_{A,B}$ may differ.
It follows that the OPEs in $\Ac_G$ are of the form
\be
 A^+(z)B^\pm(w)=\frac{f_{A,B}^\pm}{(z-w)^{\Delta_A+\Delta_B}}
 +\sum_{n=-\infty}^{\Delta_A+\Delta_B-1} \frac{[AB]_n^\pm(w)}{(z-w)^n},
 \qquad A^-(z)B^-(w)\sim0,
\label{ApBpm}
\ee
where, in $\Ac\otimes\Acb$, we have used (\ref{ABnp}) and introduced
\be
 f_{A,B,\eps}^+=f_{A,B}+\bar f_{\Ab,\bar B},\qquad f_{A,B,\eps}^-=\eps(f_{A,B}-\bar f_{\Ab,\bar B}).
\ee
The fields $A^+\in\Ac_G$ thus generate a subalgebra isomorphic to $\Ac$, but with the central
parameters $c_i$ replaced by $c_i^+$, $i\in I_\Ac$. 
This generalises the similar observation for the Galilean conformal algebra $\Vir_{_G}$ in (\ref{GVir})
and its $N=1$ superconformal extension in (\ref{GN1}).
However, this is in general not true for $\Ac$ non-simply-extended, as illustrated in Section~\ref{Sec:W}.

Due to the linear dependence of $f_{A,B}$ on the central parameters, we can write
\be
 f_{A,B}=\sum_{i\in I_\Ac}\alpha_{A,B}^ic_i,\qquad \alpha_{A,B}^i\in\mathbb{C}.
\ee
In $\Ac\otimes\Acb$, we thus have
\be
 f_{A,B,\eps}^-=\epsilon\sum_{i\in I_\Ac}\alpha_{A,B}^i(c_i-\bar c_i)=\sum_{i\in I_\Ac}\alpha_{A,B}^ic_{i,\eps}^-.
\ee
In some cases, the fields $A^-\in\Ac_G$ can therefore be renormalised such that some or all of the parameters 
$c_i^-$ do not appear in any OPE. For example, in the case of only {\em one} central parameter 
(here denoted simply by $c$) in $\Ac$, we have
\be
 f_{A,B}=\alpha_{A,B}c\qquad\Longrightarrow\qquad f_{A,B}^-=\alpha_{A,B}c^-.
\ee
For $c^-\neq0$ in $\Ac_G$, we may thus rescale all fields $A^-\in\Ac_G$ different from $\I$ as
\be
 A^-\quad\longrightarrow\quad \Ah^-=\tfrac{A^-}{c^-},
\label{hatted}
\ee
thereby obtaining the $c^-$-independent OPEs
\be
 A^+(z)\Bh^-(w)=\frac{\alpha_{A,B}}{(z-w)^{\Delta_A+\Delta_B}}
   +\sum_{n=-\infty}^{\Delta_A+\Delta_B-1}\frac{[\widehat{AB}]_n^-(w)}{(z-w)^n}.
\label{ApBm}
\ee
It follows that, for each value of the parameter $c^+$, there are exactly {\em two} possible Galilean extensions of 
(the simply-extended OPA) $\Ac$, namely the one for which $c^-=0$ and the one for which $c^-\neq0$.
For example, the nontrivial star relations of the corresponding Galilean conformal algebras $\Vir_{_G}$ 
are thus given by
\be
 T^+\gast T^+\!\simeq\frac{c^+}{2}\{\I\}+2\{T^+\},\qquad 
 T^+\gast\Th^-\!\simeq\frac{\delta}{2}\{\I\}+2\{\Th^-\},
\ee
where $\delta\in\{0,1\}$.

\section{Compatibility with other constructions}
\label{Sec:Comp}

The Galilean contraction procedure may `intertwine' a pair of maps of the form
\be
 O:\ \Ac\to\Ac',\qquad  O_G:\ \Ac_G\to\Ac'_G.
\ee
Indeed, we are interested in scenarios where the composition diagram
\begin{center}
\includegraphics[scale=1.1]{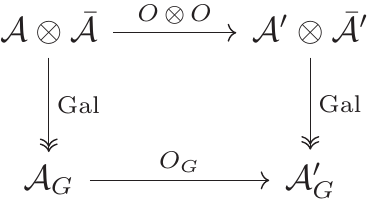}
\end{center}
%
\vspace{-.4cm}
(with Gal denoting a Galilean contraction) is commutative.
Concrete examples include the situation where $O$ encodes the topological twist of an $N=2$ superconformal 
algebra (SCA) and the situation where $O$ corresponds to an In\"on\"u-Wigner contraction of an $N=4$ SCA. 
These examples will be discussed in Section~\ref{Sec:Top} and~\ref{Sec:IW}, respectively. 
Another example is provided by the (Galilean) Sugawara construction discussed in Section~\ref{Sec:AGA}.

\subsection{$N=2$ SCA and topological twisting}
\label{Sec:Top}

The $N=2$ SCA $\Ac$ of central charge $c$, whose mode algebra was introduced in~\cite{Adeetal76a}, 
is generated by a Virasoro field $T$, a pair of super-fields $G_a$ of
conformal weight $3/2$, with $a\in\{0,1\}$, and a spin-one field $J$. 
It is a simply-extended OPA, so the corresponding Galilean algebra $\Ac_G$
generated by $T^\pm$, $G_a^\pm$ and $J^\pm$ is readily obtained. The nontrivial star relations are
\be
\begin{array}{rlll}
 &T^+\gast T^\pm\simeq\frac{c^\pm}{2}\{\I\}+2\{T^\pm\},\quad
 &T^+\gast J^\pm\simeq\{J^\pm\},\quad 
 &T^+\gast G_a^\pm\simeq\frac{3}{2}\{G_a^\pm\},
\\[.3cm]
 &G_0^+\gast G_1^\pm\simeq\frac{c^\pm}{3}\{\I\}+\{J^\pm\}+\{T^\pm\},\quad
 &J^+\gast J^\pm\simeq\frac{c^\pm}{3}\{\I\},\quad
 &J^+\gast G_a^\pm\simeq(-1)^a\{G_a^\pm\}.
\end{array}
\label{GN2}
\ee
As discussed in Section~\ref{Sec:Simply}, the parameter $c^-$ is either zero or can be absorbed in a rescaling of
the generators $A^-$.

The topologically twisted $N=2$ SCA $\Ac^{top}$\cite{EY90} is generated by
\be
 T_{top}:= T+\frac{1}{2}\pa J,\qquad 
 Q:= G_0,\qquad P:= G_1,
\label{twist}
\ee
and the spin-one field $J(z)$. The nontrivial star relations are
\be
\begin{array}{c}
 T_{top}\ast T_{top}\simeq2\{T_{top}\},\quad
 T_{top}\ast J\simeq-\frac{c}{3}\{\I\}+\{J\},\quad
 T_{top}\ast Q\simeq\{Q\},\quad
 T_{top}\ast P\simeq2\{P\},
\\[.3cm]
 Q\ast P\simeq\frac{c}{3}\{\I\}+\{J\}+\{T_{top}\},\quad
 J\ast J\simeq\frac{c}{3}\{\I\},\quad
 J\ast Q\simeq\{Q\},\quad
 J\ast P\simeq-\{P\}.
\end{array}
\ee
It follows that the twist has the effect of turning the two spin-$3/2$ superfields into a spin-one and a spin-two field
with respect to the Virasoro generator $T_{top}$. The spin-one field $J$, on the other hand, retains its conformal 
weight, but is in general not a primary field with respect to $T_{top}$; in fact, it is not even quasi-primary, but the use 
of $\{J\}$ in the star relations still applies.

The topologically twisted $N=2$ algebra $\Ac^{top}$ also belongs to the family of simply-extended algebras. 
The nontrivial OPE contractions of the corresponding Galilean algebra $\Ac^{top}_G$ are
\be
\begin{array}{c}
 T_{top}^+\gast T_{top}^\pm\simeq2\{T_{top}^\pm\},\quad
 T_{top}^+\gast J^\pm\simeq-\frac{c^\pm}{3}\{\I\}+\{J^\pm\},\quad
 T_{top}^+\gast Q^\pm\simeq\{Q^\pm\},\quad
 T_{top}^+\gast P^\pm\simeq2\{P^\pm\},
\\[.3cm]
 Q^+\gast P^\pm\simeq\frac{c^\pm}{3}\{\I\}+\{J^\pm\}+\{T_{top}^\pm\},\quad
 J^+\gast J^\pm\simeq\frac{c^\pm}{3}\{\I\},\quad
 J^+\gast Q^\pm\simeq\{Q^\pm\},\quad
 J^+\gast P^\pm\simeq-\{P^\pm\}.
\end{array}
\label{GtopN2}
\ee
In fact, applying the topological twist
\be
 T_{top}^{\pm}:= T^{\pm}+\frac{1}{2}\pa J^{\pm},\qquad 
 Q^\pm:= G^\pm_0,\qquad 
 P^\pm:= G^\pm_1 
\label{Gtwist}
\ee
to the Galilean $N=2$ SCA (\ref{GN2}) yields the {\em same} Galilean algebra (\ref{GtopN2})
as the Galilean contraction of the topologically twisted $N=2$ SCA. The diagram
\begin{center}
\includegraphics[scale=1.1]{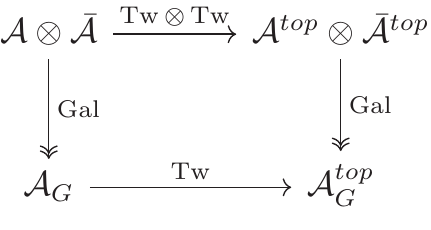}
\end{center}
%
\vspace{-.35cm}
is thus commutative. With reference to the commutative diagram in the preamble to Section~\ref{Sec:Comp}, 
the map $O$ is the upper twist (indicated above by $\mathrm{Tw}$) given in (\ref{twist}), while the map $O_G$ is 
the lower twist (likewise indicated by $\mathrm{Tw}$) given in (\ref{Gtwist}).

\subsection{$N=4$ SCAs and In\"on\"u-Wigner contractions}
\label{Sec:IW}

The Galilean contraction procedure is not only compatible with the topological twisting discussed in 
Section~\ref{Sec:Top}; it is also compatible with certain In\"on\"u-Wigner 
contractions~\cite{Seg51,IW53,Sal61,Gil06}. We illustrate this by considering $N=4$ SCAs and refer 
to~\cite{Ray15} for a more general discussion.

The known $N=4$ SCAs in two dimensions are characterised by their internal affine Lie algebra.
The small $N=4$ SCA~\cite{Adeetal76a,Adeetal76c} is based on $su(2)$; 
the large one~\cite{Sch87,Sch88,STVP88} on $su(2)\oplus su(2)\oplus u(1)$;
the middle one~\cite{AK93,Ali99} on $su(2)\oplus u(1)\oplus u(1)\oplus u(1)\oplus u(1)$;
while the non-reductive one~\cite{Ras02} is based on $su(2)\oplus\g$ where $\g$ is a four-dimensional 
non-reductive Lie algebra. The total number of generating
fields in these SCAs is $8$, $16$, $16$ and $16$, respectively.
The non-reductive $N=4$ SCA also contains an interesting $N=4$ superconformal subalgebra whose affine 
Lie algebra is based on $su(2)\oplus u(1)\oplus u(1)$~\cite{Ras00,Ras01}, 
while the total number of generating fields is $12$.

For each value of the central charge, there is a one-parameter family of large $N=4$ SCAs, usually labelled by a 
parameter $\gamma$. For fixed value of $\gamma$, the corresponding SCA is simply-extended, as are all the
other $N=4$ SCAs. Their Galilean contractions are therefore readily obtained, although not presented explicitly 
here. Details can be found in~\cite{Ray15}.

The middle and non-reductive $N=4$ SCAs can be obtained by In\"on\"u-Wigner contractions of the large
$N=4$ SCA by letting $\gamma$ approach a singular value in particular bases.
As already indicated, in both cases, the resulting algebra is simply-extended.
We have verified that the In\"on\"u-Wigner contractions employed in their construction are compatible with the 
Galilean contractions, thereby establishing the commutativity of the diagram
\begin{center}
\includegraphics[scale=1.1]{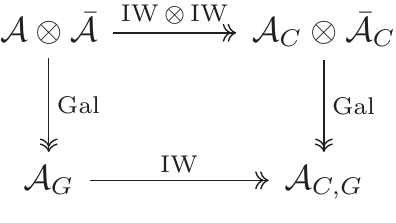}
\end{center}
%
\vspace{-.35cm}
To be clear, for this to hold in the case where $\Ac$ is the large $N=4$ SCA, the companion algebra $\Acb$ 
must be defined for the {\em same} value of $\gamma$.

\subsection{Galilean affine algebras and the Sugawara construction}
\label{Sec:AGA}

In this section, we introduce {\em Galilean affine algebras} $\gh_{_G}$ as the Galilean extensions of affine current
(super)algebras $\gh$. We then examine their compatibility with the Sugawara construction, 
analysing the commutativity of the diagram
\begin{center}
\includegraphics[scale=1.1]{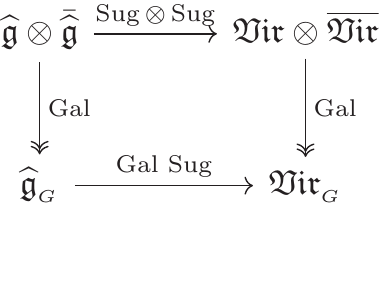}
\end{center}
%
\vspace{-1.1cm}
where ${\rm Sug}$ refers to a Sugawara construction. 

An affine current (super)algebra $\gh$ is generated by a set of fields $\{J_a\,|\,a=1,\ldots,\dim\g\}$.
With reference to the mode expansion
\be
 J_a(z)=\sum_{n\in\mathbb{Z}}J_{a,n}z^{-n-1},
\label{Jaz}
\ee
the set of zero modes $\{J_{a,0}\,|\,a=1,\ldots,\dim\g\}$ thus generates the Lie (super)algebra $\g$.
The OPE contractions of the currents are given by
\be
 J_a(z)J_b(w)\sim\frac{\kappa_{ab}k}{(z-w)^2}+\frac{{f_{ab}}^cJ_c(w)}{z-w},
\label{JaJb}
\ee
where ${f_{ab}}^c$ are the structure constants and $\kappa$, with $\kappa_{ab}=\kappa(J_{a,0},J_{b,0})$, 
the Killing form of $\g$. The central parameter $k$ is known as the level of $\gh$.
Note that the standard Einstein summation convention, summing over appropriately repeated indices, 
has been employed. 

As is evident from (\ref{JaJb}), the affine current (super)algebra is simply-extended. With the
structure constants and Killing forms on $\gh$ and its companion $\bar\gh$ the same, the Galilean algebra
$\gh_{_G}$ is readily constructed and its nontrivial OPE contractions are given by
\be
 J_a^+(z)J_b^\pm(w)\sim\frac{\kappa_{ab}k^\pm}{(z-w)^2}+\frac{{f_{ab}}^cJ_c^\pm(w)}{z-w}.
\label{JapJbpm}
\ee
The currents $J_a^+$ thus generate an affine current algebra with level $k^+$.
As before, if $k^-\neq0$, then we can renormalise the fields $J_a^-$ and thereby eliminate the parameter 
$k^-$ altogether. 

In the following, we will focus on affine current algebras based on Lie algebras and only on those
for which the Killing form $\kappa$ is non-degenerate.
Extending the considerations to superalgebras is straightforward but will not be discussed explicitly.

It is well-known that the Sugawara construction
\be
 T=\frac{\kappa^{ab}}{2(k+h^\vee)}(J_aJ_b)
\label{Sug}
\ee
provides a realisation of the Virasoro (field) algebra with central charge
\be
 c=\frac{k\,{\rm dim}\,\g}{k+h^\vee},
\label{Sugc}
\ee
where $h^\vee$ is the dual Coxeter number of the Lie algebra $\g$.
The fields $J_a$ are all primary of conformal weight $1$ with respect to the Sugawara construction, that is,
\be
 T(z)J_a(w)\sim\frac{J_a(w)}{(z-w)^2}+\frac{\pa J_a(w)}{z-w}.
\label{TJa}
\ee

With reference to the diagram just above (\ref{Jaz}), we now examine the lower
path from $\gh\otimes\bar{\gh}$ via $\gh_{_G}$ to $\Vir_{_G}$. 
The initial map, Gal, is simply the Galilean contraction producing the Galilean affine algebra $\gh_{_G}$ with OPE 
contractions (\ref{JapJbpm}). The second map, denoted by Gal Sug, is meant to 
mimic the usual Sugawara construction (\ref{Sug}), but applied to all the fields $J_a^\pm$, not just to $J_a^+$.
It thus entails constructing a realisation $T^\pm$ of the Galilean Virasoro algebra $\Vir_{_G}$ in terms of the 
Galilean affine generators $J_a^\pm$. Accordingly, we make the ansatz
\be
 T^\pm=\sum_{\alpha,\beta\in\{+,-\}}\!\!\!\lambda_{\alpha\beta}^\pm\,\kappa^{ab}(J_a^\alpha J_b^\beta),
 \qquad\lambda_{\alpha\beta}^\pm\in\mathbb{C}.
\label{TpmJJ}
\ee
Aside from realising (\ref{GVir}) for some central parameters $c^\pm$ (to be determined below), it should satisfy
\be
 T^+(z)J_a^\pm(w)\sim\frac{J_a^\pm(w)}{(z-w)^2}+\frac{\pa J_a^\pm(w)}{z-w},\qquad
 T^-(z)J_a^-(w)\sim0,
\ee
and, according to (\ref{A+B-}),
\be
 T^-(z)J_a^+(w)\sim \frac{J_a^-(w)}{(z-w)^2}+\frac{\pa J_a^-(w)}{z-w}.
\ee
Using standard Lie-algebraic relations such as
\be
 \kappa^{bc}{f_{ab}}^d{f_{dc}}^e=2h^{\vee}\delta_a^e,\qquad
 \kappa^{bd}{f_{ab}}^c+\kappa^{cb}{f_{ab}}^d=0,
\ee
we find that $T^\pm$ in (\ref{TpmJJ}) are given {\em uniquely} as
\begin{align}
 T^+&=\frac{\kappa^{ab}}{2k^-}[(J_a^+J_b^-)+(J_a^-J_b^+)]
  -\frac{k^+\!+2h^\vee}{2(k^-)^2}\,\kappa^{ab}(J_a^-J_b^-),
\label{TpJJ}
 \\[.15cm]
 T^-&=\frac{\kappa^{ab}}{2k^-}(J_a^-J_b^-),
\label{TmJJ}
\end{align}
and that they generate the Galilean Virasoro algebra $\Vir_{_G}$ with
\be
 c^+=2\dim\g,\qquad c^-=0.
\label{cpcm}
\ee

We now turn to the upper path in the diagram, the one from $\gh\otimes\bar{\gh}$ via $\Vir\otimes\Virb$ to 
$\Vir_{_G}$. The first map, Sug$\,\otimes\,$Sug, yields the tensor product of two copies of the usual Sugawara 
construction, generated by
\be
 T(z)=\frac{\kappa^{ab}}{2(k+h^\vee)}(J_aJ_b)(z),\qquad
 \bar{T}(z)=\frac{\kappa^{ab}}{2(\bar{k}+h^\vee)}(\bar{J}_a\bar{J}_b)(z),
\ee
respectively, with corresponding central charges given by
\be
 c=\frac{k\,{\rm dim}\,\g}{k+h^\vee},\qquad \bar{c}=\frac{\bar{k}\,{\rm dim}\,\g}{\bar{k}+h^\vee}.
\label{ccb}
\ee
The Galilean contraction comprising the second map in the upper path follows from
\begin{align}
 T_\eps^+&=T+\bar{T}\nonumber
 \\[.15cm]
  &=\frac{\kappa^{ab}}{2(\frac{1}{2}(k_\eps^++\frac{1}{\eps}k_\eps^-)+h^\vee)}
   \big[\tfrac{1}{2}(J_{a,\eps}^++\tfrac{1}{\eps}J_{a,\eps}^-)\tfrac{1}{2}(J_{b,\eps}^++\tfrac{1}{\eps}J_{b,\eps}^-)\big]
 \nonumber\\[.15cm]
 &+\frac{\kappa^{ab}}{2(\frac{1}{2}(k_\eps^+-\frac{1}{\eps}k_\eps^-)+h^\vee)}
  \big[\tfrac{1}{2}(J_{a,\eps}^+-\tfrac{1}{\eps}J_{a,\eps}^-)\tfrac{1}{2}(J_{b,\eps}^+-\tfrac{1}{\eps}J_{b,\eps}^-)\big]
 \nonumber\\[.15cm]
  &=\frac{\kappa^{ab}}{2k_\eps^-}[(J_{a,\eps}^+J_{b,\eps}^-)+(J_{a,\eps}^-J_{b,\eps}^+)]
  -\frac{k_\eps^+\!+2h^\vee}{2(k_\eps^-)^2}\,\kappa^{ab}(J_{a,\eps}^-J_{b,\eps}^-)
  +\Oc(\eps)
\label{Tpeps}
\end{align}
and
\be
 c_\eps^+=c+\bar{c}
 =\frac{\frac{1}{2}(k_\eps^++\frac{1}{\eps}k_\eps^-)\dim\g}{\frac{1}{2}(k_\eps^++\frac{1}{\eps}k_\eps^-)+h^\vee}
 +\frac{\frac{1}{2}(k_\eps^+-\frac{1}{\eps}k_\eps^-)\dim\g}{\frac{1}{2}(k_\eps^+-\frac{1}{\eps}k_\eps^-)+h^\vee}
 =2\dim\g+\Oc(\eps^2),
\ee
and the similar expansions of $T_\eps^-$ and $c_\eps^-$, all in $\Ac\otimes\Acb$. As $\eps\to0$, the expressions 
(\ref{TpJJ}) and (\ref{TmJJ}) for $T^\pm$ are recovered, as are the results (\ref{cpcm}) for the central parameters 
$c^\pm$. This demonstrates the commutativity of the diagram just above (\ref{Jaz}).

We note that the expressions (\ref{TpJJ}) and (\ref{TmJJ}) for $T^\pm$ are defined for $k^-\neq0$ only.
However, rescaling $J_b^-$ and $T^-$ by $k^-$, thereby introducing
\be
 \Jh_b^-=\tfrac{1}{k^-}J_b^-,\qquad \Th^-=\tfrac{1}{k^-}T^-,
\ee
allows us to eliminate $k^-$ altogether, as we then have
\be
 T^+=\tfrac{1}{2}\kappa^{ab}(J_a^+\Jh_b^-+\Jh_a^-J_b^+)
  -\tfrac{1}{2}(k^+\!+2h^\vee)\kappa^{ab}(\Jh_a^-\Jh_b^-),\qquad
 \Th^-=\tfrac{1}{2}\kappa^{ab}(\Jh_a^-\Jh_b^-).
\ee
In fact, this corresponds to basing the lower path in the diagram just above (\ref{Jaz}) on the Galilean affine algebra
(\ref{JapJbpm}) in which $k^-$ has been scaled away. Likewise, for $k^-=0$ in the original affine Lie algebra,
one merely constructs $T^\pm$ as in (\ref{TpmJJ}) using (\ref{JapJbpm}) with $k^-=0$.

\section{On the existence of $\Ac_G$}
\label{Sec:AAG}

In the computation of the series expansion (\ref{Tpeps}) of $T_\eps^+$ in powers of $\eps$, one encounters a 
cancellation of contributions proportional to $\frac{1}{\eps}$. Without this cancellation, the limit $T^+$ of 
$T_\eps^+$, as $\eps\to0$, would simply not exist. This points to a general and fundamental property: 
The existence of a Galilean counterpart, $\Ac_G$, to the OPA $\Ac$ presupposes that the coefficients in
$\Ac$ conspire appropriately.
A priori, there is no guarantee that a given OPA $\Ac$ admits a Galilean contraction resulting in $\Ac_G$.
Here, we will explore the conditions on $\Ac$ imposed by the requirement that $\Ac_G$ exists.
In the process, we find general expressions for the structure constants of $\Ac_G$ in terms of those of $\Ac$. 

Most of the conformal OPAs of our interest in Section~\ref{Sec:W} correspond to $W$-algebras generated by 
the Virasoro field $T$ and a finite linearly independent set of primary fields $P$
of integer conformal weights greater than $2$.
We will assume `non-degeneracy' of this set of generators in the sense that the corresponding matrix $(f_{P,P'})$, 
see (\ref{AB}), is non-degenerate for generic central parameters. This allows us to choose a canonical 
normalisation of the generators. In the series of $W_n$ algebras, for example, the matrix $(f_{P,P'})$ 
is diagonal and the central charge $c$ is the sole central parameter.
In these models, it is thus standard to normalise the generators such that $f_{P,P}=c/\Delta_P$. 
For generic $c$, this is equivalent to requiring that 
$[PP]_{2\Delta_P-2}=2T$; in fact, this fixes the normalisation even for $c=0$.

Let us now assume that $[AB]_n$ in the OPE contraction (\ref{ABcont}) contains the normal-ordered product 
$(CD)$ of the two non-composite fields $C$ and $D$, that $(CD)$ is linearly independent of all other terms in 
$[AB]_n$, and that $(CD)$ appears with coefficient $\phi(c)$: a function of the central charge. 
The fields $C$ and $D$ may involve linear combinations of derivatives of fields, as long 
as the ensuing expressions for $C$ and $D$ do not depend on $c$.
Then, $[AB]_{n,\eps}^+$ and $[AB]_{n,\eps}^-$ will contain the corresponding terms
\begin{align}
 \phi(c)(CD)+\phi(\bar{c})(\bar{C}\bar{D})
 &=\frac{1}{4\eps^2}\big\{\phi(\tfrac{1}{2}(c_\eps^++\tfrac{1}{\eps}c_\eps^-))
  +\phi(\tfrac{1}{2}(c_\eps^+-\tfrac{1}{\eps}c_\eps^-))\big\}
  \big\{(C_\eps^-D_\eps^-)+\eps^2(C_\eps^+D_\eps^+)\big\}
 \nonumber\\[.15cm]
 &+\frac{1}{4\eps}\big\{\phi(\tfrac{1}{2}(c_\eps^++\tfrac{1}{\eps}c_\eps^-))
  -\phi(\tfrac{1}{2}(c_\eps^+-\tfrac{1}{\eps}c_\eps^-))\big\}
  \big\{(C_\eps^+D_\eps^-)+(C_\eps^-D_\eps^+)\big\}
\label{phiCDp}
\end{align}
and
\begin{align}
 \eps\big(\phi(c)(CD)-\phi(\bar{c})(\bar{C}\bar{D})\big)
 &=\frac{1}{4\eps}\big\{\phi(\tfrac{1}{2}(c_\eps^++\tfrac{1}{\eps}c_\eps^-))
  -\phi(\tfrac{1}{2}(c_\eps^+-\tfrac{1}{\eps}c_\eps^-))\big\}
  \big\{(C_\eps^-D_\eps^-)+\eps^2(C_\eps^+D_\eps^+)\big\}
 \nonumber\\[.15cm]
 &+\frac{1}{4}\big\{\phi(\tfrac{1}{2}(c_\eps^++\tfrac{1}{\eps}c_\eps^-))
  +\phi(\tfrac{1}{2}(c_\eps^+-\tfrac{1}{\eps}c_\eps^-))\big\}
  \big\{(C_\eps^+D_\eps^-)+(C_\eps^-D_\eps^+)\big\},
\label{phiCDm}
\end{align}
respectively.

Let us further assume that $\phi(c)$ factorises as
\be
 \phi(c)=\gamma\prod_{i=1}^k(c-\lambda_i)^{\alpha_i}
\label{phic}
\ee
for some $\gamma,\lambda_i\in\mathbb{C}$ and $\alpha_i\in\frac{1}{2}\mathbb{Z}$, $i=1,\ldots,k$, where 
$k\in\mathbb{N}_0$. This is indeed the case in all the $W$-algebras considered in Section~\ref{Sec:W}. Using
\be
 \gamma\prod_{i=1}^k(c-\lambda_i)^{\alpha_i}
 =\gamma\Big(\frac{c_\eps^-}{2\eps}\Big)^{\sum_{i=1}^k\!\alpha_i}
  \Big(1+\frac{\eps}{c_\eps^-}\sum_{i=1}^k\alpha_i(c_\eps^+-2\lambda_i)+\Oc(\eps^2)\Big)
\ee
and
\be
 \gamma\prod_{i=1}^k(\bar{c}-\lambda_i)^{\alpha_i}
  =\gamma\Big(\!-\frac{c_\eps^-}{2\eps}\Big)^{\sum_{i=1}^k\!\alpha_i}
  \Big(1-\frac{\eps}{c_\eps^-}\sum_{i=1}^k\alpha_i(c_\eps^+-2\lambda_i)+\Oc(\eps^2)\Big),
\ee
we then find that, for generic central charges, the limit $\eps\to0$ is only well-defined if
\be
 \sum_{i=1}^k\alpha_i=-1\qquad\text{or}\qquad\sum_{i=1}^k\alpha_i=-2,
\label{sumalpha}
\ee
or if $\sum_{i=1}^k\alpha_i<-2$. However, in the latter case, the coefficients vanish.

More generally, if $[AB]_n$ contains the term $\phi(c)(C_1\ldots C_\ell)$,
which is linearly independent
of the other terms in $[AB]_n$ and where $C_1,\ldots,C_\ell$ are non-composite (but may involve derivatives), then
\be
 \sum_{i=1}^k\alpha_i=-(\ell-1)\qquad\text{or}\qquad\sum_{i=1}^k\alpha_i=-\ell
\label{sumalphagen}
\ee
or $\sum_{i=1}^k\alpha_i<-\ell$ for the limit $\eps\to0$ to be well-defined. Moreover, one of the conditions in
(\ref{sumalphagen}) must be satisfied for the limit to produce a non-vanishing term out of 
$\phi(c)(C_1\ldots C_\ell)$.

If the contraction prescription is well-defined, then the limits of (\ref{phiCDp}) and (\ref{phiCDm}) will make 
contributions in the Galilean OPA $\Ac_G$ to $[AB]_n^+$ and $[AB]_n^-$, respectively.
First, we describe the ensuing term in $[AB]_n^+$.
If $\sum_{i=1}^k\alpha_i=-(\ell-1)$, the term is given by 
\be
\begin{array}{c}
 \displaystyle{\frac{\gamma}{(c^-)^{\ell-1}}}\Big((C_1^+C_2^-\ldots C_\ell^-)+(C_1^-C_2^+C_3^-\ldots C_\ell^-)
  +\ldots+(C_1^-\ldots C_{\ell-1}^-C_\ell^+)\Big)
 \\[.3cm]
 -\,\displaystyle{\frac{\gamma}{(c^-)^\ell}\displaystyle{\Big((\ell-1)c^++2\sum_{i=1}^k\alpha_i \lambda_i}\Big)(C_1^-\ldots C_\ell^-)},
  \end{array}
\label{ell1}
\ee
whereas, if $\sum_{i=1}^k\alpha_i=-\ell$, the term is simply given by
\be
 \frac{2\gamma}{(c^-)^\ell}\,(C_1^-\ldots C_\ell^-).
\label{ell}
\ee
Note that these expressions only make sense if 
\be
 c^-\neq0, 
\ee
a condition we shall therefore assume here and in Section~\ref{Sec:W} and Appendix~\ref{Sec:Walg}.
Absorbing the $c^-$-dependence as in (\ref{hatted}), by introducing
\be
 \Ch_i^-=\frac{C_i}{c^-},\qquad i=1,\ldots,\ell,
\ee
the terms (\ref{ell1}) and (\ref{ell}) become
\be
 \gamma\Big((C_1^+\Ch_2^-\ldots\Ch_\ell^-)+\ldots+(\Ch_1^-\ldots\Ch_{\ell-1}^-C_\ell^+)\Big)
  -\gamma\Big((\ell-1)c^++2\sum_{i=1}^k\alpha_i \lambda_i\Big)(\Ch_1^-\ldots\Ch_\ell^-)
\label{C1a}
\ee
and
\be
 2\gamma(\Ch_1^-\ldots\Ch_\ell^-),
\ee
respectively.
In $[AB]_n^-$, on the other hand, the only term generated from (\ref{phiCDm}) is 
\be
 \gamma c^-(\Ch_1^-\ldots\Ch_\ell^-),
\label{C1c}
\ee
and this only occurs if $\sum_{i=1}^k\alpha_i=-(\ell-1)$. If $\sum_{i=1}^k\alpha_i=-\ell$, 
the contribution simply vanishes. Note that the factor of $c^-$ in (\ref{C1c}) will be absorbed on the lefthand side of 
the OPE contraction if we consider
$\bcontraction{}{A^+}{(z)}{\Bh^-} A^+(z)\Bh^-(w)$ instead of $\bcontraction{}{A^+}{(z)}{B^-} A^+(z)B^-(w)$.

We conclude this section by observing that if the contraction prescription on the quasi-primary field $Q$ with 
coefficient $\phi(c)$ is well-defined, then it is well-defined on the entire conformal tail $\{Q\}$. This follows readily 
from the $c$-independent linearity of the terms in the definition (\ref{Qtail}) of $\{Q\}$, and the fact that the analysis 
above is unaffected by the inclusion of derivatives.

\section{Galilean contractions of $W$-algebras}
\label{Sec:W}

In the following, we apply the Galilean contraction prescription to a variety of known $W$-algebras.
Details on some of the ensuing Galilean $W$-algebras are deferred to Appendix~\ref{Sec:GW}.

\subsection{Quasi-primary fields in the Galilean conformal algebra}
\label{Sec:GQP}

As discussed in Section~\ref{Sec:GCA}, the field $T^+\!\in\Vir_G$ generates a Virasoro subalgebra of the Galilean 
conformal algebra. A field $Q\in\Vir_G$ is accordingly quasi-primary with respect to $T^+$ if
\be
 [T^+Q]_3=0,\qquad [T^+Q]_2=\Delta_QQ,\qquad [T^+Q]_1=\pa Q,
\ee
where $\Delta_Q$ is the corresponding conformal weight. We denote by $\Qc_{\Vir_G}$ the set of quasi-primary 
fields in $\Vir_G$. Examples of elements of $\Qc_{\Vir_G}$ are easily constructed:
simply replace $T$ by $T^+$ in the explicit expressions $\La^{\s a,b,\ldots}$ in Section~\ref{Sec:Quasi}.
However, according to Section~\ref{Sec:AAG}, as the result of a Galilean contraction, we should only expect to 
encounter quasi-primary fields with at most one $T^+$ generator in each of their normal-ordered products, 
see (\ref{C1a})-(\ref{C1c}).
Before discussing these quasi-primary fields explicitly, let us extend some of the elements of the analysis in
Section~\ref{Sec:Quasi} to the Galilean conformal algebra $\Vir_G$.

We thus define the Galilean Virasoro Verma module of highest weight $0$ as the module generated by the free 
action of $\Vir_G$ on the highest-weight state $|0\rangle$ satisfying
\be
 L^\pm_n|0\rangle=0,\qquad n\geq0.
\ee
This module has a submodule generated from the vector $L^+_{-1}|0\rangle$. Since
\be
 L^-_{-1}|0\rangle=-[L^+_{-1},L^-_0]|0\rangle=L^-_0L^+_{-1}|0\rangle,
\ee
we see that $L^-_{-1}|0\rangle$ is an element of this submodule.
The corresponding quotient module, $\Nc$, is $\mathbb{Z}_+$-graded,
\be
 \Nc=\bigoplus_{n=0}^\infty \Nc_n,
\label{Nsum}
\ee
and its conformal character is given by
\be
 \chi_{_G}(q)=\mathrm{Tr}_{\!_\Nc} q^{L^+_0}=[\chi(q)]^2
 =\prod_{n=2}^\infty\frac{1}{(1-q^n)^2}
 =\sum_{n=0}^\infty b_nq^n,\qquad
 b_n=\dim\Nc_n. 
\ee
Concretely, the character begins as
\be
 \chi_{_G}(q)=1+2q^2+2q^3+5q^4+6q^5+13q^6+16q^7+30q^8+40q^9+66q^{10}+\Oc(q^{11}).
\ee
The adaptation to $\Vir_G$ of the decomposition (\ref{MMLM}) is given by
\be
 \Nc_n=\Nc_n'\oplus L^+_{-1}\Nc_{n-1},
\label{NNLN}
\ee
where
\be
 \Nc_n'=\{v\in\Nc_n\,|\,L^+_1v=0\},\qquad \dim\Nc_n'=b_n-b_{n-1}.
\label{Nnp}
\ee
For small $n$, these dimensions are given by
\be
\begin{array}{c}
 \dim\Nc_1'=0,\quad
 \dim\Nc_2'=2,\quad
 \dim\Nc_3'=0,\quad
 \dim\Nc_4'=3,\quad
 \dim\Nc_5'=1,
\\[.3cm]
 \dim\Nc_6'=7,\quad
 \dim\Nc_7'=3,\quad
 \dim\Nc_8'=14,\quad
 \dim\Nc_9'=10,\quad
 \dim\Nc_{10}'=26.
\end{array}
\ee
In terms of fields, the decomposition (\ref{Nsum}) reads
\be
 \Vir_G=\bigoplus_{n=0}^\infty(\Vir_G)_n,
\ee
where $(\Vir_G)_n$ consists of the descendant fields of the identity, of conformal weight $n$.
The field version of the decomposition (\ref{NNLN}) can then be expressed as
\be
 (\Vir_G)_n=(\Qc_{\Vir_G})_n\oplus\pa(\Vir_G)_{n-1},
\ee
where $(\Qc_{\Vir_G})_n\subset\Qc_{\Vir_G}$ denotes the set of quasi-primary fields of conformal weight $n$.
By construction, 
\be
 \dim(\Qc_{\Vir_G})_n=\dim\Nc_n'.
\ee

For ease of reference, we now list bases for the linear spans of the quasi-primary fields that contain
at most one copy of $T^+$ in each of their terms and have conformal weight between 
$4$ and $6$. In our notation $\Lh_\pm^{\s a,b,\ldots}$ for quasi-primary fields in $\Qc_{\Vir_G}$, 
a $+$ indicates that $T^+$ appears in at least one of the terms, whereas a $-$ indicates that the quasi-primary field 
is built using $\Th^-$ only. Of weight $4$, we thus have
\be
 \Lh_+^{2,2}=(T^+\Th^-)-\tfrac{3}{10}\pa^2\Th^-,\qquad
 \Lh_-^{2,2}=(\Th^-\Th^-).
\label{Lapm22}
\ee
Of weight $5$, we have
\be
 \Lh_+^{2',2}=(\pa T^+\Th^-)
 -\tfrac{1}{2}\pa\Lh_+^{2,2}
 -\tfrac{1}{15}\pa^3\Th^-.
\ee
Of weight $6$, we have
\begin{align}
 \Lh_+^{2,2,2}&=(T^+\Th^-\Th^-)-\tfrac{1}{6}\pa^2\Lh_-^{2,2},
\label{La222p}
\\[.15cm]
 \Lh_+^{2'',2}&=\tfrac{1}{2}[(\pa^2T^+\Th^-)+(\pa^2\Th^-T^+)]
  -\tfrac{5}{18}\pa^2\Lh_+^{2,2}-\tfrac{1}{42}\pa^4\Th^-,
\label{La2''2p}
\end{align}
and
\be
 \Lh_-^{2,2,2}=(\Th^-\Th^-\Th^-),\qquad
 \Lh_-^{2'',2}=(\pa^2\Th^-\Th^-)-\tfrac{5}{18}\pa^2\Lh_-^{2,2}.
\label{La6m}
\ee
As an aside, we note that $\dim\Nc_6'-\dim\Mc_6'=5$. As we have only listed four linearly independent 
quasi-primary fields of conformal weight $6$, this suggests that there exists a unique (up to normalisation) 
weight-$6$ quasi-primary field
involving $\Th^-$ and containing a term with more than one factor of $T^+$. Indeed, this field is given by
\be
 \Lh_{++}^{2,2,2}=(T^+T^+\Th^-)
  +\tfrac{3}{10}\pa\Lh_+^{2',2}
  -\tfrac{1}{4}\pa^2\Lh_+^{2,2}
  -\tfrac{2}{35}\pa^4\Th^-.
\ee

For any $A$ in an OPA $\Ac$, we note the field identity $(A^-A^+)=(A^+A^-)$ in $\Ac_G$. More generally, we have
\be
 (A^-A^+A^-\ldots A^-)=(A^+A^-A^-\ldots A^-).
\ee
In any Galilean conformal OPA, we thus have the identities $(\Th^-T^+)=(T^+\Th^-)$,
$(\pa\Th^-\pa T^+)=(\pa T^+\pa\Th^-)$ 
and $(\Th^-T^+\Th^-\ldots\Th^-)=(T^+\Th^-\Th^-\ldots\Th^-)$,
for example, while $(\pa\Th^-T^+)\neq(\pa T^+\Th^-)$.

Generalising the discussion at the end of Section~\ref{Sec:Quasi} to a Galilean conformal OPA $\Ac_G$, 
we now consider quasi-primary fields built from 
normal-ordered products of $T^\pm$ and $A^\pm$, where $A$ is a primary field of conformal weight $\Delta$ in 
$\Ac$. In addition to the primary fields $A^+$ and $\Ah^-$ of conformal weight $\Delta$, 
we thus find that the following fields are quasi-primary with respect to $T^+$:
\begin{align}
 \Lh_+^{2,\Delta}&=\tfrac{1}{2}[(T^+\Ah^-)+(\Th^-A^+)]
 -\tfrac{3}{2(1+2\Delta)}\pa^2\Ah^-,
\label{LhTAp}
\\[.15cm]
 \Lh_+^{2',\Delta}&=\tfrac{1}{2}[(\pa T^+\Ah^-)+(\pa\Th^-A^+)]
 -\tfrac{2}{2+\Delta}\pa\Lh_+^{2,\Delta}
 -\tfrac{1}{(1+\Delta)(1+2\Delta)}\pa^3\Ah^-,
\label{LhT1Ap}
\\[.15cm]
 \Lh_+^{2,2,\Delta}&=\tfrac{1}{3}[(T^+\Th^-\Ah^-)+(\Th^-T^+\Ah^-)+(\Th^-\Th^-A^+)]
 +\tfrac{1}{3+\Delta}\pa\Lh_-^{2',\Delta}
 -\tfrac{2(1+\Delta)}{(2+\Delta)(5+2\Delta)}\pa^2\Lh_-^{\s2,\Delta},
\label{LhTTAp}
\\[.15cm]
 \Lh_+^{2'',\Delta}&=\tfrac{1}{2}[(\pa^2T^+\Ah^-)+(\pa^2\Th^-A^+)]
  -\tfrac{5}{3+\Delta}\pa\Lh_+^{2',\Delta}
 -\tfrac{10}{(2+\Delta)(5+2\Delta)}\pa^2\Lh_+^{2,\Delta}
\nonumber\\[.15cm]
 & 
 -\tfrac{5}{2(1+\Delta)(1+2\Delta)(3+2\Delta)}\pa^4\Ah^-,
\\[.15cm]
 \Lh_+^{2',2,\Delta}&=\tfrac{1}{3}[(\pa T^+\Th^-\Ah^-)+(\pa\Th^-T^+\Ah^-)+(\pa\Th^-\Th^-A^+)]
 -\tfrac{2}{4+\Delta}\pa\Lh_+^{2,2,\Delta}
 +\tfrac{1}{4+\Delta}\pa\Lh_-^{2'',\Delta}
\nonumber\\[.15cm]
 & 
 -\tfrac{-4+\Delta}{(3+\Delta)(7+2\Delta)}\pa^2\Lh_-^{2',\Delta}
 -\tfrac{4(1+2\Delta)}{3(2+\Delta)(3+\Delta)(5+2\Delta)}\pa^3\Lh_-^{2,\Delta},
\label{LhT1TAp}
\\[.15cm]
 \Lh_+^{2,2',\Delta}&=\tfrac{1}{3}[(T^+\pa\Th^-\Ah^-)+(\Th^-\pa T^+\Ah^-)+(\Th^-\pa\Th^-A^+)]
 -\tfrac{2}{4+\Delta}\pa\Lh_+^{2,2,\Delta}
 -\tfrac{1+\Delta}{(3+\Delta)(7+2\Delta)}\pa^2\Lh_-^{2',\Delta}
\nonumber\\[.15cm]
 & 
 -\tfrac{2(7+4\Delta)}{3(2+\Delta)(3+\Delta)(5+2\Delta)}\pa^3\Lh_-^{2,\Delta},
\label{LhTT1Ap}
\\[.15cm]
 \Lh_+^{2''',\Delta}&=\tfrac{1}{2}[(\pa^3T^+\Ah^-)+(\pa^3\Th^-A^+)]
 -\tfrac{9}{4+\Delta}\pa\Lh_+^{2'',\Delta}
 -\tfrac{45}{(3+\Delta)(7+2\Delta)}\pa^2\Lh_+^{2',\Delta}
\nonumber\\[.15cm]
 & 
 -\tfrac{30}{(2+\Delta)(3+\Delta)(5+2\Delta)}\pa^3\Lh_+^{2,\Delta}
 -\tfrac{9}{2(1+\Delta)(2+\Delta)(1+2\Delta)(3+2\Delta)}\pa^5\Ah^-,
\label{LhT3Ap}
\end{align}
and
\begin{align}
 \Lh_-^{2,\Delta}&=(\Th^-\Ah^-),
\label{LhTAm}
\\[.15cm]
 \Lh_-^{2',\Delta}&=(\pa\Th^-\Ah^-)
 -\tfrac{2}{2+\Delta}\pa\Lh_-^{2,\Delta},
\label{LhT1Am}
\\[.15cm]
 \Lh_-^{2,2,\Delta}&=(\Th^-\Th^-\Ah^-),
\label{LhTTAm}
\\[.15cm]
 \Lh_-^{2'',\Delta}&=(\pa^2\Th^-\Ah^-)
  -\tfrac{5}{3+\Delta}\pa\Lh_-^{2',\Delta}
 -\tfrac{10}{(2+\Delta)(5+2\Delta)}\pa^2\Lh_-^{2,\Delta},
\\[.15cm]
 \Lh_-^{2',2,\Delta}&=(\pa\Th^-\Th^-\Ah^-)
 -\tfrac{2}{4+\Delta}\pa\Lh_-^{2,2,\Delta},
\label{LhT1TAm}
\\[.15cm]
 \Lh_-^{2,2',\Delta}&=(\Th^-\pa\Th^-\Ah^-)
 -\tfrac{2}{4+\Delta}\pa\Lh_-^{2,2,\Delta},
\label{LhTT1Am}
\\[.15cm]
 \Lh_-^{2''',\Delta}&=(\pa^3\Th^-\Ah^-)
 -\tfrac{9}{4+\Delta}\pa\Lh_-^{2'',\Delta}
 -\tfrac{45}{(3+\Delta)(7+2\Delta)}\pa^2\Lh_-^{2',\Delta}
 -\tfrac{30}{(2+\Delta)(3+\Delta)(5+2\Delta)}\pa^3\Lh_-^{2,\Delta}.
\label{LhT3Am}
\end{align}

\subsection{Galilean $W_3$ algebra}
\label{Sec:GW3}

The $W_3$ algebra~\cite{Zam85}
is generated by a Virasoro field $T$ and an even primary field $W$ of conformal weight $3$.
The corresponding star products are
\be
 T\ast T\simeq\tfrac{c}{2}\{\I\}+2\{T\},\qquad T\ast W\simeq3\{W\},\qquad 
 W\ast W\simeq \tfrac{c}{3}\{\I\}+2\{T\}+\tfrac{32}{22+5c}\{\La^{\s2,2}\},
\label{W3}
\ee
where the quasi-primary field $\La^{\s2,2}$ is given in (\ref{La22}).
The vacuum module is generated from the highest-weight vector $|0\rangle$ subject to
\be
 L_n|0\rangle=W_{n'}|0\rangle=0,\qquad n\geq-1,\quad n'\geq-2,
\ee
and its Virasoro character is given by
\be
 \chi(q)=\prod_{\Delta=2,3}\,\prod_{n=\Delta}^\infty\frac{1}{1-q^n}
   =1+q^2+2q^3+3q^4+4q^5+8q^6+10q^7+17q^8+\Oc(q^{9}).
\ee

This conformal OPA is readily seen to respect the conditions, outlined in Section~\ref{Sec:AAG}, for admitting a 
Galilean contraction. Indeed, the only term preventing the OPA from being simply-extended is the composite 
quasi-primary field $\La^{\s2,2}=(TT)-\frac{3}{10}\pa^2T$, whose terms have $\ell=2$ and $\ell=1$, respectively, 
multiplied by $\phi(c)=\frac{32}{22+5c}$ satisfying $\sum_i\alpha_i=-1$.
We thus find that the {\em Galilean $W_3$ algebra} is generated by the fields $\{T^+,W^+,\Th^-,\Wh^-\}$,
and that their nontrivial star products are given by
\be
 T^+\gast T^+\!\simeq\tfrac{c^+}{2}\{\I\}+2\{T^+\},\quad T^+\gast W^+\!\simeq3\{W^+\},\quad 
 T^+\gast\Th^-\!\simeq\tfrac{1}{2}\{\I\}+2\{\Th^-\},\quad T^+\gast\Wh^-\!\simeq3\{\Wh^-\},
\label{GW3a}
\ee
and
\begin{align}
 W^+\gast W^+&\simeq\tfrac{c^{+}}{3}\{\I\}+2\{T^+\}+\tfrac{64}{5}\{\Lh_+^{2,2}\}
  -\tfrac{32(44+5c^+)}{25}\{\Lh_-^{2,2}\},
\\[.15cm]
 W^+\gast\Wh^-&\simeq\tfrac{1}{3}\{\I\}+2\{\Th^-\}+\tfrac{32}{5}\{\Lh_-^{2,2}\},
\label{GW3c}
\end{align}
where $\Lh_\pm^{2,2}$ are given in (\ref{Lapm22}).
We have verified that the star relations (\ref{GW3a})-(\ref{GW3c}) indeed define an associative OPA.

We conclude this subsection by noting that the Galilean $W_3$ algebra discussed above is not the only 
possible extension of the Galilean Virasoro algebra $\Vir_G$ by a pair of primary fields of conformal weight $3$. 
For example, letting the nonzero star products of the fields $\{T^\pm,W^\pm\}$ be defined by
\be
 T^+\gast T^+\!\simeq\tfrac{c^+}{2}\{\I\}+2\{T^+\},\quad T^+\gast W^+\!\simeq3\{W^+\},\quad 
 T^\pm\gast T^\mp\!\simeq2\{T^-\},\quad T^\pm\gast W^\mp\!\simeq3\{W^-\},
\ee
and
\be
 W^+\gast W^+\!\simeq \sigma\{\Lh_-^{2,2}\},\qquad \sigma\in\mathbb{C},
\ee
yields a well-defined OPA, where $\{T^\pm\}$ generates a Galilean Virasoro subalgebra with $c^-=0$.

\subsection{Galilean $W(2,4)$ algebra}
\label{Sec:GW24}

The $W(2,4)$ algebra~\cite{Bou88,HT88} 
is generated by a Virasoro field $T$ and an even primary field $W$ of conformal weight $4$.
The corresponding star relations are
\be
 T \ast T \simeq \tfrac{c}{2}\{\I\} + 2\{T \}, \qquad 
 T \ast W \simeq 4\{W\},
\ee
and
\be
 W \ast W\simeq\tfrac{c}{4}\{\I\}+2\{T \}
  +C_{4,4}^{2,2}\{\La^{\s2,2}\}
  +C_{4,4}^{2,2,2}\{\La^{\s2,2,2}\}
  +C_{4,4}^{2'',2}\{\La^{\s2'',2}\} 
  +C_{4,4}^{4}\{W\}  
  +C_{4,4}^{2,4}\{\La^{\s2,4}\},
\label{W24WW}
\ee
where the quasi-primary fields $\La^{\s2,2}$, $\La^{\s2,2,2}$ and $\La^{\s2'',2}$ are given in 
(\ref{La22}) and (\ref{La222})-(\ref{La2''2}), while $\La^{\s2,4}$ follows from (\ref{LTA}) and is given by 
\be
 \La^{\s2,4}=(TW)-\tfrac{1}{6}\pa^2W.
\ee
The structure constants in (\ref{W24WW}) are given by
\be
 C_{4,4}^{2,2}=\tfrac{42}{22+5c},\qquad
 C_{4,4}^{2,2,2}=\tfrac{24(13+72c)}{(-1+2c)(22+5c)(68+7c)},\qquad
 C_{4,4}^{2'',2}=\tfrac{3(-2368-786c+19c^2)}{2(-1+2c)(22+5c)(68+7c)},
\ee
and
\be
\begin{array}{c}
 C_{4,4}^4=3\sqrt{6}\sqrt{\tfrac{(24+c)(196-172c+c^2)}{(-1+2c)(22+5c)(68+7c)}},\qquad
 C_{4,4}^{2,4}=\tfrac{28}{3(24+c)}C_{4,4}^4.
\end{array}
\ee
The vacuum module is generated from the highest-weight vector $|0\rangle$ subject to
\be
 L_n|0\rangle=W_{n'}|0\rangle=0,\qquad n\geq-1,\quad n'\geq-3,
\ee
and its Virasoro character is given by
\be
 \chi(q)=\prod_{\Delta=2,4}\,\prod_{n=\Delta}^\infty\frac{1}{1-q^n}
   =1+q^2+q^3+3q^4+3q^5+6q^6+7q^7+13q^8+\Oc(q^{9}).
\ee

As in the case of the $W_3$ algebra in Section~\ref{Sec:GW3}, this conformal OPA is seen to admit a Galilean 
contraction. We thus find that the {\em Galilean $W(2,4)$ algebra} is generated by the fields 
$\{T^+,W^+,\Th^-,\Wh^-\}$, and that their nontrivial star products are given by
\be
 T^+\gast T^+\!\simeq\tfrac{c^+}{2}\{\I\}+2\{T^+\},\quad T^+\gast W^+\!\simeq4\{W^+\},\quad 
 T^+\gast\Th^-\!\simeq\tfrac{1}{2}\{\I\}+2\{\Th^-\},\quad T^+\gast\Wh^-\!\simeq4\{\Wh^-\},
\label{GW4a}
\ee
and
\begin{align}
 W^+\gast W^+\!&\simeq
  \tfrac{c^+}{4}\{\I\}+2\{T^+\}
  +\tfrac{84}{5}\{\Lh_+^{2,2}\}
  +\tfrac{2592}{35}\{\Lh_+^{2,2,2}\}
  +\tfrac{57}{70}\{\Lh_+^{2'',2}\}
  +\tfrac{3\sqrt{3}}{\sqrt{35}}\{W^+\}
  +\tfrac{8\sqrt{21}}{\sqrt{5}}\{\Lh_+^{2,4}\},
\nonumber\\[.15cm]
 &
  -\tfrac{42(44+5c^+)}{25}\{\Lh_-^{2,2}\}
  -\tfrac{24(33853+2520c^+)}{1225}\{\Lh_-^{2,2,2}\}
  -\tfrac{3(73127+665c^+)}{4900}\{\Lh_-^{2'',2}\}
  -\tfrac{33939\sqrt{3}}{70\sqrt{35}}\{\Wh^-\}
\nonumber\\[.15cm]
 &
  -\tfrac{2\sqrt{3}(14673+70c^+)}{5\sqrt{35}}\{\Lh_-^{2,4}\},
\\[.15cm]
 W^+\gast\Wh^-\!&\simeq
  \tfrac{1}{4}\{\I\}+2\{\Th^-\}
 +\tfrac{42}{5}\{\Lh_-^{2,2}\}
 +\tfrac{864}{35}\{\Lh_-^{2,2,2}\}
 +\tfrac{57}{140}\{\Lh_-^{2'',2}\}
 +\tfrac{3\sqrt{3}}{\sqrt{35}}\{\Wh^-\}
 +\tfrac{4\sqrt{21}}{\sqrt{5}}\{\Lh_-^{2,4}\},
\end{align}
where $\Lh_\pm^{2,2}$, $\Lh_\pm^{2,2,2}$ and $\Lh_\pm^{2'',2}$ are given in (\ref{Lapm22}) 
and (\ref{La222p})-(\ref{La6m}), while
\be
 \Lh_+^{2,4}=\tfrac{1}{2}[(T^+\Wh^-)+(\Th^-W^+)]-\tfrac{1}{6}\pa^2\Wh^-,\qquad
 \Lh_-^{2,4}=(\Th^-\Wh^-),
\ee
follow from (\ref{LhTAp}) and (\ref{LhTAm}).
We have verified that these star relations indeed define an associative OPA.

We conclude this subsection by noting that the Galilean $W(2,4)$ algebra discussed above is not the only 
possible extension of the Galilean Virasoro algebra $\Vir_G$ by a pair of primary fields of conformal weight $4$. 
For example, letting the nonzero star products of the fields $\{T^\pm,W^\pm\}$ be defined by
\be
 T^+\gast T^+\!\simeq\tfrac{c^+}{2}\{\I\}+2\{T^+\},\quad T^+\gast W^+\!\simeq4\{W^+\},\quad 
 T^\pm\gast T^\mp\!\simeq2\{T^-\},\quad T^\pm\gast W^\mp\!\simeq4\{W^-\},
\ee
and
\begin{align}
 W^+\gast W^+\!&\simeq 
 \tfrac{7\sigma_1(8+c^+)}{20}\{\Lh_-^{2,2}\}
 +3\sigma_1\{\Lh_+^{2,2,2}\}
 +\sigma_2\{\Lh_-^{2,2,2}\}
 + \tfrac{\sigma_1(-172+c^+)}{40}\{\Lh_-^{2'',2}\}
 +\sigma_3\{\Lh_-^{2,4}\},
\\[.15cm]
 W^+\gast W^-\!&\simeq \sigma_1\{\Lh_-^{2,2,2}\},
\end{align}
where $\sigma_1,\sigma_2,\sigma_3\in\mathbb{C}$,
yields a well-defined OPA, where $\{T^\pm\}$ generates a Galilean Virasoro subalgebra with $c^-=0$.

\subsection{More Galilean $W$-algebras}
\label{Sec:GW4W5}

The $W$-algebras $W(2,6)$, $W_4$ and $W_5$ are reviewed in Appendix~\ref{Sec:Walg};
their Galilean counterparts are given in Appendix~\ref{Sec:GW}.
Although these Galilean $W$-algebras are straightforward to work out following Section~\ref{Sec:Galilean} 
and Section~\ref{Sec:AAG}, the computations are lengthy and the results rather involved.
In each of the three cases, we have verified explicitly that the corresponding OPA is indeed associative.

One can also apply Galilean contractions to {\em infinitely} generated $W$-algebras. For example, the
algebra $W_{\infty}$~\cite{PRS90} contains a field, $V^{i}$, of each conformal weight 
$\Delta_{V^{i}} = i+2$, where $i\in\mathbb{N}_0$. 
The field $V^{0}$ generates a Virasoro subalgebra, and all other fields are primary fields with respect to $V^0$. 
The following expressions for the mode algebra are taken from~\cite{BS92}:
\be
[V^{i}_{m},V^{j}_{n}] = \sum_{\ell \geq 0} g^{ij}_{2\ell}(m,n)V^{i+j-2\ell}_{m+n} + c_{i}(m)\delta^{ij}\delta_{m+n},
\ee
where the structure constants $g^{ij}_{2\ell}(m,n)$ are independent of $c$, while
\be
c_i(m)= m(m^{2}-1)(m^{2}-4)\ldots(m^{2}-(i+1)^{2})c_{i}, \qquad
c_i= \frac{2^{2i-3}i!(i+2)!}{(2i+1)!!(2i+3)!!}c.
\ee
The summation over $\ell$ is finite since $g^{ij}_{2\ell}(m,n) = 0$ if $2\ell > i+j$.
The corresponding star relations are given by
\be
V^{i}\ast V^{j} \simeq c_{i}\delta_{ij}\{\I\} + \sum_{\ell \geq 0}\tilde{g}^{ij}_{2\ell}\{V^{i+j-2\ell}\},
\ee
where the structure constants $\tilde{g}^{ij}$ are the mode-label independent versions of $g^{ij}_{2\ell}(m,n)$. 
This algebra is readily seen to be simply-extended,
and its Galilean counterpart, $(W_{\infty})_G$, is generated by the fields $\{V^{i,\pm}\}$, with nontrivial star relations
given by
\be
 V^{i,+}\ast V^{j,\pm} \simeq c^\pm_{i}\delta_{ij}\{\I\} + \sum_{\ell \geq 0}\tilde{g}^{ij}_{2\ell}\{V^{i+j-2\ell,\pm}\}.
\ee
This Galilean algebra has also appeared in~\cite{GRR15}.
Other examples of infinitely generated $W$-algebras are the $W_{1+\infty}$ 
algebra~\cite{PRS90a}, various supersymmetric extensions of 
$W_{\infty}$, see~\cite{BPRSS90,BDZ96,GS01},
and a family of nonlinear $W_\infty$ algebras~\cite{FMR93}.
More recent examples, with applications to higher-spin theory, are discussed in~\cite{CG12,BCGG13}.
We have not explicitly calculated Galilean contractions of these algebras.

\section{Discussion}
\label{Sec:Discussion}

We have developed a general Galilean contraction prescription of OPAs. It yields nontrivial
extensions of known symmetry algebras, including $W$-algebras. The results generalise and, where possible, 
match the ones found in the literature, obtained using a similar contraction but in terms of mode algebras.
In particular, several new Galilean $W$-algebras have been constructed, providing evidence for the
existence of a whole new class of such algebras. 
Compatibility between Galilean contractions and certain other operations on OPAs has also been demonstrated.

We have explored what is required of an OPA for it to admit a Galilean contraction.
Thus, assuming that all OPAs {\em do} admit a Galilean contraction, these results have significant implications 
for the possible structure constants a $W$-algebra can have. Indeed, under certain simple and 
rather non-restrictive assumptions about 
the dependence of the structure constants on the central charge (assumptions satisfied by all the $W$-algebras 
we have considered), we have determined very concrete conditions these structure constants must satisfy. 
Expecting universal applicability of the Galilean contraction prescription, we accordingly conjecture that these 
conditions must be respected by all $W$-algebras defined for generic central charge. 
The structure constants in the corresponding 
Galilean $W$-algebras have been characterised explicitly, allowing us to construct subsequently
the new Galilean $W$-algebras mentioned above.
 
We note that the Takiff algebra~\cite{BR13}, see also~\cite{HSSU12}, 
associated to an affine Lie (super)algebra is equivalent to the 
corresponding Galilean affine algebra discussed in Section~\ref{Sec:AGA}. The generalised Sugawara 
construction of the generator $T^+$ of the Virasoro algebra of $\Vir_G$ also appeared in~\cite{BR13},
whereas the similar construction of the generator $T^-$ appears to be new. Extensions of the Takiff construction 
to $W$-algebras were not addressed in~\cite{BR13}.

Due to their central role in a raft of applications, free-field realisations are ubiquitous in conformal field theory, 
see~\cite{DF84,Wak86,FMS86,FL88,FF90,BS92,Fre94,BS95,Ras96,DiFMSbook,Ras98,Kau00,Schbook} 
and references therein. It is therefore of great interest to determine to what extent free fields can be
utilised when Galilean conformal symmetries are present.
Work in this direction has recently been undertaken~\cite{BJMN16,AR16,BJLMN16}, 
constructing free-field realisations of the Galilean Virasoro algebra and some of its superconformal extensions.
It would be very interesting to perform a systematic analysis of the various Galilean algebras in terms of free fields,
and we hope to return with a discussion of this elsewhere.

As discussed in~\cite{RR17}, the Galilean contraction prescription can be generalised from tensor products 
of {\em pairs} of identical OPAs (up to their central parameters) to {\em higher-order} tensor products. 
For each OPA (or equivalently, each VOA) admitting a Galilean contraction, 
this generalisation gives rise to an infinite hierarchy of higher-order Galilean algebras, 
naturally termed higher-order Galilean conformal algebras and, more specifically, 
higher-order Galilean $W$-algebras.

\subsection*{Acknowledgements}

JR was supported by the Australian Research Council under the Future Fellowship scheme, project number 
FT100100774, and under the Discovery Project scheme, project number DP160101376. 
CR was funded by an Australian Postgraduate Award from the Australian Government and by a University of 
Queensland Research Scholarship. This work is largely based on CR's thesis\cite{Ray15} submitted in July 
2015 for the degree of MPhil at the University of Queensland.
JR thanks Costas Zoubos for stimulating discussions during the formative stages of the project,
and the Niels Bohr Institute for making them possible with their hospitality in July 2013.
The authors thank Shashank Kanade, Thomas Quella, Eric Ragoucy, David Ridout, Philippe Ruelle, Aiden Suter, 
and Simon Wood for helpful discussions and comments.

\appendix

\section{$W$-algebras}
\label{Sec:Walg}

In this appendix, we present the defining star relations for the $W$-algebras $W(2,6)$, $W_4$ and $W_5$.
Since all generators are even in these algebras, it must hold that $\Delta_A+\Delta_B-\Delta_Q\in2\mathbb{N}$
for $C_{A,B}^Q\neq0$.
The structure constants are related in various ways, for example by~\cite{BFKNRV91,Hor93}
\be
 \Delta_BC_{A,B}^Q=\Delta_QC_{A,Q}^B,
\ee
but it is beyond the scope of the present work to discuss these relations more generally.

\subsection{$W(2,6)$ algebra}
\label{Sec:W26}

The $W(2,6)$ algebra~\cite{Bou88,FoFS90} 
is generated by a Virasoro field $T$ and an even primary field $W$ of conformal weight $6$.
The corresponding star relations are
\be
 T\ast T \simeq \tfrac{c}{2}\{\I\} + 2\{T \}, \qquad T \ast W \simeq 6\{W\},
\ee
and
\begin{align}
 W \ast W&\simeq 
 \tfrac{c}{6}\{\I\}
 +2\{T\}
 +C_{6,6}^{2,2}\{\La^{\s2,2}\} 
 +C_{6,6}^{2,2,2}\{\La^{\s2,2,2}\}
 +C_{6,6}^{2'',2}\{\La^{\s2'',2}\}
 +C_{6,6}^{2,2,2,2}\{\La^{\s2,2,2,2}\}
\nonumber\\[.15cm]
 &
 +C_{6,6}^{2'',2,2}\{\La^{\s2'',2,2}\}
 +C_{6,6}^{2'''',2}\{\La^{\s2'''',2}\}
 +C_{6,6}^{2,2,2,2,2}\{\La^{\s2,2,2,2,2}\}
 +C_{6,6}^{2'',2,2,2}\{\La^{\s2'',2,2,2}\}
\nonumber\\[.15cm]
 &
 +C_{6,6}^{2'''',2,2}\{\La^{\s2'''',2,2}\}
 +C_{6,6}^{2'''''',2}\{\La^{\s2'''''',2}\}
\nonumber\\[.15cm]
 &
 +C_{6,6}^6\{W\}
 +C_{6,6}^{2,6}\{\La^{\s2,6}\}
 +C_{6,6}^{2,2,6}\{\La^{\s2,2,6}\}
 +C_{6,6}^{2'',6}\{\La^{\s2'',6}\} ,
\end{align}
where $\La^{\s2,2}$, $\La^{\s2,2,2}$ and $\La^{\s2'',2}$ are given in (\ref{La22}) and (\ref{La222})-(\ref{La2''2}); 
$\La^{\s2,2,2,2}$, $\La^{\s2'',2,2}$ and $\La^{\s2'''',2}$ in (\ref{TTTT})-(\ref{T4T}); 
$\La^{\s2,2,2,2,2}$, $\La^{\s2'',2,2,2}$, $\La^{\s2'''',2,2}$ and $\La^{\s2'''''',2}$ in (\ref{TTTTT})-(\ref{T6T});
while 
\begin{align}
 \La^{\s2,6}&=(TW)-\tfrac{3}{26}\pa^2W,
\\[.15cm]
 \La^{\s2',6}&=(\pa TW)-\tfrac{1}{4}\pa\La^{\s2,6}-\tfrac{1}{91}\pa^3W,
\\[.15cm]
 \La^{\s2,2,6}&=(TTW)+\tfrac{1}{6}\pa\La^{\s2',6}
 -\tfrac{21}{136}\pa^2\La^{\s2,6}
 -\tfrac{1}{91}\pa^4W,
\\[.15cm]
 \La^{\s2'',6}&=(\pa^2TW)-\tfrac{5}{9}\pa\La^{\s2',6}
 -\tfrac{5}{68}\pa^2\La^{\s2,6}
 -\tfrac{1}{546}\pa^4W
\end{align}
follow from (\ref{LTA})-(\ref{LT2A}).
The structure constants are given by
\be
\begin{array}{c}
C_{6,6}^{6}=\tfrac{20}{\sqrt{3}}\sqrt{\tfrac{(2+c)(47+c)^2(516+13c)^2(4-388c+c^2)}{(-1+2c)(46+3c)(286+3c)(3+5c)(22+5c)(68+7c)(232+11c)}},
\end{array}
\nonumber
\ee
\begin{align}
\begin{split}
 C_{6,6}^{2,2}&=\tfrac{62}{22+5c},\\[.15cm]
 C_{6,6}^{2,2,2}&=\tfrac{80(35+139c)}{3(-1+2c)(22+5c)(68+7c)},\\[.15cm] 
 C_{6,6}^{2,2,2,2}&=\tfrac{8(-2179+57652c+22992c^2)}{(-1+2c)(46+3c)(3+5c)(22+5c)(68+7c)},\\[.15cm]
 C_{6,6}^{2,2,2,2,2}&=\tfrac{32(15707+874936c+172800c^2)}{(-1+2c)(46+3c)(3+5c)(22+5c)(68+7c)(232+11c)},\\[.15cm] 
 C_{6,6}^{2'',2,2,2}&=\tfrac{20(-1165464-19972498c-1570617c^2+29188c^3)}{3(-1+2c)(46+3c)(3+5c)(22+5c)(68+7c)(232+11c)},
\end{split}
\ \ 
\begin{split}
 C_{6,6}^{2'',2}&=\tfrac{(-740+17c)(-16840-5550c+85c^2)}{2(-1+2c)(68+7c)(-16280-3326c+85c^2)},\\[.15cm]
 C_{6,6}^{2'',2,2}&=\tfrac{436944-3525368c-483594c^2+8093c^3}{2(-1+2c)(46+3c)(3+5c)(22+5c)(68+7c)},\\[.15cm]
 C_{6,6}^{2,6}&=\tfrac{186}{516+13c}C_{6,6}^{6},\\[.15cm]
 C_{6,6}^{2,2,6}&=\tfrac{12(2089+572c)}{5(2+c)(47+c)(516+13c)}C_{6,6}^{6},\\[.15cm]
 C_{6,6}^{2'',6}&=\tfrac{9(-28832-530c+13c^2)}{20(2+c)(47+c)(516+13c)}C_{6,6}^{6},
\end{split}
\nonumber
\end{align}
\begin{align}
 C_{6,6}^{2'''',2}
 &=\tfrac{-7169952+4735580c-119168c^2-24161c^3+326 c^4}{24(-1+2c)(46+3c)(3+5c)(22+5c)(68+7c)},
\nonumber\\[.15cm]
 C_{6,6}^{2'''''',2}
 &=\tfrac{-1049672832-233820912c+7931836c^2+478664c^3-28759c^4+265c^5}{144 (-1+2c)(46+3c)(3+5c)(22+5c)(68+7c)(232+11c)},
\nonumber\\[.15cm]
 C_{6,6}^{2'''',2,2}
 &=\tfrac{10(28554672+39310136c-2532744c^2-142124c^3+2259c^4)}{24(-1+2c)(46+3c)(3+5c)(22+5c)(68+7c)(232+11c)}.
\end{align}
The vacuum module is generated from the highest-weight vector $|0\rangle$ subject to
\be
 L_n|0\rangle=W_{n'}|0\rangle=0,\qquad n\geq-1,\quad n'\geq-5,
\ee
and its Virasoro character is given by
\be
 \chi(q)=\prod_{\Delta=2,6}\,\prod_{n=\Delta}^\infty\frac{1}{1-q^n}
   =1+q^2+q^3+2q^4+2q^5+5q^6+5q^7+9q^8+\Oc(q^{9}).
\ee

\subsection{$W_4$ algebra}
\label{Sec:W4}

The $W_{4}$ algebra~\cite{KW91,BFKNRV91,Zhu93} is generated by the following even fields: 
the Virasoro field $T$, a primary field $W$ of conformal weight $3$, and a primary field $U$ of conformal weight $4$. 
The nontrivial star relations are
\be
 T\ast T\simeq\tfrac{c}{2}\{\I\}+2\{T\},\qquad T\ast W\simeq3\{W\},\qquad T\ast U\simeq4\{U\},
\ee
and
\begin{align}
 W\ast W&\simeq
  \tfrac{c}{3}\{\I\}
  +2\{T\}
  +C_{3,3}^{2,2}\{\La^{\s2,2}\}
  +C_{3,3}^4\{U\}
 ,\\[.15cm]
 W\ast U&\simeq
  C_{3,4}^3\{W\}
 +C_{3,4}^{2,3}\{\La^{\s2,3}\}
 +C_{3,4}^{2',3}\{\La^{\s2',3}\}
 ,\\[.15cm]
 U\ast U&\simeq
  \tfrac{c}{4}\{\I\}+ 2\{T\}
  +C_{4,4}^{2,2}\{\La^{\s2,2}\}
  +C_{4,4}^{2,2,2}\{\La^{\s2,2,2}\}
  +C_{4,4}^{2'',2}\{\La^{\s2'',2}\}
  \nonumber\\[.15cm]
 &
  +C_{4,4}^{3,3}\{\La^{\s3,3}\}
  +C_{4,4}^{4}\{U\} 
  +C_{4,4}^{2,4}\{\La^{\s2,4}\} 
 ,
\end{align}
where $\La^{\s2,2}$,
$\La^{\s2,2,2}$ and $\La^{\s2'',2}$ are given in (\ref{La22}) and (\ref{La222})-(\ref{La2''2});
\begin{align}
 \La^{\s2,3}&=(TW)-\tfrac{3}{14}\pa^2W,
\\[.15cm]
 \La^{\s2',3}&=(\pa TW)-\tfrac{2}{5}\pa\La^{\s2,3}-\tfrac{1}{28}\pa^3W,
\\[.15cm]
 \La^{\s2,4}&=(TU)-\tfrac{1}{6}\pa^2U
\end{align}
follow from (\ref{LTA}) and (\ref{LT1A}); while
\be
 \La^{\s3,3}=(WW)-\tfrac{5}{36}\pa^2\big[C_{3,3}^{2,2}\La^{\s2,2}+C_{3,3}^4U\big]-\tfrac{1}{84}\pa^4T.
\ee
The structure constants are given by
\be
\begin{array}{rclrclrcl}
 C_{3,3}^{2,2}&\!\!\!\!=\!\!\!\!&\tfrac{32}{22+5c},
 &\qquad
 C_{3,3}^4&\!\!\!\!=\!\!\!\!&\tfrac{4}{\sqrt{3}}\sqrt{\frac{(2+c)(114+7c)}{(7+c)(22+5c)}},
 &\qquad
 &\!\!\!&
\\[.3cm]
 C_{3,4}^3&\!\!\!\!=\!\!\!\!&\tfrac{3}{4}C_{3,3}^4,
 &\qquad
 C_{3,4}^{2,3}&\!\!\!\!=\!\!\!\!&\tfrac{39}{114+7c}C_{3,3}^4,
 &\qquad
 C_{3,4}^{2',3}&\!\!\!\!=\!\!\!\!&\tfrac{3}{4(2+c)}C_{3,3}^4,
\\[.3cm]
 C_{4,4}^{2,2}&\!\!\!\!=\!\!\!\!&\tfrac{42}{22+5c},
 &\qquad
 C_{4,4}^{2,2,2}&\!\!\!\!=\!\!\!\!&\tfrac{96(-2+9c)}{(2+c)(22+5c)(114+7c)},
 &\qquad
 C_{4,4}^{2'',2}&\!\!\!\!=\!\!\!\!&\tfrac{3(-2484-844c+19c^2)}{4(2+c)(22+5c)(114+7c)},
\\[.3cm]
 C_{4,4}^{3,3}&\!\!\!\!=\!\!\!\!&\tfrac{45(22+5c)}{2(2+c)(114+7c)},
 &\qquad
 C_{4,4}^{4}&\!\!\!\!=\!\!\!\!&\tfrac{-9(218+c+c^2)}{4(2+c)(114+7c)}C_{3,3}^4,
 &\qquad
 C_{4,4}^{2,4}&\!\!\!\!=\!\!\!\!&\tfrac{-3}{2+c}C_{3,3}^4.
\end{array}
\ee
The vacuum module is generated from the highest-weight vector $|0\rangle$ subject to
\be
 L_n|0\rangle=W_{n'}|0\rangle=U_{n''}|0\rangle=0,\qquad n\geq-1,\quad n'\geq-2,\quad n''\geq-3,
\ee
and its Virasoro character is given by
\be
 \chi(q)=\prod_{\Delta=2}^4\,\prod_{n=\Delta}^\infty\frac{1}{1-q^n}
   =1+q^2+2q^3+4q^4+5q^5+10q^6+14q^7+25q^8+\Oc(q^{9}).
\ee

\subsection{$W_5$ algebra}
\label{Sec:W5}

The $W_5$ algebra~\cite{Hor93,Zhu93} 
is generated by the Virasoro field $T$ and the three even primary fields 
$W$, $U$ and $V$ of conformal weight $3$,  $4$ and $5$, respectively. The nontrivial star relations are
\be
 T\ast T\simeq\tfrac{c}{2}\{\I\}+2\{T\},
 \qquad 
 T\ast W\simeq3\{W\},
 \qquad 
 T\ast U\simeq4\{U\},
 \qquad 
 T\ast V\simeq5\{V\},
\ee
and
\begin{align}
 W\ast W &\simeq 
  \tfrac{c}{3}\{\I\}
  +2\{T\} 
 +C_{3,3}^{2,2}\{\La^{\s2,2}\} 
 +C_{3,3}^4\{ U\}
 ,\\[.15cm]
 W\ast U&\simeq\
 C_{3,4}^3\{W\}  
 +C_{3,4}^{2,3}\{\La^{\s2,3}\} 
 +C_{3,4}^{2',3}\{\La^{\s2',3}\} 
 +C_{3,4}^5\{V\}
 ,\\[.15cm]
 W\ast V&\simeq 
 C_{3,5}^{2,2,2}\{\La^{\s2,2,2}\}
 +C_{3,5}^{2'',2}\{\La^{\s2'',2}\}
 +C_{3,5}^{3,3}\{\La^{\s3,3}\}
 +C_{3,5}^4\{U\} 
 +C_{3,5}^{2,4}\{ \La^{\s2,4}\}
 +C_{3,5}^{2',4}\{\La^{\s2',4}\}
 ,\\[.15cm]
 U\ast U&\simeq
 \tfrac{c}{4}\{\I\}
 +2\{T\} 
 +C_{4,4}^{2,2}\{\La^{\s2,2}\} 
 +C_{4,4}^{2,2,2}\{\La^{\s2,2,2}\} 
 +C_{4,4}^{2'',2}\{\La^{\s2'',2}\}
 \nonumber\\[.15cm]
&
 +C_{4,4}^{3,3}\{\La^{\s3,3}\}
 +C_{4,4}^{4}\{U\} 
 +C_{4,4}^{2,4}\{\La^{\s2,4}\} 
 ,\\[.15cm]
 U\ast V&\simeq
 C_{4,5}^3\{W \} 
 +C_{4,5}^{2,3}\{\La^{\s2,3}\} 
 +C_{4,5}^{2',3}\{\La^{\s2',3}\}
 +C_{4,5}^{2'',3}\{\La^{\s2'',3}\}
 +C_{4,5}^{2''',3}\{\La^{\s2''',3}\} 
 +C_{4,5}^{2,2,3}\{\La^{\s2,2,3}\} 
 \nonumber\\[.15cm]
 &
 +C_{4,5}^{2',2,3}\{\La^{\s2',2,3}\} 
 +C_{4,5}^{5}\{V\} 
 +C_{4,5}^{2,5}\{\La^{\s2,5}\}
 +C_{4,5}^{2',5}\{\La^{\s2',5}\}
 +C_{4,5}^{3,4}\{\La^{\s3,4}\} 
 +C_{4,5}^{3',4}\{\La^{\s3',4}\} 
 ,\\[.15cm]
 V\ast V&\simeq
 \tfrac{c}{5}\{\I\}
 +2\{T\} 
 +C_{5,5}^{2,2}\{\La^{\s2,2}\}
 +C_{5,5}^{2,2,2}\{ \La^{\s2,2,2}\}
 +C_{5,5}^{2'',2}\{\La^{\s2'',2}\}
 +C_{5,5}^{2,2,2,2}\{\La^{\s2,2,2,2}\}
 \nonumber\\[.15cm]
&
 +C_{5,5}^{2'',2,2}\{\La^{\s2'',2,2}\} 
 +C_{5,5}^{2'''',2}\{\La^{\s2'''',2}\} 
 +C_{5,5}^{3,3}\{\La^{\s3,3}\}
 +C_{5,5}^{2,3,3}\{\La^{\s2,3,3}\}
 +C_{5,5}^{3'',3}\{\La^{\s3'',3}\} 
 \nonumber\\[.15cm]
&
 +C_{5,5}^{4}\{U\}
 +C_{5,5}^{2,4}\{\La^{\s2,4}\} 
 +C_{5,5}^{2,2,4}\{\La^{\s2,2,4}\} 
 +C_{5,5}^{2'',4}\{\La^{\s2'',4}\} 
 +C_{5,5}^{4,4}\{\La^{\s4,4}\} 
 +C_{5,5}^{3,5}\{\La^{\s3,5}\}
,
\end{align}
where $\La^{\s2,2}$, $\La^{\s2,2,2}$, $\La^{\s2'',2}$, $\La^{\s2,2,2,2}$, $\La^{\s2'',2,2}$ and $\La^{\s2'''',2}$
are given in (\ref{La22}), (\ref{La222})-(\ref{La2''2}) and (\ref{TTTT})-(\ref{T4T});
\begin{align}
 \La^{\s2,3}&=(TW)-\tfrac{3}{14}\pa^2W,
\nonumber\\[.15cm]
 \La^{\s2',3}&=(\pa TW)-\tfrac{2}{5}\pa\La^{\s2,3}-\tfrac{1}{28}\pa^3W,
\nonumber\\[.15cm]
 \La^{\s2'',3}&=(\pa^2TW)-\tfrac{5}{6}\pa\La^{\s2',3}-\tfrac{2}{11}\pa^2\La^{\s2,3}-\tfrac{5}{504}\pa^4W,
\nonumber\\[.15cm]
 \La^{\s2''',3}&=(\pa^3TW)-\tfrac{9}{7}\pa\La^{\s2'',3}-\tfrac{15}{26}\pa^2\La^{\s2',3}
  -\tfrac{1}{11}\pa^3\La^{\s2,3}-\tfrac{1}{280}\pa^5W
\nonumber\\[.15cm]
 \La^{\s2,2,3}&=(TTW)+\tfrac{1}{4}\pa\La^{\s2',3}-\tfrac{12}{55}\pa^2\La^{\s2,3}-\tfrac{11}{336}\pa^4W,
\nonumber\\[.15cm]
 \La^{\s2',2,3}&=(\pa TTW)-\tfrac{2}{7}\pa\La^{\s2,2,3}+\tfrac{3}{14}\pa\La^{\s2'',3}+\tfrac{1}{52}\pa^2\La^{\s2',3}
  -\tfrac{7}{165}\pa^3\La^{\s2,3}-\tfrac{1}{240}\pa^5W,
\nonumber\\[.15cm]
 \La^{\s2,4}&=(TU)-\tfrac{1}{6}\pa^2U,
\nonumber\\[.15cm]
 \La^{\s2',4}&=(\pa TU)-\tfrac{1}{3}\pa\La^{\s2,4}-\tfrac{1}{45}\pa^3U,
\nonumber\\[.15cm]
 \La^{\s2,2,4}&=(TTU)+\tfrac{3}{14}\pa\La^{\s2',4}-\tfrac{5}{26}\pa^2\La^{\s2,4}-\tfrac{7}{330}\pa^4U,
\nonumber\\[.15cm]
 \La^{\s2'',4}&=(\pa^2TU)-\tfrac{5}{7}\pa\La^{\s2',4}-\tfrac{5}{39}\pa^2\La^{\s2,4}-\tfrac{1}{198}\pa^4U,
\nonumber\\[.15cm]
 \La^{\s2,5}&=(TV)-\tfrac{3}{22}\pa^2V,
\nonumber\\[.15cm]
 \La^{\s2',5}&=(\pa TV)-\tfrac{2}{7}\pa\La^{\s2,5}-\tfrac{1}{66}\pa^3V
\end{align}
follow from (\ref{LTA})-(\ref{LT3A}); while
\begin{align}
 \La^{\s3,3}&=(WW)-\tfrac{5}{36}\pa^2\big[C_{3,3}^{2,2}\La^{\s2,2}+C_{3,3}^4U\big]-\tfrac{1}{84}\pa^4T,
\nonumber\\[.15cm]
 \La^{\s2,3,3}&=(TWW)+\tfrac{5}{28}C_{3,3}^4\pa\La^{\s2',4}-\tfrac{5}{78}C_{3,3}^4\pa^2\La^{\s2,4}
  -\tfrac{3}{26}\pa^2\La^{\s3,3}-\tfrac{80}{39(22+5c)}\pa^2\La^{\s2,2,2}-\tfrac{5(-2+c)}{52(22+5c)}\pa^2\La^{\s2'',2}
\nonumber\\[.15cm]
 &-\tfrac{5(94+c)}{25344}C_{3,3}^{2,2}\pa^4\La^{\s2,2}-\tfrac{7}{396}C_{3,3}^4\pa^4U-\tfrac{5}{3024}\pa^6T,
\nonumber\\[.15cm]
 \La^{\s3'',3}&=(\pa^2WW)-\tfrac{7}{26}\pa^2\La^{\s3,3}
 -\tfrac{7}{792}\pa^4\big[C_{3,3}^{2,2}\La^{\s2,2}+C_{3,3}^4U\big]-\tfrac{1}{2160}\pa^6T,
\nonumber\\[.15cm]
 \La^{\s3,4}&=(WU)-\tfrac{5}{12}C_{3,4}^{2',3}\pa\La^{\s2',3}
  -\tfrac{1}{11}\pa^2\big[C_{3,4}^{2,3}\La^{\s2,3}+C_{3,4}^5V\big]
  -\tfrac{5}{3024}C_{3,4}^3\pa^4W,
\nonumber\\[.15cm]
 \La^{\s3',4}&=(\pa WU)
  -\tfrac{3}{7}\pa\La^{\s3,4}
  -\tfrac{5}{52}C_{3,4}^{2',3}\pa^2\La^{\s2',3}
  -\tfrac{1}{66}\pa^3\big[C_{3,4}^{2,3}\La^{\s2,3}+C_{3,4}^5V\big]
  -\tfrac{1}{5040}C_{3,4}^3\pa^5W,
\nonumber\\[.15cm]
 \La^{\s3,5}&=(WV)
  -\tfrac{5}{14}C_{3,5}^{2',4}\pa\La^{\s2',4}
  -\tfrac{5}{78}\pa^2\big[C_{3,5}^{2,2,2}\La^{\s2,2,2}
   +C_{3,5}^{2'',2}\La^{\s2'',2}+C_{3,5}^{3,3}\La^{\s3,3}+C_{3,5}^{2,4}\La^{\s2,4}\big]
  -\tfrac{1}{1584}C_{3,5}^4\pa^4U,
\nonumber\\[.15cm]
 \La^{\s4,4}&=(UU)
  -\tfrac{7}{52}\pa^2\big[C_{4,4}^{2,2,2}\La^{\s2,2,2}+C_{4,4}^{2'',2}\La^{\s2'',2}
   +C_{4,4}^{3,3}\La^{\s3,3}+C_{4,4}^{2,4}\La^{\s2,4}\big]
  -\tfrac{7}{1584}\pa^4\big[C_{4,4}^{2,2}\La^{\s2,2}+C_{4,4}^4U\big]
\nonumber\\[.15cm]
 &
  -\tfrac{1}{4320}\pa^6T
.
\end{align}
The structure constants are given by
\be
\begin{array}{rclrclrcl}
 C_{3,3}^{2,2}&\!\!\!\!=\!\!\!\!&\tfrac{32}{22+5c},
 &\qquad
 C_{3,3}^4&\!\!\!\!=\!\!\!\!&\tfrac{32}{\sqrt{3}}\sqrt{\frac{(2+c)(23+c)}{(22+5c)(68+7c)}},
 &\qquad
 C_{3,4}^5&\!\!\!\!=\!\!\!\!&5\sqrt{\frac{(116+3c)(22+5c)}{(68+7c)(114+7c)}},
\\[.3cm]
 C_{3,4}^3&\!\!\!\!=\!\!\!\!&\tfrac{3}{4}C_{3,3}^4,
 &\qquad
 C_{3,4}^{2,3}&\!\!\!\!=\!\!\!\!&\tfrac{39}{114+7c}C_{3,3}^4,
 &\qquad
 C_{3,4}^{2',3}&\!\!\!\!=\!\!\!\!&\tfrac{3}{4(2+c)}C_{3,3}^4,
\\[.3cm]
 C_{3,5}^4&\!\!\!\!=\!\!\!\!&\tfrac{4}{5}C_{3,4}^5,
 &\qquad
 C_{3,5}^{2,4}&\!\!\!\!=\!\!\!\!&\tfrac{32(22+3c)}{(116+3c)(22+5c)}C_{3,4}^5,
 &\qquad
 C_{3,5}^{2',4}&\!\!\!\!=\!\!\!\!&\tfrac{32}{5(22+5c)}C_{3,4}^5,
\end{array}
\ee
\be
\begin{array}{rclrclrcl}
 C_{3,5}^{2,2,2}&\!\!\!\!=\!\!\!\!&\tfrac{-6(22+191c)}{5(2+c)(23+c)(116+3c)(22+5c)}C_{3,3}^4C_{3,4}^5,
 &\quad
 C_{3,5}^{2'',2}&\!\!\!\!=\!\!\!\!&\tfrac{-9(-3784-1266c+43c^2)}{80(2+c)(23+c)(116+3c)(22+5c)}C_{3,3}^4C_{3,4}^5,
\\[.3cm]
 C_{3,5}^{3,3}&\!\!\!\!=\!\!\!\!&\tfrac{9(-1+2c)(68+7c)}{40(2+c)(23+c)(116+3c)}C_{3,3}^4C_{3,4}^5,
 &\quad
 C_{4,4}^{2,2}&\!\!\!\!=\!\!\!\!&\tfrac{42}{22+5c},
\\[.3cm]
 C_{4,4}^{2,2,2}&\!\!\!\!=\!\!\!\!&\tfrac{72(38+3c)(-1+4c)}{(2+c)(23+c)(22+5c)(68+7c)},
 &\quad
 C_{4,4}^{2'',2}&\!\!\!\!=\!\!\!\!&\tfrac{3(-31888-13380c-624c^2+19c^3)}{4(2+c)(23+c)(22+5c)(68+7c)},
\\[.3cm]
 C_{4,4}^{3,3}&\!\!\!\!=\!\!\!\!&\tfrac{9(22+5c)}{2(2+c)(23+c)},
 &\quad
 C_{4,4}^{4}&\!\!\!\!=\!\!\!\!&\tfrac{9(-128+70c+c^2)}{32(2+c)(23+c)}C_{3,3}^4,
\\[.3cm]
 C_{4,4}^{2,4}&\!\!\!\!=\!\!\!\!&\tfrac{3(-118+7c)}{8(2+c)(23+c)}C_{3,3}^4,
 &\quad
 C_{4,5}^3&\!\!\!\!=\!\!\!\!&\tfrac{3}{5}C_{3,4}^5,
\\[.3cm]
 C_{4,5}^{2,3}&\!\!\!\!=\!\!\!\!&\tfrac{198}{5(114+7c)}C_{3,4}^5,
 &\quad
 C_{4,5}^{2',3}&\!\!\!\!=\!\!\!\!&\tfrac{4}{5(2+c)}C_{3,4}^5,
\\[.3cm]
 C_{4,5}^{2'',3}&\!\!\!\!=\!\!\!\!&\tfrac{-3617880-434632c-6392c^2+297c^3}{10(23+c)(116+3c)(22+5c)(114+7c)}C_{3,4}^5,
 &\quad
 C_{4,5}^{2''',3}&\!\!\!\!=\!\!\!\!&\tfrac{6(-120-38c+c^2)}{5(2+c)(22+5c)(114+7c)}C_{3,4}^5,
\\[.3cm]
 C_{4,5}^{2,2,3}&\!\!\!\!=\!\!\!\!&\tfrac{12(-28834+23921c+1224c^2)}{5(23+c)(116+3c)(22+5c)(114+7c)}C_{3,4}^5,
 &\quad
 C_{4,5}^{2',2,3}&\!\!\!\!=\!\!\!\!&\tfrac{144(-2+9c)}{5(2+c)(22+5c)(114+7c)}C_{3,4}^5,
\\[.3cm]
 C_{4,5}^{5}&\!\!\!\!=\!\!\!\!&\tfrac{-15(70272+9340c+204c^2+11c^3)}{64(2+c)(23+c)(114+7c)}C_{3,3}^4,
 &\quad
 C_{4,5}^{2,5}&\!\!\!\!=\!\!\!\!&\tfrac{-15(7796+1196c+29c^2)}{16(2+c)(23+c)(114+7c)}C_{3,3}^4,
\\[.3cm]
 C_{4,5}^{2',5}&\!\!\!\!=\!\!\!\!&\tfrac{-3(13320+262c+11c^2)}{16(2+c)(23+c)(114+7c)}C_{3,3}^4,
 &\quad
 C_{4,5}^{3,4}&\!\!\!\!=\!\!\!\!&\tfrac{3(68+7c)(334+37c)}{80(2+c)(23+c)(116+3c)}C_{3,3}^4C_{3,4}^5,
\\[.3cm]
 C_{4,5}^{3',4}&\!\!\!\!=\!\!\!\!&\tfrac{9(68+7c)}{80(2+c)(23+c)}C_{3,3}^4C_{3,4}^5,
 &\quad
  C_{5,5}^{2,2}&\!\!\!\!=\!\!\!\!&\tfrac{52}{22+5c},
\end{array}
\ee
\be
\begin{array}{rclrclrcl}
 C_{5,5}^{3,3}&\!\!\!\!=\!\!\!\!&\tfrac{3(1507824+248948c+14880c^2+181c^3)}{2(2+c)(23+c)(116+3c)(114+7c)},
 &\quad
 C_{5,5}^{2,3,3}&\!\!\!\!=\!\!\!\!&\tfrac{48(-1+2c)(572+31c)}{(2+c)(23+c)(116+3c)(114+7c)},
\\[.3cm]
 C_{5,5}^{3'',3}&\!\!\!\!=\!\!\!\!&\tfrac{2(-13656-306c+11c^2)}{(2+c)(116+3c)(114+7c)},
&\quad
 C_{5,5}^{4}&\!\!\!\!=\!\!\!\!&\tfrac{-3(70272+9340c+204c^2+11c^3)}{16(2+c)(23+c)(114+7c)}C_{3,3}^4,
\\[.3cm]
 C_{5,5}^{2,4}&\!\!\!\!=\!\!\!\!&\tfrac{-3(3767568+452876c+11520c^2+187c^3)}{8(2+c)(23+c)(116+3c)(114+7c)}C_{3,3}^4,
&\quad
 C_{5,5}^{2,2,4}&\!\!\!\!=\!\!\!\!&\tfrac{-48(-5266+23131c+2393c^2+43c^3)}{(2+c)(23+c)(116+3c)(22+5c)(114+7c)}C_{3,3}^4,
\\[.3cm]
 C_{5,5}^{4,4}&\!\!\!\!=\!\!\!\!&\tfrac{64(114+7c)}{(116+3c)(22+5c)},
&\quad
 C_{5,5}^{3,5}&\!\!\!\!=\!\!\!\!&\tfrac{-3(68+7c)}{8(2+c)(23+c)}C_{3,3}^4C_{3,4}^5,
\end{array}
\ee
and
\be
\begin{array}{rclrclrcl}
 C_{5,5}^{2,2,2}&\!\!\!\!=\!\!\!\!&\tfrac{24(-3744688+14490156c+1942364c^2+86853c^3+1148c^4)}{(2+c)(23+c)(116+3c)(22+5c)(68+7c)(114+7c)},
\\[.3cm]
 C_{5,5}^{2'',2}&\!\!\!\!=\!\!\!\!&\tfrac{-3179356800-1514441152c-143786712c^2-4143084c^3+28774c^4+1491c^5}{4(2+c)(23+c)(116+3c)(22+5c)(68+7c)(114+7c)},
\\[.3cm]
 C_{5,5}^{2,2,2,2}&\!\!\!\!=\!\!\!\!&\tfrac{768(10972-84704c+171793c^2+17652c^3+504c^4)}{(2+c)(23+c)(116+3c)(22+5c)^2(68+7c)(114+7c)},
\\[.3cm]
 C_{5,5}^{2'',2,2}&\!\!\!\!=\!\!\!\!&\tfrac{4(148360896-564291840c-139424540c^2-9031284c^3-142659c^4+2492c^5)}{(2+c)(23+c)(116+3c)(22+5c)^2(68+7c)(114+7c)},
\\[.3cm]
 C_{5,5}^{2'''',2}&\!\!\!\!=\!\!\!\!&\tfrac{208376832+1627582784c+197099280c^2-160532c^3-738044c^4-20289c^5+119c^6}{3(2+c)(23+c)(116+3c)(22+5c)^2(68+7c)(114+7c)},
\\[.3cm]
 C_{5,5}^{2'',4}&\!\!\!\!=\!\!\!\!&\tfrac{-(-155818176-17252736c-139972c^2+17732c^3+203c^4)}{16(2+c)(23+c)(116+3c)(22+5c)(114+7c)}C_{3,3}^4
 .
\end{array}
\ee
The vacuum module is generated from the highest-weight vector $|0\rangle$ subject to
\be
 L_n|0\rangle=W_{n'}|0\rangle=U_{n''}|0\rangle=V_{n'''}|0\rangle=0,\qquad 
 n\geq-1,\quad n'\geq-2,\quad n''\geq-3,\quad n'''\geq-4,
\ee
and its Virasoro character is given by
\be
 \chi(q)=\prod_{\Delta=2}^5\,\prod_{n=\Delta}^\infty\frac{1}{1-q^n}
   =1+q^2+2q^3+4q^4+6q^5+11q^6+16q^7+29q^8+\Oc(q^{9}).
\ee

\section{Galilean $W$-algebras}
\label{Sec:GW}

In this appendix, we present the Galilean contracted algebras constructed from the $W$-algebras discussed in 
Appendix~\ref{Sec:Walg}. The results are presented using star relations, and
we have verified that the ensuing algebras are indeed associative OPAs.
Supplementing the discussion of the Galilean
Virasoro algebra in Section~\ref{Sec:GQP}, the following quasi-primary fields are used in the decomposition of 
some of the star products:
\begin{align}
 \Lh_+^{2,2,2,2}&=(T^+\Th^-\Th^-\Th^-)-\tfrac{3}{26}\pa^2\Lh_-^{2,2,2},
\label{Lh2222p}
\\[.15cm]
 \Lh_+^{2'',2,2}&=\tfrac{1}{3}[(\pa^2T^+\Th^-\Th^-)+(\pa^2\Th^-T^+\Th^-)+(\pa^2\Th^-\Th^-T^+)]
  -\tfrac{5}{39}\pa^2\Lh_+^{2,2,2}
  +\tfrac{4}{39}\pa^2\Lh_-^{2'',2}
  -\tfrac{4}{297}\pa^4\Lh_-^{2,2},
\\[.15cm]
 \Lh_+^{2'''',2}&=\tfrac{1}{2}[(\pa^4T^+\Th^-)+(\pa^4\Th^-T^+)]
  -\tfrac{21}{13}\pa^2\Lh_+^{2'',2}
  -\tfrac{7}{66}\pa^4\Lh_+^{2,2}
  -\tfrac{1}{180}\pa^6\Th^-,
\label{Lh211112p}
\\[.15cm]
 \Lh_+^{2''',2,2}&=\tfrac{1}{3}[(\pa^3T^+\Th^-\Th^-)+(\pa^3\Th^-T^+\Th^-)+(\pa^3\Th^-\Th^-T^+)]
  -\tfrac{9}{8}\pa\Lh_+^{2'',2,2}
  +\tfrac{1}{8}\pa\Lh_-^{2'''',2}
  -\tfrac{5}{91}\pa^3\Lh_+^{2,2,2}
\nonumber\\[.15cm] 
  &+\tfrac{9}{182}\pa^3\Lh_-^{2'',2}
   -\tfrac{1}{165}\pa^5\Lh_-^{2,2},
\\[.15cm]
 \Lh_+^{2,2,2,2,2}&=(T^+\Th^-\Th^-\Th^-\Th^-)
  -\tfrac{3}{34}\pa^2\Lh_-^{2,2,2,2},
\\[.15cm]
 \Lh_+^{2'',2,2,2}&=\tfrac{1}{4}[(\pa^2T^+\Th^-\Th^-\Th^-)+(\pa^2\Th^-T^+\Th^-\Th^-)+(\pa^2\Th^-\Th^-T^+\Th^-)
  +(\pa^2\Th^-\Th^-\Th^-T^+)]
\nonumber\\[.15cm]
 &+\tfrac{7}{36}\pa\Lh_-^{2''',2,2}
  -\tfrac{5}{68}\pa^2\Lh_+^{2,2,2,2}
  +\tfrac{27}{544}\pa^2\Lh_-^{2'',2,2}
  -\tfrac{1}{156}\pa^4\Lh_-^{2,2,2},
\\[.15cm]
 \Lh_+^{2'''',2,2}&=\tfrac{1}{3}[(\pa^4T^+\Th^-\Th^-)+(\pa^4\Th^-T^+\Th^-)+(\pa^4\Th^-\Th^-T^+)]
  -\tfrac{14}{9}\pa\Lh_+^{2''',2,2}-\tfrac{63}{68}\pa^2\Lh_+^{2'',2,2}+\tfrac{13}{68}\pa^2\Lh_-^{2'''',2}
\nonumber\\[.15cm]
 &-\tfrac{1}{39}\pa^4\Lh_+^{2,2,2}
  +\tfrac{4}{195}\pa^4\Lh_-^{2'',2}-\tfrac{119}{38610}\pa^6\Lh_-^{2,2},
\\[.15cm]
 \Lh_+^{2'''''',2}&=\tfrac{1}{2}[(\pa^6T^+\Th^-)+(\pa^6\Th^-T^+)]-\tfrac{135}{34}\pa^2\Lh_+^{2'''',2}
  -\tfrac{18}{13}\pa^4\Lh_+^{2'',2}-\tfrac{7}{143}\pa^6\Lh_+^{2,2}-\tfrac{3}{1540}\pa^8\Th^-,
\end{align} 
and
\begin{align}
 \Lh_-^{2,2,2,2}&=(\Th^-\Th^-\Th^-\Th^-),
\label{Lh2222m}
\\[.15cm]
 \Lh_-^{2'',2,2}&=(\pa^2\Th^-\Th^-\Th^-)-\tfrac{5}{39}\pa^2\Lh_-^{2,2,2},
\\[.15cm]
 \Lh_-^{2'''',2}&=(\pa^4\Th^-\Th^-)-\tfrac{21}{13}\pa^2\Lh_-^{2'',2}-\tfrac{7}{66}\pa^4\Lh_-^{2,2},
\label{Lh211112m}
\\[.15cm]
 \Lh_-^{2''',2,2}&=(\pa^3\Th^-\Th^-\Th^-)-\tfrac{9}{8}\pa\Lh_-^{2'',2,2}-\tfrac{5}{91}\pa^3\Lh_-^{2,2,2},
\\[.15cm]
 \Lh_-^{2,2,2,2,2}&=(\Th^-\Th^-\Th^-\Th^-\Th^-),
\\[.15cm]
 \Lh_-^{2'',2,2,2}&=(\pa^2\Th^-\Th^-\Th^-\Th^-)-\tfrac{5}{68}\pa^2\Lh_-^{2,2,2,2},
\\[.15cm]
 \Lh_-^{2'''',2,2}&=(\pa^4\Th^-\Th^-\Th^-)-\tfrac{14}{9}\pa\Lh_-^{2''',2,2}-\tfrac{63}{68}\pa^2\Lh_-^{2'',2,2}
  -\tfrac{1}{39}\pa^4\Lh_-^{2,2,2},
\\[.15cm]
 \Lh_-^{2'''''',2}&=(\pa^6\Th^-\Th^-)-\tfrac{135}{34}\pa^2\Lh_-^{2'''',2}-\tfrac{18}{13}\pa^4\Lh_-^{2'',2}
  -\tfrac{7}{143}\pa^6\Lh_-^{2,2}.
\label{Lh21111112m}
\end{align}

\subsection{Galilean $W(2,6)$ algebra}
\label{Sec:GW26}

The Galilean $W(2,6)$ algebra $(W(2,6))_G$ is generated by the fields $\{T^{+},W^{+},\Th^{-},\Wh^{-}\}$. 
The fields $\{T^+,\Th^-\}$ generate a subalgebra isomorphic to $\Vir_G$, while $\{W^+,\Wh^-\}$ are primary fields 
of conformal weight $6$ with respect to the Virasoro generator $T^+$. 
The nontrivial star products not involving $T^+$ nor $\Th^-$ are given by
\begin{align}
W^+\gast W^+\!&\simeq
   \tfrac{c^+}{6}\{\I\}
  +2\{T^+\}
  +\tfrac{124}{5}\{\Lh_+^{2,2}\}
  +\tfrac{1112}{7}\{\Lh_+^{2,2,2}\}
  +\tfrac{17}{14}\{\Lh_+^{2'',2}\}
  +\tfrac{122624}{175}\{\Lh_+^{2,2,2,2}\}
  +\tfrac{8093}{700}\{\Lh_+^{2'',2,2}\}
 \nonumber\\[.15cm]
 &
  +\tfrac{163}{6300}\{\Lh_+^{2'''',2}\}
  +\tfrac{184320}{77}\{\Lh_+^{2,2,2,2,2}\}
  +\tfrac{233504}{3465}\{\Lh_+^{2'',2,2,2}\}
  +\tfrac{753}{3080}\{\Lh_+^{2'''',2,2}\}
  +\tfrac{53}{166320}\{\Lh_+^{2'''''',2}\}
 \nonumber\\[.15cm]
 &
  +\tfrac{26\sqrt{2}}{3\sqrt{231}}\{W^+\}
  +\tfrac{248\sqrt{2}}{\sqrt{231}}\{\Lh_+^{2,6}\}
  +\tfrac{416\sqrt{66}}{5\sqrt{7}}\{\Lh_+^{2,2,6}\}
  +\tfrac{13\sqrt{6}}{5\sqrt{77}}\{\Lh_+^{2'',6}\}
  -\tfrac{62(44+5c^+)}{25}\{\Lh_-^{2,2}\}
 \nonumber\\[.15cm]
 &
  -\tfrac{8(130017+9730c^+)}{735}\{\Lh_-^{2,2,2}\}
  -\tfrac{93901+595c^+}{980}\{\Lh_-^{2'',2}\}
  -\tfrac{32(1087987+60354c^+)}{3675}\{\Lh_-^{2,2,2,2}\}
 \nonumber\\[.15cm]
 &
  -\tfrac{30354361+339906c^+}{44100}\{\Lh_-^{2'',2,2}\}
  -\tfrac{709664+3423c^+}{264600}\{\Lh_-^{2'''',2}\}
  -\tfrac{256(75800741+3326400c^+)}{444675}\{\Lh_-^{2,2,2,2,2}\}
 \nonumber\\[.15cm]
 &
  -\tfrac{4(704239157+10113642c^+)}{800415}\{\Lh_-^{2'',2,2,2}\}
  -\tfrac{39503531+347886c^+}{2134440}\{\Lh_-^{2'''',2,2}\}
  -\tfrac{19486333+61215c^+}{384199200}\{\Lh_-^{2'''''',2}\}
 \nonumber\\[.15cm]
 &
  -\tfrac{2153675\sqrt{2}}{693\sqrt{231}}\{\Wh^-\}
  -\tfrac{62\sqrt{2}(202343 + 462c^+)}{231\sqrt{231}}\{\Lh_-^{2,6}\}
  -\tfrac{32\sqrt{2}(1587589+6006c^+)}{105\sqrt{231}}\{\Lh_-^{2,2,6}\}
 \nonumber\\[.15cm]
 &
  -\tfrac{3708767+6006c^+}{770\sqrt{462}}\{\Lh_-^{2'',6}\}
\end{align}
and
\begin{align}
W^+\gast\Wh^-\!&\simeq
  \tfrac{1}{6}\{\I\}
  +2\{\Th^-\}
  +\tfrac{62}{5}\{\Lh_-^{2,2}\}
  +\tfrac{1112}{21}\{\Lh_-^{2,2,2}\}
  +\tfrac{17}{28}\{\Lh_-^{2'',2}\}
  +\tfrac{30656}{175}\{\Lh_-^{2,2,2,2}\}
  +\tfrac{8093}{2100}\{\Lh_-^{2'',2,2}\}
 \nonumber\\[.15cm]
 &
  +\tfrac{163}{12600}\{\Lh_-^{2'''',2}\}
  +\tfrac{36864}{77}\{\Lh_-^{2,2,2,2,2}\}
  +\tfrac{58376}{3465}\{\Lh_-^{2'',2,2,2}\}
  +\tfrac{251}{3080}\{\Lh_-^{2'''',2,2}\}
  +\tfrac{53}{332640}\{\Lh_-^{2'''''',2}\}
 \nonumber\\[.15cm]
 &
  +\tfrac{26\sqrt{2}}{3\sqrt{231}}\{\Wh^-\}
  +\tfrac{124\sqrt{2}}{\sqrt{231}}\{\Lh_-^{2,6}\}
  +\tfrac{416\sqrt{22}}{5\sqrt{21}}\{\Lh_-^{2,2,6}\}
  +\tfrac{13\sqrt{3}}{5\sqrt{154}}\{\Lh_-^{2'',6}\},
\end{align}
where the quasi-primary fields
$\Lh_\pm^{2,2}$, $\Lh_\pm^{2,2,2}$ and $\Lh_\pm^{2'',2}$ are given in (\ref{Lapm22}) and (\ref{La222p})-(\ref{La6m});
$\Lh_\pm^{2,2,2,2}$, $\Lh_\pm^{2'',2,2}$, $\Lh_\pm^{2'''',2}$, $\Lh_\pm^{2,2,2,2,2}$, $\Lh_\pm^{2'',2,2,2}$, $\Lh_\pm^{2'''',2,2}$, 
and $\Lh_\pm^{2'''''',2}$ in (\ref{Lh2222p})-(\ref{Lh21111112m}); while
\begin{align}
 \Lh_+^{2,6}&=\tfrac{1}{2}[(T^+\Wh^-)+(\Th^-W^+)]-\tfrac{3}{26}\pa^2\Wh^-,
\\[.15cm]
 \Lh_+^{2',6}&=\tfrac{1}{2}[(\pa T^+\Wh^-)+(\pa\Th^-W^+)]-\tfrac{1}{4}\pa\Lh_+^{2,6}-\tfrac{1}{91}\pa^3\Wh^-,
\\[.15cm]
 \Lh_+^{2,2,6}&=\tfrac{1}{3}[(T^+\Th^-\Wh^-)+(\Th^-T^+\Wh^-)+(\Th^-\Th^-W^+)]+\tfrac{1}{9}\pa\Lh_-^{2',6}
  -\tfrac{7}{68}\pa^2\Lh_-^{2,6},
\\[.15cm]
 \Lh_+^{2'',6}&=\tfrac{1}{2}[(\pa^2T^+\Wh^-)+(\pa^2\Th^-W^+)]
  -\tfrac{5}{9}\pa\Lh_+^{2',6}
  -\tfrac{5}{68}\pa^2\Lh_+^{2,6}
  -\tfrac{1}{546}\pa^4\Wh^-,
\end{align}
and
\be
\begin{array}{rclrclrcl}
 \Lh_-^{2,6}&\!\!\!\!=\!\!\!\!&(\Th^-\Wh^-),
 &\quad
 \Lh_-^{2',6}&\!\!\!\!=\!\!\!\!&(\pa\Th^-\Wh^-)-\tfrac{1}{4}\pa\Lh_-^{2,6},
\\[.3cm]
 \Lh_-^{2,2,6}&\!\!\!\!=\!\!\!\!&(\Th^-\Th^-\Wh^-),
 &\quad
  \Lh_-^{2'',6}&\!\!\!\!=\!\!\!\!&(\pa^2\Th^-\Wh^-)-\tfrac{5}{9}\pa\Lh_-^{2',6}-\tfrac{5}{68}\pa^2\Lh_-^{2,6},
\end{array}
\ee
follow from (\ref{LhTAp})-(\ref{LhT3Am}).

\subsection{Galilean $W_4$ algebra}
\label{Sec:GW4}

The Galilean $W_4$ algebra $(W_4)_G$ is generated by the fields $\{T^+,W^+,U^+,\Th^-,\Wh^-,\Uh^-\}$.
The fields $\{T^+,\Th^-\}$ generate a subalgebra isomorphic to $\Vir_G$, while $\{W^+,\Wh^-\}$ and $\{U^+,\Uh^-\}$ 
are primary fields of conformal weight $3$ and $4$ with respect to the Virasoro generator $T^+$. 
The nontrivial star products not involving $T^+$ nor $\Th^-$ are given by
\begin{align}
 W^+\gast W^+\!&\simeq
  \tfrac{c^+}{3}\{\I\}
  +2\{T^+\} 
  +\tfrac{64}{5}\{\Lh_+^{2,2}\}
  +\tfrac{4\sqrt{7}}{\sqrt{15}}\{U^+\}
  -\tfrac{32(44+5c^+)}{25}\{\Lh_-^{2,2}\}
  +\tfrac{964}{5\sqrt{105}}\{\Uh^-\}
,
\\[.15cm]
 W^+\gast U^+\!&\simeq
  \tfrac{\sqrt{21}}{\sqrt{5}}\{W^+\}
  +\tfrac{104\sqrt{3}}{\sqrt{35}}\{\Lh_+^{2,3}\}
  +\tfrac{2\sqrt{21}}{\sqrt{5}}\{\Lh_+^{2',3}\}
 \nonumber\\[.15cm]
 &
  +\tfrac{241\sqrt{3}}{5\sqrt{35}}\{\Wh^-\}
  -\tfrac{52\sqrt{3}(899+35c^+)}{35\sqrt{35}}\{\Lh_-^{2,3}\}
  -\tfrac{\sqrt{3}(-101+35c^+)}{5\sqrt{35}}\{\Lh_-^{2',3}\}
,
\\[.15cm]
 U^+\gast U^+\!&\simeq
  \tfrac{c^+}{4}\{\I\}
  +2\{T^+\}
  +\tfrac{84}{5}\{\Lh_+^{2,2}\}
  +\tfrac{2592}{35}\{\Lh_+^{2,2,2}\} 
  +\tfrac{57}{70}\{\Lh_+^{2'',2}\}
  +\tfrac{225}{7}\{\Lh_+^{3,3}\}
  -\tfrac{3\sqrt{3}}{\sqrt{35}}\{U^+\}
 \nonumber\\[.15cm]
 &
  -\tfrac{8\sqrt{21}}{\sqrt{5}}\{\Lh_+^{2,4}\}
  -\tfrac{42(44+5c^+)}{25}\{\Lh_-^{2,2}\}
  -\tfrac{192(7216+315c^+)}{1225}\{\Lh_-^{2,2,2}\}
  -\tfrac{3(89252+665c^+)}{4900}\{\Lh_-^{2'',2}\}
 \nonumber\\[.15cm]
 &
  -\tfrac{45(972+35c^+)}{98}\{\Lh_-^{3,3}\}
  +\tfrac{2907\sqrt{3}}{35\sqrt{35}}\{\Uh^-\}
  +\tfrac{4\sqrt{3}(-101+35c^+)}{5\sqrt{35}}\{\Lh_-^{2,4}\}
,
\end{align}
and
\begin{align}
 W^+\gast\Wh^-\!&\simeq
  \tfrac{1}{3}\{\I\}+ 2\{\Th^-\}
  +\tfrac{32}{5}\{\Lh_-^{2,2}\}
  +\tfrac{4\sqrt{7}}{\sqrt{15}}\{\Uh^-\},
\\[.15cm]
 W^+\gast\Uh^-\!&\simeq
  \tfrac{\sqrt{21}}{\sqrt{5}}\{\Wh^-\}
  +\tfrac{52\sqrt{3}}{\sqrt{35}}\{\Lh_-^{2,3}\}
  +\tfrac{\sqrt{21}}{\sqrt{5}}\{\Lh_-^{2',3}\},
\\[.15cm]
 U^+\gast\Uh^-\!&\simeq
  \tfrac{1}{4}\{\I\}+2\{\Th^-\}
  +\tfrac{42}{5}\{\Lh_-^{2,2}\}
  +\tfrac{864}{35}\{\Lh_-^{2,2,2}\}
  +\tfrac{57}{140}\{\Lh_-^{2'',2}\} 
  +\tfrac{225}{14}\{\Lh_-^{3,3}\}
 \nonumber\\[.15cm]
 &
  -\tfrac{3\sqrt{3}}{\sqrt{35}}\{\Uh^-\}
  -\tfrac{4\sqrt{21}}{\sqrt{5}}\{\Lh_-^{2,4}\}
,
\end{align}
where the quasi-primary fields $\Lh_\pm^{2,2}$, $\Lh_\pm^{2,2,2}$ and $\Lh_\pm^{2'',2}$ 
are given in (\ref{Lapm22}) and (\ref{La222p})-(\ref{La6m});
\begin{align}
 \Lh_+^{2,3}&=\tfrac{1}{2}[(T^+\Wh^-)+(\Th^-W^+)]-\tfrac{3}{14}\pa^2\Wh^-,
\\[.15cm]
 \Lh_+^{2',3}&=\tfrac{1}{2}[(\pa T^+\Wh^-)+(\pa\Th^-W^+)]-\tfrac{2}{5}\pa\Lh_+^{2,3}-\tfrac{1}{28}\pa^3\Wh^-,
\\[.15cm]
 \Lh_+^{2,4}&=\tfrac{1}{2}[(T^+\Uh^-)+(\Th^-U^+)]-\tfrac{1}{6}\pa^2\Uh^-,
\end{align}
and
\be
 \Lh_-^{2,3}=(\Th^-\Wh^-),\qquad
 \Lh_-^{2',3}=(\pa\Th^-\Wh^-)-\tfrac{2}{5}\pa\Lh_-^{2,3},\qquad
 \Lh_-^{2,4}=(\Th^-\Uh^-)
\ee
follow from (\ref{LhTAp})-(\ref{LhT1Ap}) and (\ref{LhTAm})-(\ref{LhT1Am}); while
\be
 \Lh_+^{3,3}=(W^+\Wh^-)
  -\tfrac{8}{9}\pa^2\Lh_-^{2,2}
  -\tfrac{\sqrt{35}}{9\sqrt{3}}\pa^2\Uh^-
  -\tfrac{1}{84}\pa^4\Th^-,\qquad
 \Lh_-^{3,3}=(\Wh^-\Wh^-).
\ee

\subsection{Galilean $W_5$ algebra}
\label{Sec:GW5}

The Galilean $W_5$ algebra $(W_5)_G$ is generated by the fields $\{T^+,W^+,U^+,V^+,\Th^-,\Wh^-,\Uh^-,\Vh^-\}$.
The fields $\{T^+,\Th^-\}$ generate a subalgebra isomorphic to $\Vir_G$, while $\{W^+,\Wh^-\}$, $\{U^+,\Uh^-\}$ 
and $\{V^+,\Vh^-\}$ are 
primary fields of conformal weight $3$, $4$ and $5$ with respect to the Virasoro generator $T^+$. 
The nontrivial star products not involving $T^+$ nor $\Th^-$ are given by
\begin{align}
 W^+\gast W^+\!&\simeq
  \tfrac{c^+}{3}\{\I\}
  +2\{T^+\} 
  +\tfrac{64}{5}\{\Lh_+^{2,2}\}
  +\tfrac{32}{\sqrt{105}}\{U^+\}
  -\tfrac{32(44+5c^+)}{25}\{\Lh_-^{2,2}\}
  +\tfrac{4064\sqrt{3}}{35\sqrt{35}}\{\Uh^-\}
, \\[.15cm]
 W^+\gast U^+\!&\simeq
  \tfrac{8\sqrt{3}}{\sqrt{35}}\{W^+\}
  +\tfrac{832\sqrt{3}}{7\sqrt{35}}\{\Lh_+^{2,3}\}
  +\tfrac{16\sqrt{3}}{\sqrt{35}}\{\Lh_+^{2',3}\}
  +\tfrac{5\sqrt{15}}{7}\{V^+\} 
  +\tfrac{3048\sqrt{3}}{35\sqrt{35}}\{\Wh^-\}
 \nonumber\\[.15cm]
 &
  -\tfrac{416\sqrt{3}(759+35c^+)}{245\sqrt{35}}\{\Lh_-^{2,3}\}
  -\tfrac{8\sqrt{3}(-241+35c^+)}{35\sqrt{35}}\{\Lh_-^{2',3}\}
  +\tfrac{256\sqrt{5}}{7\sqrt{3}}\{\Vh^-\}
,\\[.15cm]
 W^+\gast V^+\!&\simeq
  \tfrac{48}{\sqrt{7}}\{\Lh_+^{3,3}\}
  +\tfrac{4\sqrt{15}}{7}\{U^+\}
  +\tfrac{64\sqrt{15}}{7}\{\Lh_+^{2,4}\}
  +\tfrac{64\sqrt{3}}{7\sqrt{5}}\{\Lh_+^{2',4}\}
  -\tfrac{24448}{35\sqrt{7}}\{\Lh_-^{2,2,2}\}
  -\tfrac{516}{35\sqrt{7}}\{\Lh_-^{2'',2}\}
 \nonumber\\[.15cm]
 &
  -\tfrac{8(1700+21c^+)}{7\sqrt{7}}\{\Lh_-^{3,3}\}
  +\tfrac{1024}{7\sqrt{15}}\{\Uh^-\}
  -\tfrac{32\sqrt{3}(272+5c^+)}{7\sqrt{5}}\{\Lh_-^{2,4}\}
  -\tfrac{32(-124+15c^+)}{35\sqrt{15}}\{\Lh_-^{2',4}\}
,
\end{align}
\begin{align}
U^+\gast U^+\!&\simeq
  \tfrac{c^+}{4}\{\I\}
  +2\{T^+\} 
  +\tfrac{84}{5}\{\Lh_+^{2,2}\}
  +\tfrac{2592}{35}\{\Lh_+^{2,2,2}\}
  +\tfrac{57}{70}\{\Lh_+^{2'',2}\}
  +45\{\Lh_+^{3,3}\}
  +\tfrac{3\sqrt{3}}{\sqrt{35}}\{U^+\}
 \nonumber\\[.15cm]
 &
  +\tfrac{8\sqrt{21}}{\sqrt{5}}\{\Lh_+^{2,4}\}
  -\tfrac{42(44+5c^+)}{25}\{\Lh_-^{2,2}\}
  -\tfrac{144(11213+420c^+)}{1225}\{\Lh_-^{2,2,2}\}
  -\tfrac{3(95702+133c^+)}{245}\{\Lh_-^{2'',2}\}
  \nonumber\\[.15cm]
 &
  -\tfrac{9(206+5c^+)}{2}\{\Lh_-^{3,3}\}
  +\tfrac{10593\sqrt{3}}{35\sqrt{35}}\{\Uh^-\}
  -\tfrac{4\sqrt{3}(2549+35c^+)}{5\sqrt{35}}\{\Lh_-^{2,4}\}
,\\[.15cm]
U^+\gast V^+\!&\simeq
  \tfrac{3\sqrt{15}}{7}\{W^+\} 
  +\tfrac{396\sqrt{15}}{49}\{\Lh_+^{2,3}\}
  +\tfrac{8\sqrt{15}}{7}\{\Lh_+^{2',3}\}
  +\tfrac{99\sqrt{3}}{49\sqrt{5}}\{\Lh_+^{2'',3}\}
  +\tfrac{12\sqrt{3}}{49\sqrt{5}}\{\Lh_+^{2''',3}\}
 \nonumber\\[.15cm]
 &
  +\tfrac{14688\sqrt{3}}{49\sqrt{5}}\{\Lh_+^{2,2,3}\}
  +\tfrac{3888\sqrt{3}}{49\sqrt{5}}\{\Lh_+^{2',2,3}\} 
  +\tfrac{148}{\sqrt{7}}\{\Lh_+^{3,4}\}
  +\tfrac{36}{\sqrt{7}}\{\Lh_+^{3',4}\}
  -\tfrac{11\sqrt{15}}{14\sqrt{7}}\{V^+\} 
  -\tfrac{116\sqrt{15}}{7\sqrt{7}}\{\Lh_+^{2,5}\}
 \nonumber\\[.15cm]
 &
  -\tfrac{44\sqrt{3}}{7\sqrt{35}}\{\Lh_+^{2',5}\}
  +\tfrac{256\sqrt{3}}{7\sqrt{5}}\{\Wh^-\} 
  -\tfrac{66\sqrt{3}(1628+105c^+)}{343\sqrt{5}}\{\Lh_-^{2,3}\}
  -\tfrac{4(-196+15c^+)}{7\sqrt{15}}\{\Lh_-^{2',3}\}
 \nonumber\\[.15cm]
 &
  -\tfrac{8\sqrt{3}(2325173+42840c^+)}{1715\sqrt{5}}\{\Lh_-^{2,2,3}\}
  -\tfrac{1982138+10395c^+}{3430\sqrt{15}}\{\Lh_-^{2'',3}\}
  -\tfrac{2\sqrt{3}(10952+105c^+)}{1715\sqrt{5}}\{\Lh_-^{2''',3}\}
 \nonumber\\[.15cm]
 &
  -\tfrac{288\sqrt{3}(4528+315c^+)}{1715\sqrt{5}}\{\Lh_-^{2',2,3}\}
  -\tfrac{2(48095+777c^+)}{21\sqrt{7}}\{\Lh_-^{3,4}\}
  -\tfrac{6(55+21c^+)}{7\sqrt{7}}\{\Lh_-^{3',4}\}
 \nonumber\\[.15cm]
 &
  +\tfrac{13319\sqrt{3}}{98\sqrt{35}}\{\Vh^-\} 
  +\tfrac{2\sqrt{3}(-10959+1015c^+)}{49\sqrt{35}}\{\Lh_-^{2,5}\}
  +\tfrac{2\sqrt{3}(9259+385c^+)}{245\sqrt{35}}\{\Lh_-^{2',5}\}
,
\end{align}
\begin{align}
 V^+\gast V^+\!&\simeq
  \tfrac{c^+}{5}\{\I\}
  +2\{T^+\} 
  +\tfrac{104}{5}\{\Lh_+^{2,2}\}
  +\tfrac{3936}{35}\{\Lh_+^{2,2,2}\}
  +\tfrac{71}{70}\{\Lh_+^{2'',2}\}
  +\tfrac{73728}{175}\{\Lh_+^{2,2,2,2}\}
  +\tfrac{1424}{175}\{\Lh_+^{2'',2,2}\}
 \nonumber\\[.15cm]
 &
  +\tfrac{34}{1575}\{\Lh_+^{2'''',2}\}
  +\tfrac{181}{7}\{\Lh_+^{3,3}\}
  +\tfrac{44}{21}\{\Lh_+^{3'',3}\}
  +\tfrac{2976}{7}\{\Lh_+^{2,3,3}\}
  -\tfrac{22\sqrt{3}}{7\sqrt{35}}\{U^+\}
  -\tfrac{1496}{7\sqrt{105}}\{\Lh_+^{2,4}\}
 \nonumber\\[.15cm]
 &
  -\tfrac{22016\sqrt{3}}{35\sqrt{35}}\{\Lh_+^{2,2,4}\}
  -\tfrac{116}{15\sqrt{105}}\{\Lh_+^{2'',4}\}
  +\tfrac{896}{15}\{\Lh_+^{4,4}\}
  -\tfrac{120}{\sqrt{7}}\{\Lh_+^{3,5}\}
  -\tfrac{52(44+5c^+)}{25}\{\Lh_-^{2,2}\}
 \nonumber\\[.15cm]
 &
  -\tfrac{16(317033+17220c^+)}{3675}\{\Lh_-^{2,2,2}\}
  -\tfrac{371598+2485c^+}{4900}\{\Lh_-^{2'',2}\}
  -\tfrac{18432(4441+105c^+)}{6125}\{\Lh_-^{2,2,2,2}\}
 \nonumber\\[.15cm]
 &
  -\tfrac{8(5820569+37380c^+)}{55125}\{\Lh_-^{2'',2,2}\}
  -\tfrac{960196+1785c^+}{165375}\{\Lh_-^{2'''',2}\}
  -\tfrac{-17162+3801c^+}{294}\{\Lh_-^{3,3}\}
 \nonumber\\[.15cm]
 &
  -\tfrac{2(39164+231c^+)}{441}\{\Lh_-^{3'',3}\}
  -\tfrac{32(80725+1302c^+)}{147}\{\Lh_-^{2,3,3}\}
  +\tfrac{26638\sqrt{3}}{245\sqrt{35}}\{\Uh^-\}
 \nonumber\\[.15cm]
 &
  +\tfrac{4(506789+19635c^+)}{735\sqrt{105}}\{\Lh_-^{2,4}\}
  +\tfrac{512(210023+9030c^+)}{3675\sqrt{105}}\{\Lh_-^{2,2,4}\}
  +\tfrac{2(-7343+435c^+)}{225\sqrt{105}}\{\Lh_-^{2'',4}\}
 \nonumber\\[.15cm]
 &
  -\tfrac{64(5624+105c^+)}{225}\{\Lh_-^{4,4}\}
  +\tfrac{20(55+21c^+)}{7\sqrt{7}}\{\Lh_-^{3,5}\}
,
\end{align}
\begin{align}
 W^+\gast\Wh^-\!&\simeq
  \tfrac{1}{3}\{\I\}
  +2\{\Th^-\} 
  +\tfrac{32}{5}\{\Lh_-^{2,2}\}
  +\tfrac{32}{\sqrt{105}}\{\Uh^-\}
,\\[.15cm]
 W^+\gast \Uh^-\!&\simeq
  \tfrac{8\sqrt{3}}{\sqrt{35}}\{\Wh^-\}
  +\tfrac{416\sqrt{3}}{7\sqrt{35}}\{\Lh_-^{2,3}\}
  +\tfrac{8\sqrt{3}}{\sqrt{35}}\{\Lh_-^{2',3}\}
  +\tfrac{5\sqrt{15}}{7}\{\Vh^-\}
,\\[.15cm]
 W^+\gast\Vh^-\!&\simeq
  \tfrac{24}{\sqrt{7}}\{\Lh_-^{3,3}\}
  +\tfrac{4\sqrt{15}}{7}\{\Uh^-\}
  +\tfrac{32\sqrt{15}}{7}\{\Lh_-^{2,4}\}
  +\tfrac{32\sqrt{3}}{7\sqrt{5}}\{\Lh_-^{2',4}\}
,
\end{align}
\begin{align}
 U^+\gast\Uh^-\!&\simeq
  \tfrac{1}{4}\{\I\}
  +2\{\Th^{-}\} 
  +\tfrac{42}{5}\{\Lh_-^{2,2}\}
  +\tfrac{864}{35}\{\Lh_-^{2,2,2}\}
  +\tfrac{57}{140}\{\Lh_-^{2'',2}\}
 \nonumber\\[.15cm]
 &
  +\tfrac{45}{2}\{\Lh_-^{3,3}\}
  +\tfrac{3\sqrt{3}}{\sqrt{35}}\{\Uh^-\}
  +\tfrac{4\sqrt{21}}{\sqrt{5}}\{\Lh_-^{2,4}\}
,\\[.15cm]
 U^+\gast\Vh^-\!&\simeq
  \tfrac{3\sqrt{15}}{7}\{\Wh^-\} 
  +\tfrac{198\sqrt{15}}{49}\{\Lh_-^{2,3}\} 
  +\tfrac{4\sqrt{15}}{7}\{\Lh_-^{2',3}\}
  +\tfrac{4896\sqrt{3}}{49\sqrt{5}}\{\Lh_-^{2,2,3}\}
  +\tfrac{99\sqrt{3}}{98\sqrt{5}}\{\Lh_-^{2'',3}\}
  +\tfrac{6\sqrt{3}}{49\sqrt{5}}\{\Lh_-^{2''',3}\}
 \nonumber\\[.15cm]
 &
  +\tfrac{1296\sqrt{3}}{49\sqrt{5}}\{\Lh_-^{2',2,3}\}
  +\tfrac{74}{\sqrt{7}}\{\Lh_-^{3,4}\}
  +\tfrac{18}{\sqrt{7}}\{\Lh_-^{3',4}\}
  -\tfrac{11\sqrt{15}}{14\sqrt{7}}\{\Vh^-\}
  -\tfrac{58\sqrt{15}}{7\sqrt{7}}\{\Lh_-^{2,5}\}
  -\tfrac{22\sqrt{3}}{7\sqrt{35}}\{\Lh_-^{2',5}\},
\end{align}
and
\begin{align}
 V^+\gast\Vh^-\!&\simeq
  \tfrac{1}{5}\{\I\}
  +2\{\Th^-\} 
  +\tfrac{52}{5}\{\Lh_-^{2,2}\}
  +\tfrac{1312}{35}\{\Lh_-^{2,2,2}\}
  +\tfrac{71}{140}\{\Lh_-^{2'',2}\}
  +\tfrac{18432}{175}\{\Lh_-^{2,2,2,2}\}
  +\tfrac{1424}{525}\{\Lh_-^{2'',2,2}\}
 \nonumber\\[.15cm]
 &
  +\tfrac{17}{1575}\{\Lh_-^{2'''',2}\}
  +\tfrac{181}{14}\{\Lh_-^{3,3}\}
  +\tfrac{992}{7}\{\Lh_-^{2,3,3}\}
  +\tfrac{22}{21}\{\Lh_-^{3'',3}\}
  -\tfrac{22\sqrt{3}}{7\sqrt{35}}\{\Uh^-\} 
  -\tfrac{748}{7\sqrt{105}}\{\Lh_-^{2,4}\}
 \nonumber\\[.15cm]
 &
  -\tfrac{22016}{35\sqrt{105}}\{\Lh_-^{2,2,4}\}
  -\tfrac{58}{15\sqrt{105}}\{\Lh_-^{2'',4}\}
  +\tfrac{448}{15}\{\Lh_-^{4,4}\}
  -\tfrac{60}{\sqrt{7}}\{\Lh_-^{3,5}\}
,
\end{align}
where the quasi-primary fields
$\Lh_\pm^{2,2}$, $\Lh_\pm^{2,2,2}$ and $\Lh_\pm^{2'',2}$ are given in (\ref{Lapm22}) and (\ref{La222p})-(\ref{La6m});
$\Lh_\pm^{2,2,2,2}$, $\Lh_\pm^{2'',2,2}$ and $\Lh_\pm^{2'''',2}$ in (\ref{Lh2222p})-(\ref{Lh211112p}) and
(\ref{Lh2222m})-(\ref{Lh211112m});
\begin{align}
 \Lh_+^{2,3}&=\tfrac{1}{2}[(T^+\Wh^-)+(\Th^-W^+)]
  -\tfrac{3}{14}\pa^2\Wh^-,
\\[.15cm]
 \Lh_+^{2',3}&=\tfrac{1}{2}[(\pa T^+\Wh^-)+(\pa\Th^-W^+)]
  -\tfrac{2}{5}\pa\Lh_+^{2,3}
  -\tfrac{1}{28}\pa^3\Wh^-,
\\[.15cm]
 \Lh_+^{2,2,3}&=\tfrac{1}{3}[(T^+\Th^-\Wh^-)+(\Th^-T^+\Wh^-)+(\Th^-\Th^-W^+)]
  +\tfrac{1}{6}\pa\Lh_-^{2',3}
  -\tfrac{8}{55}\pa^2\Lh_-^{2,3},
\\[.15cm]
 \Lh_+^{2'',3}&=\tfrac{1}{2}[(\pa^2T^+\Wh^-)+(\pa^2\Th^-W^+)]
  -\tfrac{5}{6}\pa\Lh_+^{2',3}
  -\tfrac{2}{11}\pa^2\Lh_+^{2,3}
  -\tfrac{5}{504}\pa^4\Wh^-,
\\[.15cm]
 \Lh_+^{2',2,3}&=\tfrac{1}{3}[(\pa T^+\Th^-\Wh^-)+(\pa\Th^-T^+\Wh^-)+(\pa\Th^-\Th^-W^+)]
  -\tfrac{2}{7}\pa\Lh_+^{2,2,3}
  +\tfrac{1}{7}\pa\Lh_-^{2'',3}
\nonumber\\[.15cm]
&
  +\tfrac{1}{78}\pa^2\Lh_-^{2',3}
  -\tfrac{14}{495}\pa^3\Lh_-^{2,3},
\\[.15cm]
 \Lh_+^{2''',3}&=\tfrac{1}{2}[(\pa^3T^+\Wh^-)+(\pa^3\Th^-W^+)]
  -\tfrac{9}{7}\pa\Lh_+^{2'',3}
  -\tfrac{15}{26}\pa^2\Lh_+^{2',3}
  -\tfrac{1}{11}\pa^3\Lh_+^{2,3}
  -\tfrac{1}{280}\pa^5\Wh^-,
\end{align}
\begin{align}
 \Lh_+^{2,4}&=\tfrac{1}{2}[(T^+\Uh^-)+(\Th^-U^+)]
  -\tfrac{1}{6}\pa^2\Uh^-,
\\[.15cm]
 \Lh_+^{2',4}&=\tfrac{1}{2}[(\pa T^+\Uh^-)+(\pa\Th^-U^+)]
  -\tfrac{1}{3}\pa\Lh_+^{2,4}
  -\tfrac{1}{45}\pa^3\Uh^-,
\\[.15cm]
 \Lh_+^{2,2,4}&=\tfrac{1}{3}[(T^+\Th^-\Uh^-)+(\Th^-T^+\Uh^-)+(\Th^-\Th^-U^+)]
  +\tfrac{1}{7}\pa\Lh_-^{2',4}
  -\tfrac{5}{39}\pa^2\Lh_-^{2,4},
\\[.15cm]
 \Lh_+^{2'',4}&=\tfrac{1}{2}[(\pa^2T^+\Uh^-)+(\pa^2\Th^-U^+)]
  -\tfrac{5}{7}\pa\Lh_+^{2',4}
  -\tfrac{5}{39}\pa^2\Lh_+^{2,4}
  -\tfrac{1}{198}\pa^4\Uh^-,
\\[.15cm]
 \Lh_+^{2,5}&=\tfrac{1}{2}[(T^+\Vh^-)+(\Th^-V^+)]
  -\tfrac{3}{22}\pa^2\Vh^-,
\\[.15cm]
 \Lh_+^{2',5}&=\tfrac{1}{2}[(\pa T^+\Vh^-)+(\pa\Th^-V^+)]
  -\tfrac{2}{7}\pa\Lh_+^{2,5}
  -\tfrac{1}{66}\pa^3\Vh^-,
\end{align}
and
\be
\begin{array}{rclrclrcl}
\Lh_-^{2,3}&\!\!\!\!=\!\!\!\!&(\Th^-\Wh^-),
 &\quad
 \Lh_-^{2',3}&\!\!\!\!=\!\!\!\!&(\pa\Th^-\Wh^-)
  -\tfrac{2}{5}\pa\Lh_-^{2,3},
\\[.3cm]
 \Lh_-^{2,2,3}&\!\!\!\!=\!\!\!\!&(\Th^-\Th^-\Wh^-),
 &\quad
 \Lh_-^{2'',3}&\!\!\!\!=\!\!\!\!&(\pa^2\Th^-\Wh^-)
  -\tfrac{5}{6}\pa\Lh_-^{2',3}
  -\tfrac{2}{11}\pa^2\Lh_-^{2,3},
\\[.3cm]
 \Lh_-^{2',2,3}&\!\!\!\!=\!\!\!\!&(\pa\Th^-\Th^-\Wh^-)
  -\tfrac{2}{7}\pa\Lh_-^{2,2,3},
 &\quad
 \Lh_-^{2''',3}&\!\!\!\!=\!\!\!\!&(\pa^3\Th^-\Wh^-)
  -\tfrac{9}{7}\pa\Lh_-^{2'',3}
  -\tfrac{15}{26}\pa^2\Lh_-^{2',3}
  -\tfrac{1}{11}\pa^3\Lh_-^{2,3},
\\[.3cm]
 \Lh_-^{2,4}&\!\!\!\!=\!\!\!\!&(\Th^-\Uh^-),
 &\quad
 \Lh_-^{2',4}&\!\!\!\!=\!\!\!\!&(\pa\Th^-\Uh^-)
  -\tfrac{1}{3}\pa\Lh_-^{2,4},
\\[.3cm]
 \Lh_-^{2,2,4}&\!\!\!\!=\!\!\!\!&(\Th^-\Th^-\Uh^-),
 &\quad
 \Lh_-^{2'',4}&\!\!\!\!=\!\!\!\!&(\pa^2\Th^-\Uh^-)
  -\tfrac{5}{7}\pa\Lh_-^{2',4}
  -\tfrac{5}{39}\pa^2\Lh_-^{2,4},
\\[.3cm]
 \Lh_-^{2,5}&\!\!\!\!=\!\!\!\!&(\Th^-\Vh^-),
 &\quad
 \Lh_-^{2',5}&\!\!\!\!=\!\!\!\!&(\pa\Th^-\Vh^-)
  -\tfrac{2}{7}\pa\Lh_-^{2,5},
\end{array}
\ee

follow from (\ref{LhTAp})-(\ref{LhT3Am});
while
\begin{align}
 \Lh_+^{3,3}&=(W^+\Wh^-)
  -\tfrac{8}{9}\pa^2\Lh_-^{2,2}
  -\tfrac{8\sqrt{5}}{9\sqrt{21}}\pa^2\Uh^-
  -\tfrac{1}{84}\pa^4\Th^-,
\\[.15cm]
 \Lh_+^{2,3,3}&=\tfrac{1}{3}[(T^+\Wh^-\Wh^-)+(\Th^-W^+\Wh^-)+(\Th^-\Wh^-T^+)]
  +\tfrac{16\sqrt{5}}{21\sqrt{21}}\pa\Lh_-^{2',4}
  -\tfrac{32\sqrt{5}}{117\sqrt{21}}\pa^2\Lh_-^{2,4}
  -\tfrac{1}{13}\pa^2\Lh_-^{3,3}
\nonumber\\[.15cm]
&
  -\tfrac{32}{117}\pa^2\Lh_-^{2,2,2}
  -\tfrac{1}{78}\pa^2\Lh_-^{2'',2}
  -\tfrac{1}{1188}\pa^4\Lh_-^{2,2},
\\[.15cm]
 \Lh_+^{3'',3}&=\tfrac{1}{2}[(\pa^2W^+\Wh^-)+(\pa^2\Wh^-W^+)]
  -\tfrac{7}{26}\pa^2\Lh_+^{3,3}
  -\tfrac{28}{495}\pa^4\Lh_-^{2,2}
  -\tfrac{4\sqrt{7}}{99\sqrt{15}}\pa^4\Uh^-
  -\tfrac{1}{2160}\pa^6\Th^-,
\\[.15cm]
 \Lh_+^{3,4}&=\tfrac{1}{2}[(W^+\Uh^-)+(\Wh^-U^+)]
  -\tfrac{2\sqrt{5}}{\sqrt{21}}\pa\Lh_-^{2',3}
  -\tfrac{416\sqrt{3}}{77\sqrt{35}}\pa^2\Lh_-^{2,3}
  -\tfrac{5\sqrt{15}}{77}\pa^2\Vh^-
  -\tfrac{\sqrt{5}}{126\sqrt{21}}\pa^4\Wh^-,
\\[.15cm]
 \Lh_+^{3',4}&=\tfrac{1}{2}[(\pa W^+\Uh^-)+(\pa\Wh^-U^+)]
  -\tfrac{3}{7}\pa\Lh_+^{3,4}
  -\tfrac{2\sqrt{15}}{13\sqrt{7}}\pa^2\Lh_-^{2',3}
  -\tfrac{208}{77\sqrt{105}}\pa^3\Lh_-^{2,3}
  -\tfrac{5\sqrt{5}}{154\sqrt{3}}\pa^3\Vh^-
\nonumber\\[.15cm]
&
  -\tfrac{1}{210\sqrt{105}}\pa^5\Wh^-,
\\[.15cm]
 \Lh_+^{3,5}&=\tfrac{1}{2}[(W^+\Vh^-)+(\Wh^-V^+)]
  -\tfrac{16\sqrt{15}}{49}\pa\Lh_-^{2',4}
  -\tfrac{20}{13\sqrt{7}}\pa^2\Lh_-^{3,3}
  -\tfrac{80\sqrt{5}}{91\sqrt{3}}\pa^2\Lh_-^{2,4}
  -\tfrac{\sqrt{5}}{924\sqrt{3}}\pa^4\Uh^-,
\\[.15cm]
 \Lh_+^{4,4}&=(U^+\Uh^-)
  -\tfrac{216}{65}\pa^2\Lh_-^{2,2,2}
  -\tfrac{57}{1040}\pa^2\Lh_-^{2'',2}
  -\tfrac{315}{104}\pa^2\Lh_-^{3,3}
  -\tfrac{7\sqrt{21}}{13\sqrt{5}}\pa^2\Lh_-^{2,4}
  -\tfrac{49}{1320}\pa^4\Lh_-^{2,2}
\nonumber\\[.15cm]
&
  -\tfrac{\sqrt{7}}{176\sqrt{15}}\pa^4\Uh^-
  -\tfrac{1}{4320}\pa^6\Th^-,
\end{align}
and
\be
\begin{array}{rclrclrclrcl}
 \Lh_-^{3,3}&\!\!\!\!=\!\!\!\!&(\Wh^-\Wh^-),
 &\quad
 \Lh_-^{2,3,3}&\!\!\!\!=\!\!\!\!&(\Th^-\Wh^-\Wh^-),
 &\quad
 \Lh_-^{3'',3}&\!\!\!\!=\!\!\!\!&(\pa^2\Wh^-\Wh^-)
  -\tfrac{7}{26}\pa^2\Lh_-^{3,3},
\\[.3cm]
 \Lh_-^{3,4}&\!\!\!\!=\!\!\!\!&(\Wh^-\Uh^-),
 &\quad
 \Lh_-^{3',4}&\!\!\!\!=\!\!\!\!&(\pa\Wh^-\Uh^-)
  -\tfrac{3}{7}\pa\Lh_-^{3,4},
 &\quad
 \Lh_-^{3,5}&\!\!\!\!=\!\!\!\!&(\Wh^-\Vh^-),
\\[.3cm]
 \Lh_-^{4,4}&\!\!\!\!=\!\!\!\!&(\Uh^-\Uh^-).
 &\quad
 &&&&
\end{array}
\ee

%

\end{document}